\documentclass[nofootinbib,preprint,preprintnumbers,a4paper,11pt]{revtex4}
\usepackage{epsfig}
\usepackage{footnote}
\usepackage{ulem}
\usepackage{color}
\usepackage{array}
\usepackage{amssymb}
\usepackage{amsmath}
\usepackage{graphicx}
\usepackage{subfigure}
\usepackage{longtable}
\usepackage{verbatim}
\usepackage{amsfonts}
\usepackage{hyperref}
\usepackage{multirow}
\usepackage{cancel}

\begin{document}

\title{A self-consistent framework of topological amplitude \\ and its $SU(N)$ decomposition}

\author{Di Wang$^{1,2}$\footnote{Corresponding author.}}\email{wangdi@hunnu.edu.cn}
\author{Cai-Ping Jia$^{2}$}\email{jiacp17@lzu.edu.cn}
\author{Fu-Sheng Yu$^{2}$\footnote{Corresponding author.}}\email{yufsh@lzu.edu.cn}

\address{%
$^1$Department of Physics, Hunan Normal University, Changsha 410081, People's Republic of China \\
$^2$School of Nuclear Science and Technology,  Lanzhou University,  Lanzhou 730000,  People's Republic of China
}

\begin{abstract}

We propose a systematic theoretical framework for the topological amplitudes of the heavy meson decays and their $SU(N)$ decomposition.
In the framework, the topologies are expressed in invariant tensors and classified into tree- and penguin-operator-induced diagrams according to which four-quark operators, tree or penguin, being inserted into their effective weak vertexes.
The number of possible topologies contributing to one type of decay can be counted by permutations and combinations.
The Wigner-Eckhart theorem ensures the topological amplitudes under flavor symmetry are the same for different decay channels.
By decomposing the four-quark operators into irreducible representations of $SU(N)$ group, one can get the $SU(N)$ irreducible amplitudes.
Taking the $D\to PP$ decay ($P$ denoting a pseudoscalar meson) with $SU(3)_F$ symmetry as an example, we present our framework in detail.
The linear correlation of topologies in the $SU(3)_F$ limit is clarified in group theory.
It is found there are only nine independent topologies in all tree- and penguin-operator-induced diagrams contributing to the $D\to PP$ decays in the Standard Model.
If a large quark-loop diagram, named $T^{LP}$, is assumed, the large $\Delta A_{CP}$ and the very different $D^0\to K^+K^-$ and $D^0\to \pi^+\pi^-$ branching fractions can be explained with a normal $U$-spin breaking.
Moreover, our framework provides a simple way to analyze the $SU(N)$ breaking effects.
The linear $SU(3)_F$ breaking and the high order $U$-spin breaking in charm decays are re-investigated in our framework, which are consistent with literature.
Analogous to the degeneracy and splitting of energy levels, we propose the concepts of degeneracy and splitting of topologies to describe the flavor symmetry breaking effects in decay.
As applications, we analyze the strange-less $D$ decays in $SU(3)_F$ symmetry breaking into Isospin symmetry and the charm-less $B$ decays in $SU(4)_F$ symmetry breaking into $SU(3)_F$ symmetry.

\end{abstract}

\maketitle
{{\tableofcontents}}

\newpage
\section{Introduction}

Heavy quark non-leptonic decays provide an ideal platform to test the Standard Model (SM) and search for new physics.
A tremendous amount of data on the heavy hadron (especially $B/D$ meson) decays have been collected by experiments in the last few decades \cite{Tanabashi:2018oca}.
In particular, the LHCb Collaboration observed the $CP$ violation in the charm
sector with $5.3\sigma$ in 2019 \cite{Aaij:2019kcg}.
It is a milestone of heavy flavor physics since it fills the last piece of the puzzle of the Kobayashi-Maskawa (KM) mechanism \cite{Cabibbo:1963yz,Kobayashi:1973fv}.
In theory, several QCD-inspired approaches are established to calculate the non-leptonic $B$ meson decays, such as
QCD factorization (QCDF) \cite{Beneke:1999br,Beneke:2000ry,Beneke:2003zv,Beneke:2001ev}, perturbative QCD approach (PQCD) \cite{Keum:2000ph,Keum:2000wi,Lu:2000em,Lu:2000hj}, and soft-collinear effective theory (SCET) \cite{Bauer:2001cu,Bauer:2001yt}.
However, the QCD-inspired approaches do not work well in the $D$ meson decays because the expansion parameters $\alpha_s(m_c)$ and $\Lambda_{\rm QCD}/m_c$ are bigger than the ones in the $B$ meson decays.
It remains to be seen if the heavy quark expansion (HQE) can be applied to the charm sector \cite{Lenz:2013aua,Lenz:2015dra}.

An alternative way to investigate the heavy meson decays is the flavor symmetry analysis.
This method bypasses form the dynamic details, widely used in studying charm/bottom meson \cite{Rizzo:1980yh,ChauWang:1980ex,Bhattacharya:2012ah,Yu:2017oky,Grinstein:1996us,Wang:2017ksn,Gronau:1994bn,ReyLeLorier:2011ww,
Bhattacharya:2013cla,Bertholet:2018tmx,Zeppenfeld:1980ex,Gaillard:1974mw,
Einhorn:1975fw,Kingsley:1975fe,Wang:1979dx,Eilam:1979mn,FAT1,FAT2,Wang:2017hxe,
Zhou:2016jkv,Zhou:2015jba,Bhattacharya:2008ke,Bhattacharya:2008ss,Savage:1989ub,
Deshpande:1994ii,He:1998rq,He:2000ys,Hsiao:2015iiu,Chau:1990ay,Gronau:1994rj,
Gronau:1995hm,Cheng:2014rfa,Chau:1982da,Grossman:2012ry,Chau:1986du,Chau:1987tk,
Cheng:2012wr,Cheng:2012xb,Cheng:2016ejf,Bhattacharya:2009ps,Cheng:2010ry}, baryon \cite{Roy:2020nyx,Roy:2019cky,Geng:2020zgr,Dery:2020lbc,Geng:2019xbo,Geng:2018upx,Geng:2019bfz,Wang:2019dls,Wang:2018utj,Shi:2017dto,
Wang:2017azm,Yu:2017zst,Geng:2018rse,Grossman:2018ptn,Zhao:2018mov,Geng:2018bow,
Jiang:2018iqa,Cheng:2018hwl,Geng:2018plk,Wang:2017gxe,Geng:2017mxn,Geng:2017esc,
Lu:2016ogy,Jia:2019zxi,Cen:2019ims} and even stable tetraquark \cite{Li:2018bkh,Xing:2018bqt,Xing:2019hjg,Xing:2019wil} decays.
There are two popular approaches based on the flavor symmetry.
One is the topological diagram amplitude (TDA) \cite{Rizzo:1980yh,Bhattacharya:2012ah,Wang:2017ksn,Yu:2017oky,
Zeppenfeld:1980ex,FAT1,FAT2,Wang:2017hxe,Zhou:2016jkv,Zhou:2015jba,
Bhattacharya:2008ke,Bhattacharya:2008ss,
Chau:1990ay,Gronau:1994rj,Gronau:1995hm,Cheng:2014rfa,Chau:1982da,
Chau:1986du,Chau:1987tk,Cheng:2012wr,Cheng:2012xb,Cheng:2016ejf,
Bhattacharya:2009ps,Cheng:2010ry} approach, in which the topological diagrams are classified according to the topologies in the flavor flow of weak decay diagrams, with all strong interaction effects induced implicitly.
It is intuitive and helpful for understanding the internal dynamics of hadron decays, providing a framework in which we cannot only do the model-dependent data analysis but also make evaluations of theoretical model calculations.
The other method is the $SU(3)$ irreducible representation amplitude (IRA) \cite{Grinstein:1996us,Savage:1989ub,Deshpande:1994ii,He:1998rq,He:2000ys,Hsiao:2015iiu,Grossman:2012ry} approach which is blind to the dynamic mechanics.
The $SU(3)$ irreducible representation amplitudes are expressed in the tensor form \cite{Savage:1989ub} and the Wigner-Eckhart theorem \cite{Eckart30,Wigner59} ensures that there is one constant for each invariant tensor.
Both the TDA and IRA approaches can include the flavor symmetry breaking effects.
The first order flavor $SU(3)$ breaking has been analyzed in the irreducible amplitude \cite{Grossman:2012ry} and the topological amplitude \cite{Muller:2015rna,Muller:2015lua} approaches.

The TDA and IRA approaches  seem to be equivalent in the $SU(3)_F$ limit. The equivalence between them was discussed as early as in 1980 \cite{Rizzo:1980yh,ChauWang:1980ex,Zeppenfeld:1980ex}, and followed by other literature such as Refs.~\cite{Chau:1995gk,Paz:2002ev}.
But in some literature, such as Refs.~\cite{Chau:1987tk,Chau:1995gk}, the relation between topological amplitudes and irreducible amplitudes is extracted by expanding the decay amplitudes of some channels in the TDA and IRA approaches and comparing them.
The equivalence of TDA and IRA is used like a priori presumption rather than a derived conclusion.
An instructive attempt to solve the relation between the topological amplitude and the $SU(3)$ irreducible amplitude was done in Refs.~\cite{He:2018php,He:2018joe}.
In their work, a bridge between topological diagrams and invariant tensors constructed by the four-quark operators and the initial and final states was built.
It shows that the difference between the $SU(3)$ irreducible amplitude and the topological amplitude is whether the four-quark operators are decomposed into the $SU(3)$ irreducible representations or not, and the equivalence relation between them can be derived.
However, there are some mistakes in the $SU(3)$ decomposition, resulting in an incorrect relation between  topological and irreducible amplitudes \cite{He:2018php,He:2018joe}.
Their treatment of charm-quark loop and classification of topologies in the $B$ meson decays are ambiguous.
They did not explain why not all topologies are independent.
A complete and self-consistent framework of the topologies and their $SU(N)$ decomposition for the heavy hadron decays has not been established yet.

\begin{figure}
  \centering
  \includegraphics[width=12cm]{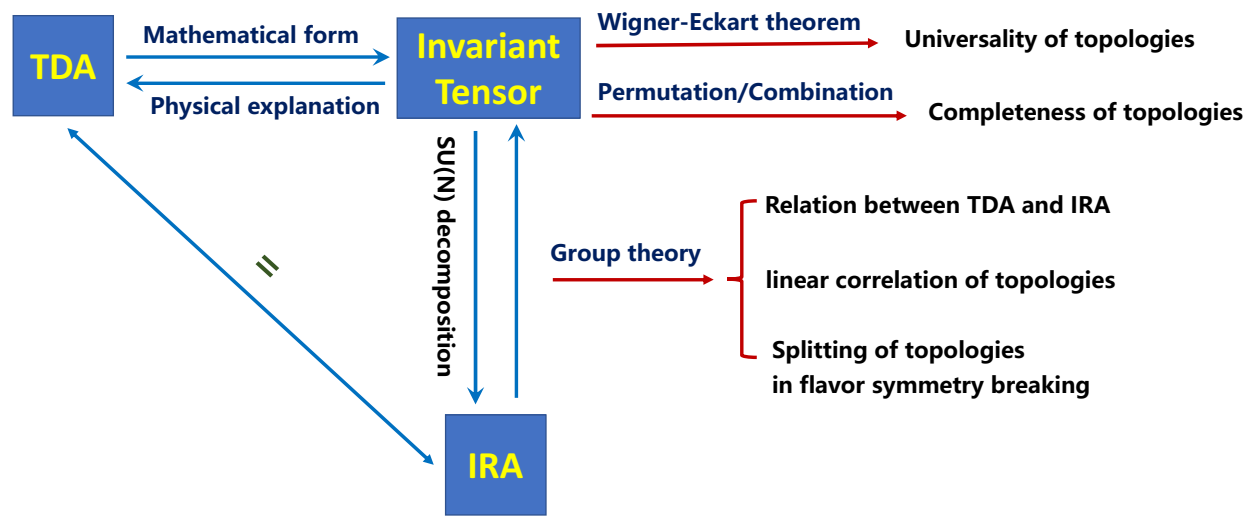}
  \caption{Flow chart for the main point of topology in the tensor form.}\label{flow}
\end{figure}

The goal of this work is to propose a systematic theoretical framework for the topological amplitudes of heavy meson decays.
For this purpose, a one-to-one mapping between the topological diagram and the invariant tensor is set up.
Then some mathematical techniques can be introduced to study the topological amplitudes. For example,
the number of possible topologies contributing to one type of decay is counted in permutations and combinations.
And the $SU(N)$ irreducible amplitudes are derived from the tensor form of the topologies
by decomposing the four-quark operators into irreducible representations.
Taking the $D\to PP$ decay as an example, we will show our framework in detail. And Fig.~\ref{flow} is a sketch.

An attractive achievement of our framework is the linear correlation of the topological amplitudes.
In some of earlier literature, such as Refs.~\cite{Rizzo:1980yh,ChauWang:1980ex,Zeppenfeld:1980ex}, it has been noticed that one of the topologies in the $D$ and $B$ meson decays is not independent in the $SU(3)_F$ limit.
This conclusion still holds when the diagrams with quark loop are included \cite{He:2018php,He:2018joe}.
Moreover, Ref.~\cite{Muller:2015lua} pointed out that matrix linking $T$, $C$, $E$ and $A$ diagrams to the physical amplitudes has only rank three in the case of only the $D\to PP$ modes without $\eta^{(\prime)}$ being analyzed.
But it is no longer correct when $\eta$ and $\eta^{\prime}$ are taken into account \cite{Bhattacharya:2009ps,Hiller:2012xm}. In this work, it is found that above conclusions can be be explained coherently in group theory, in which some specialities of $SU(3)$ group play a crucial role.

In order to match the tensor form of topology, we suggest to classify the topologies in the Standard Model into tree- and penguin-operator-induced diagrams according to which operators, tree or penguin, being inserted into the effective vertexes, no matter whether the topologies involving quark loop or not.
It is found that once the tree-operator-induced amplitudes are given, the penguin-operator-induced amplitudes are completely determined.
There are nine independent tree-operator-induced diagrams contributing to the $D\to PP$ decays.
Five of them, namely $T$, $C$, $E$, $A$ and $T^{LP}$, are not suppressed by the hard gluon exchanges.
If we assume the quark-loop diagram $T^{LP}$ is comparable to other four diagrams,
the large $\Delta A_{CP}$ and the very different $D^0\to K^+K^-$ and $D^0\to \pi^+\pi^-$ branching fractions can be explained together with a normal $U$-spin breaking.
Analogous to the charm meson decays, we deduce that a sizeable $CP$ violation might exist in the $\Xi^+_c\to pK^-\pi^+$ mode.

In addition, our framework provides a simple and systematic way to analyze the $SU(N)_F$ breaking effects.
The linear $SU(3)_F$ breaking \cite{Muller:2015lua} and the high-order $U$-spin breaking \cite{Gronau:2013xba,Gronau:2015rda} in charm decays are re-investigated in the tensor form of topology, which are consistent with the original literature.
Analogous to the degeneracy and splitting of energy levels, we propose the concepts of degeneracy and splitting of topological diagrams.
As an application, we analyze the charm-less $B$/strange-less $D$ decays in the $SU(4)_F$/$SU(3)_F$ symmetry breaking into $SU(3)_F$/$SU(2)_F$ symmetry.

The rest of this paper is organized as follows.
In Sec.~\ref{ga}, we introduce our theoretical framework with a model-independent analysis of the $D\to PP$ decays.
The linear correlation of topologies under $SU(3)$ symmetry will be clarified.
In Sec.~\ref{ppv}, we shall discuss the topologies in the Standard Model and analyze the observed $CP$ violation.
In Sec.~\ref{break}, we will generalize our framework to the broken flavor symmetry.
And Sec.~\ref{sum} is a short summary.
Besides, the topological and irreducible amplitudes in the $D\to PV$ decays will be presented in Appendix \ref{pv}.
And the $SU(3)$ decomposition of operators in the $b$-quark decays and the decay amplitudes of the charmless $\overline B\to PV$ modes will be discussed in Appendixes \ref{b} and \ref{b-pv} respectively.

\section{Model-independent analysis}\label{ga}

In this section, we study the topological amplitudes and the $SU(3)$ irreducible amplitudes model-independently, taking the $D\to PP$ decay as an example.

\subsection{Topological amplitude}\label{tda}

The weak Hamiltonian of charm decay in a general effective theory can be written as
 \begin{equation}\label{hg}
 \mathcal H_{\rm eff}= \sum_p \sum_{i,j,k=1}^3 V_{\rm CKM}O_{ij}^{(p)k},
 \end{equation}
in which
\begin{equation}\label{a5}
O_{ij}^{(p)k} = \frac{G_F}{\sqrt{2}} \sum_{\rm color} \sum_{\rm current}C_p(\overline q_iq_k)(\overline q_jc).
\end{equation}
$O_{ij}^{(p)k}$ denotes the four-quark operator with the Fermi coupling constant $G_F$ and the Wilson coefficient $C_p$.
$V_{\rm CKM}$ is the product of the CKM matrix elements.
The indices $i,j,k$ of $O_{ij}^{(p)k}$ are flavor indices, $1=u$, $2=d$, $3=s$.
Superscript $p$ in $O_{ij}^{(p)k}$ denotes the order of perturbation in an effective theory.
For example, one can set $p=0$ for tree operators and $p=1$ for penguin operators in the effective Hamiltonian in the SM.
For certain $p$, $O_{ij}^{(p)k}$ has $3\times3\times3=27$ possible flavor structures.
For each operator $O_{ij}^{(p)k}$, there are two "vertexes".
The light quark $q_j$ exits from the vertex that $c$ quark annihilates, and quark $ q_i$ and anti-quark $\overline q_k$ exit from the other vertex.
The color indices and current structures of four quark operators are summed because once the flavor structure of operator is determined, the operators with different color indices and current structures are always appear simultaneously and their contributions can be absorbed into one parameter.
If the product of CKM matrix elements corresponding to operator $O_{ij}^{(p)k}$ is labeled by $(H^{(p)})^{ij}_{k}$, the effective Hamiltonian of charm decay can be written as
\begin{equation}\label{h}
 \mathcal H_{\rm eff}= \sum_p \sum_{i,j,k=1}^3 (H^{(p)})^{ij}_{k}O_{ij}^{(p)k}.
\end{equation}
In above notation, $(H^{(p)})$ is a $3\times 3\times 3$ complex matrix and $(H^{(p)})^{ij}_{k}$ is a component of matrix.
In the rest of paper, we will call $(H^{(p)})^{ij}_{k}$ as the "CKM component".

To illustrate above convention more clearly, we re-write the effective Hamiltonian of charm decay in the SM into the form of Eq.~\eqref{h}.
The effective Hamiltonian of charm decay in the SM is written as
\cite{Buchalla:1995vs}:
 \begin{equation}\label{hsm}
 \mathcal H_{\rm eff}^{\rm SM}={G_F\over \sqrt 2}
 \left[\sum_{q=d,s}V_{cq_1}^*V_{uq_2}\left(\sum_{q=1}^2C_i(\mu)O_i(\mu)\right)
 -V_{cb}^*V_{ub}\left(\sum_{i=3}^6C_i(\mu)O_i(\mu)+C_{8g}(\mu)O_{8g}(\mu)\right)\right],
 \end{equation}
where the tree operators are
\begin{eqnarray}
O_1=(\bar{u}_{\alpha}q_{2\beta})_{V-A}
(\bar{q}_{1\beta}c_{\alpha})_{V-A},\qquad
O_2=(\bar{u}_{\alpha}q_{2\alpha})_{V-A}
(\bar{q}_{1\beta}c_{\beta})_{V-A},
\end{eqnarray}
with $\alpha,\beta$ being color indices, and $q_{1,2}$ being the $d$
or $s$ quark.
The QCD penguin operators are
 \begin{align}
 O_3&=\sum_{q'=u,d,s}(\bar u_\alpha c_\alpha)_{V-A}(\bar q'_\beta
 q'_\beta)_{V-A},~~~
 O_4=\sum_{q'=u,d,s}(\bar u_\alpha c_\beta)_{V-A}(\bar q'_\beta q'_\alpha)_{V-A},
 \nonumber\\
 O_5&=\sum_{q'=u,d,s}(\bar u_\alpha c_\alpha)_{V-A}(\bar q'_\beta
 q'_\beta)_{V+A},~~~
 O_6=\sum_{q'=u,d,s}(\bar u_\alpha c_\beta)_{V-A}(\bar q'_\beta
 q'_\alpha)_{V+A}.
 \end{align}
The chromomagnetic penguin operator is
\begin{eqnarray}
O_{8g}=\frac{g}{8\pi^2}m_c{\bar
u}\sigma_{\mu\nu}(1+\gamma_5)T^aG^{a\mu\nu}c.
\end{eqnarray}
The magnetic-penguin contributions can be included into the Wilson coefficients for the penguin operators following the substitutions
\cite{Beneke:2001ev,Beneke:2003zv,Beneke:2000ry,Beneke:1999br}
$C_{3,5}(\mu)\to C_{3,5}(\mu) + \frac{\alpha_s(\mu)}{8\pi N_c}
\frac{2m_c^2}{\langle l^2\rangle}C_{8g}^{\rm eff}(\mu)$,
$C_{4,6}(\mu)\to C_{4,6}(\mu) - \frac{\alpha_s(\mu)}{8\pi }
\frac{2m_c^2}{\langle l^2\rangle}C_{8g}^{\rm eff}(\mu)$,
with the effective Wilson coefficient $C_{8g}^{\rm eff}=C_{8g}+C_5$ and $\langle l^2\rangle$ being the averaged invariant mass squared of the virtual gluon emitted from the magnetic penguin operator.
In the notation \eqref{h}, the tree and penguin operators can be written as,
\begin{align}\label{s1}
  O^{(0)k}_{1j} & = \frac{G_F}{\sqrt{2}} \big[ C_1 (\bar{u}_{\alpha}q_{k,\beta})_{V-A}
(\bar{q}_{j,\beta}c_{\alpha})_{V-A} + C_2 (\bar{u}_{\alpha}q_{k,\alpha})_{V-A}
(\bar{q}_{j,\beta}c_{\beta})_{V-A}\big], \\ \label{s2}
  O^{(1)k}_{i1}  & = \frac{G_F}{\sqrt{2}} \big[  C_3 (\bar{q}_{i,\alpha}q_{k,\alpha})_{V-A}
(\bar{u}_{\beta}c_{\beta})_{V-A}  +  C_4 (\bar{q}_{i,\alpha}q_{k,\beta})_{V-A}
(\bar{u}_{\alpha}c_{\beta})_{V-A} \nonumber\\
&~~~~~~~~~~~~~~~~+  C_5 (\bar{q}_{i,\alpha}q_{k,\alpha})_{V+A}
(\bar{u}_{\beta}c_{\beta})_{V-A}  +  C_6 (\bar{q}_{i,\alpha}q_{k,\beta})_{V+A}
(\bar{u}_{\alpha}c_{\beta})_{V-A}  \big].
\end{align}
The corresponding CKM components of operators $O^{(0)k}_{1j}$ and $O^{(1)k}_{i1}$ are
\begin{align}\label{CKM}
  (H^{(0)})^{1j}_{k}  = V_{cq_j}^*V_{uq_k}, \qquad (H^{(1)})^{i1}_{k}  = - V_{cb}^*V_{ub},
\end{align}
and the other $(H^{(0,1)})^{ij}_{k}$ are zero.

We use the general effective Hamiltonian Eq.~\eqref{h} to construct the model-independent amplitude of the $D\to PP$ decay.
To achieve this goal, an algebraic tool, tensor analysis, is needed.
According to Ref.~\cite{Georgi:1999wka}, an arbitrary state in the tensor product space can be written as
\begin{align}
  |v\rangle = v_{i_1...i_m}^{j_1...j_n}|v^{i_1...i_m}_{j_1...j_n} \rangle.
\end{align}
Tensor $v$ is a "wave-function", because one can get tensor component $v_{i_1...i_m}^{j_1...j_n}$ by taking the matrix element of $|v\rangle$ with the tensor product state,
\begin{align}
 v_{i_1...i_m}^{j_1...j_n}  = \langle v^{i_1...i_m}_{j_1...j_n}|v\rangle.
\end{align}
Applying to physics, a light pseudoscalar meson state is expressed as
\begin{align}\label{a3}
  |P_\alpha\rangle = (P_\alpha)_{i}^{j}|P^{i}_{j} \rangle,
\end{align}
in which $|P^{i}_{j} \rangle$ is the quark composition of meson state, $|P^{i}_{j} \rangle = |q_i\bar q_j\rangle$.
$(P_\alpha)_{i}^{j}$ is the coefficient of the quark composition $|P^{i}_{j} \rangle$.
In the $SU(3)$ picture, pseudoscalar meson notet $|P^{i}_{j} \rangle$ is expressed as
\begin{eqnarray}\label{a1}
 |P^i_j\rangle =  \left( \begin{array}{ccc}
   \frac{1}{\sqrt 2} |\pi^0\rangle +  \frac{1}{\sqrt 6} |\eta_8\rangle,    & |\pi^+\rangle,  & |K^+\rangle \\
   | \pi^-\rangle, &   - \frac{1}{\sqrt 2} |\pi^0\rangle+ \frac{1}{\sqrt 6} |\eta_8\rangle,   & |K^0\rangle \\
   | K^- \rangle,& |\overline K^0\rangle, & -\sqrt{2/3}|\eta_8\rangle \\
  \end{array}\right) +  \frac{1}{\sqrt 3} \left( \begin{array}{ccc}
   |\eta_1\rangle,    & 0,  & 0 \\
    0, &  |\eta_1\rangle,   & 0 \\
   0, & 0, & |\eta_1\rangle \\
  \end{array}\right),
\end{eqnarray}
where $i$ is row index and $j$ is column index.
According to Eq.~\eqref{a1}, one can derive
\begin{align}
|\pi^+\rangle = (\pi^+)^2_1| P^1_2\rangle = |u\bar d\rangle,\qquad |\pi^0\rangle = (\pi^0)^1_1| P^1_1\rangle - (\pi^0)^2_2| P^2_2\rangle = \frac{1}{\sqrt{2}}|u\bar u\rangle - \frac{1}{\sqrt{2}}|d\bar d\rangle, \quad ...
\end{align}
The bar state of Eq.~\eqref{a3} is
\begin{align}\label{h1}
  \langle P_\alpha| = (\overline P_\alpha)_{i}^{j}\langle P^{i}_{j} |.
\end{align}
Since $ ( P_\alpha)_{i}^{j}$ is a real number, $ (\overline P_\alpha)_{i}^{j} =  (P_\alpha)_{i}^{j}$.
A charmed meson state is expressed as
\begin{align}\label{a4}
|D_\alpha\rangle = (D_\alpha)_i |D^{i}\rangle,
\end{align}
and
\begin{align}
|D^{i}\rangle = (|D^0\rangle, \,|D^+\rangle, \,|D_s^+\rangle) = (|c\bar u\rangle, \,|c\bar d\rangle, \,|c\bar s\rangle).
\end{align}

The decay amplitude of $D_\gamma\to P_\alpha P_\beta$ can be constructed to be
\begin{align}\label{a6}
\mathcal{A}(D_\gamma\to P_\alpha P_\beta)& =  \langle P_\alpha P_\beta |\mathcal{H}_{\rm eff}| D_\gamma\rangle \nonumber\\
&~~=\sum_p \sum_{\rm Per.}\,(D_\gamma)_i (H^{(p)})^{jk}_{l}(P_\alpha)_{m}^{n}(P_\beta)_{r}^{s}\times \langle P^{m}_{n} P_s^r |O_{jk}^{(p)l} |D^{i}\rangle,
\end{align}
in which $\sum_{\rm Per.}$ present summing over all the possible full contractions of $\langle P^{m}_{n} P_s^r |O_{jk}^{(p)l} |D^{i}\rangle$.
Under the flavor symmetry, decay amplitude is a complex number without flavor indices, i.e., a $SU(N)$ invariant.
Then $\langle P^{m}_{n} P_s^r |O_{jk}^{(p)l} |D^{i}\rangle$ is a invariant tensor in which all the indices either contract with each other \cite{Georgi:1999wka}.
Once the contraction of $\langle P^{m}_{n} P_s^r |O_{jk}^{(p)l} |D^{i}\rangle$ is determined, the form of $(D_\gamma)_i(H^{(p)})^{jk}_{l}(P_\alpha)^n_m(P_\beta)^s_r$ is determined too, vice versa.
For more simplicity, the decay amplitude of $D_\gamma\to P_\alpha P_\beta$ in $p$-order is expressed as \cite{Grossman:2012ry}
\begin{align}
\mathcal{A}^{(p)}(D_\gamma\to P_\alpha P_\beta) = \sum_\omega X^{(p)}_{\omega}(C^{(p)}_\omega)_{\alpha\beta\gamma},
\end{align}
where $\omega$ labels the different contractions of the $SU(3)$ indices. $X^{(p)}_\omega$ is the reduced matrix element and $(C^{(p)}_\omega)_{\alpha\beta\gamma}$ is the Clebsch-Gordan coefficient calculated by $(D_\gamma)_i(H^{(p)})^{jk}_{l}(P_\alpha)^n_m(P_\beta)^s_r$.
According to the Wigner-Eckhart theorem \cite{Eckart30,Wigner59}, $X^{(p)}_\omega$ is independent of decay channels, i.e., indices $\alpha$, $\beta$ and $\gamma$.
All the information of initial/final states is absorbed into the Clebsch-Gordan coefficient $(C^{(p)}_\omega)_{\alpha\beta\gamma}$.

\begin{figure}
  \centering
  \includegraphics[width=5cm]{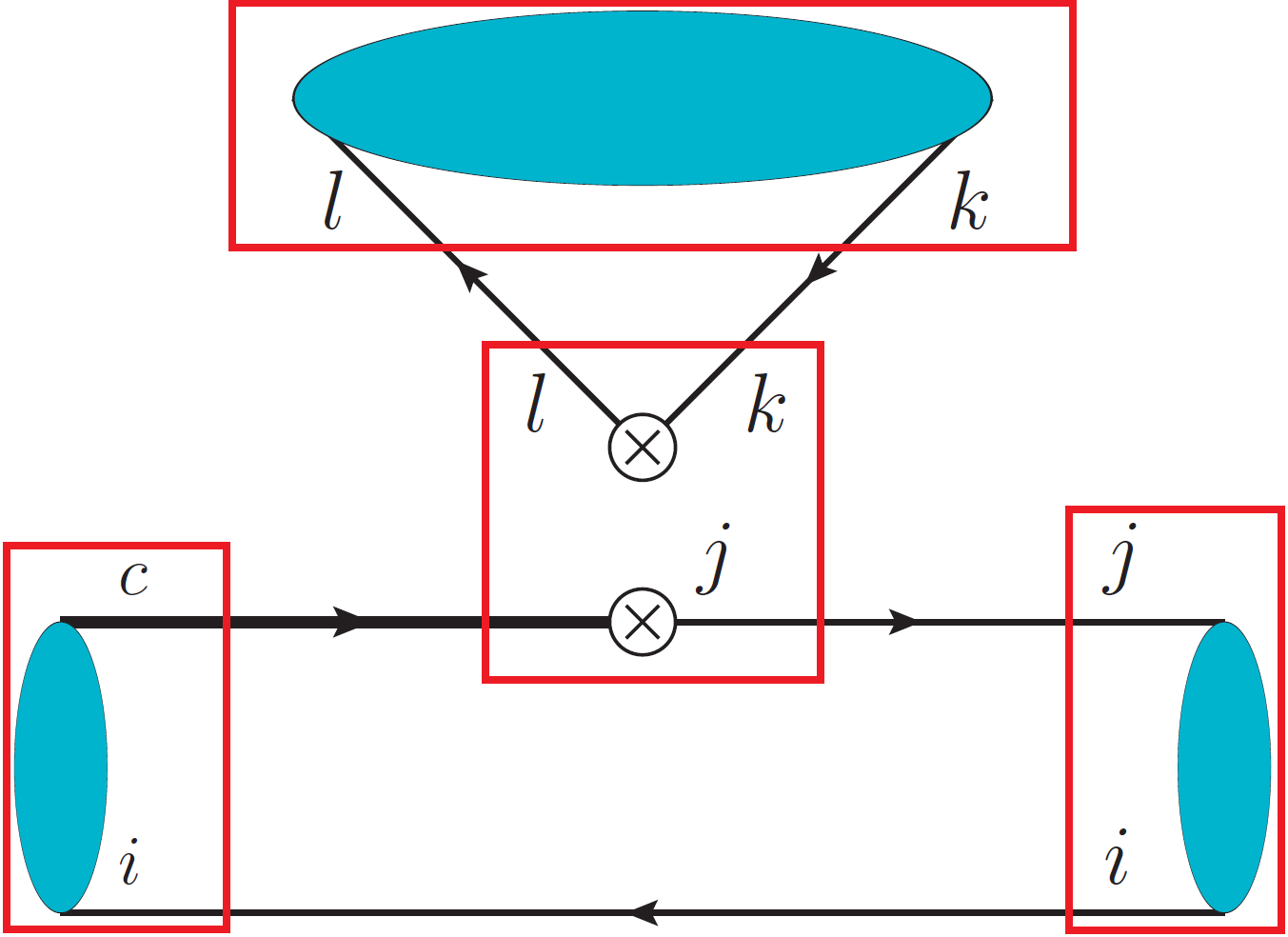}
  \caption{Schematic description of the index-contraction of $T$ diagram.}\label{t}
\end{figure}
If the index-contraction is understood as quark flowing, the reduced matrix element $X^{(p)}_\omega$ is a topological amplitude.
The contraction maps the topological diagram by following rules.
\begin{itemize}
  \item The contraction between the final-state meson $P$ and the four-quark operator indicates that the quark or anti-quark produced in one effective vertex of operator $O^{(p)k}_{ij}$ enters the final state-meson $P$.
  \item  The contraction between the initial-state $D$ meson and the four-quark operator indicates that the light anti-quark in $D$ meson annihilates in the vertex of four-quark operator.
\item The contraction between the initial-state $D$ meson and the final-state meson $P$ indicates that the light anti-quark in $D$ meson, as a spectator quark, enters the final state meson $P$.
\item   The contraction between two indices of the four-quark operator presents the quark loop.
$O_{il}^{(p)l}$ presents the quark loop that connects the two effective vertexes in the topological diagram.
While $O_{lj}^{(p)l}$ presents the quark loop induced in one effective vertex in the topological diagram.
\end{itemize}

According to these rules, one can set up a one-to-one mapping between the topological diagram and the invariant tensor.
For example, the reduced matrix element $\langle P_{i}^j P_k^l |O_{jl}^{(0)k} |D^{i}\rangle$ presents the $T$ diagram in literature.
The schematic description of index-contraction in $T$ diagram is shown in Fig.~\ref{t}.
By substituting Eq.~\eqref{s1} into $\langle P_{i}^j P_k^l |O_{jl}^{(0)k} |D^{i}\rangle$, amplitude $T$ can be calculated via the naive factorization:
\begin{align}
T &= \langle P_{i}^j P_k^l |O_{jl}^{(0)k} |D^{i}\rangle \nonumber\\
&~= \frac{G_F}{\sqrt{2}}[C_1(\mu) \langle P_{i}^j P_k^u| (\bar{u}_{\alpha}q_{k,\beta})_{V-A}
(\bar{q}_{j,\beta}c_{\alpha})_{V-A} |D^{i} \rangle + C_2(\mu)\langle P_{i}^j P_k^u|  (\bar{u}_{\alpha}q_{k,\alpha})_{V-A}
(\bar{q}_{j,\beta}c_{\beta})_{V-A}|D^{i} \rangle]\nonumber\\
&~~= \frac{G_F}{\sqrt{2}} [C_1(\mu) \langle P_k^u| (\bar{u}_{\alpha}q_{k,\beta})_{V-A}
|0\rangle\langle P_{i}^j |(\bar{q}_{j,\beta}c_{\alpha})_{V-A} |D^{i} \rangle \nonumber\\
& ~~~~~~~~~~~~~~~~+ C_2(\mu)\langle P_k^l|  (\bar{u}_{\alpha}q_{k,\alpha})_{V-A}|0 \rangle \langle P_{i}^j|
(\bar{q}_{j,\beta}c_{\beta})_{V-A}|D^{i} \rangle ]\nonumber\\
&~~~= \frac{G_F}{\sqrt{2}} \Big(C_2(\mu)+\frac{C_1(\mu)}{N_c} \Big) \langle P_k^u| (\bar{u}q_{k})_{V-A}
|0\rangle\langle P_{i}^j |(\bar{q}_{j}c)_{V-A} |D^{i} \rangle \nonumber\\
&~~~~= \frac{G_F}{\sqrt{2}} \Big(C_2(\mu)+\frac{C_1(\mu)}{N_c}\Big) f_{P^u_k}(m^2_{D^i}-m^2_{P_i^j})F_0^{D^i\to P_i^j}(m^2_{P^u_k}).
\end{align}

\begin{figure}
  \centering
  \includegraphics[width=14cm]{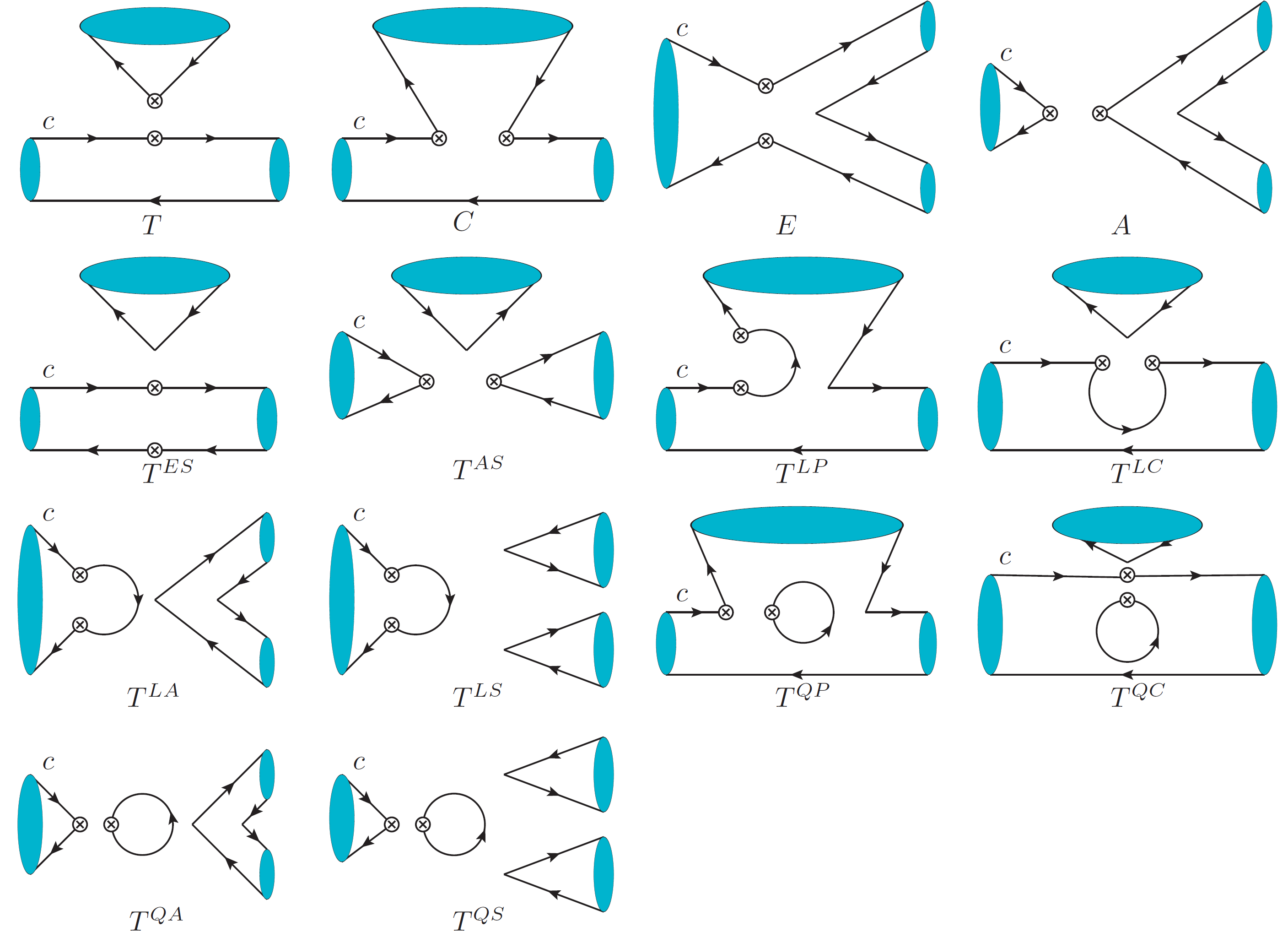}
  \caption{Topological diagrams in the $D\to PP$ decays.}\label{top1}
\end{figure}
There are four upper/lower indices in $\langle P^{m}_{n} P_s^r |O_{jk}^{(p)l} |D^{i}\rangle$. The number of all possible contractions is $N=A^4_4 = 24$. Considering that we cannot distinguish the two pseudoscalar nonet in the $D\to PP$ decay, some repeated count should be subtracted and then $N=A^4_4 -2\times A^3_3 +2 = 14$.
Amplitude of the $D\to PP$ decay can be written as
\begin{align}\label{ha}
 \mathcal{A}^{\rm TDA}_{D_\gamma \to P_\alpha P_\beta}&=  T  (D_\gamma)_i  (H)^{lj}_k(P_\alpha)^{i}_j  (P_\beta)^k_l + C (D_\gamma)_i  (H)^{jl}_k  (P_\alpha)^{i}_j(P_\beta)^k_l+E  (D_\gamma)_i  (H)^{il}_j (P_\alpha)^j_k (P_\beta)^{k}_l\nonumber \\ & + A  (D_\gamma)_i (H)^{li}_j   (P_\alpha)^j_k (P_\beta)^{k}_l +T^{ES} (D_\gamma)_i   (H)^{ij}_{l}   (P_\alpha)^{l}_j    (P_\beta)^k_k+T^{AS} (D_\gamma)_i  (H)^{ji}_{l}  (P_\alpha)^{l}_j  (P_\beta)^k_k \nonumber\\& +T^{LP} (D_\gamma)_i (H)^{kl}_{l}(P_\alpha)^{i}_j   (P_\beta)^j_k + T^{LC} (D_\gamma)_i  (H)^{jl}_{l} (P_\alpha)^{i}_j    (P_\beta)^k_k   \nonumber\\& + T^{LA} (D_\gamma)_i  (H)^{il}_{l}  (P_\alpha)^j_k (P_\beta)^{k}_j + T^{LS} (D_\gamma)_i   (H)^{il}_{l} (P_\alpha)^{j}_j (P_\beta)^k_k \nonumber\\&  +T^{QP}(D_\gamma)_i(H)^{lk}_{l}  (P_\alpha)^{i}_j   (P_\beta)^j_k  + T^{QC} (D_\gamma)_i   (H)^{lj}_{l}  (P_\alpha)^i_j (P_\beta)_{k}^{k}   \nonumber\\
& + T^{QA} (D_\gamma)_i  (H)^{li}_{l}  (P_\alpha)^j_k (P_\beta)^{k}_j + T^{QS} (D_\gamma)_i (H)^{li}_{l}  (P_\alpha)_j^j (P_\beta)_{k}^{k} + \alpha \leftrightarrow \beta.
\end{align}
Each term and the corresponding interchange $\alpha \leftrightarrow \beta$ in Eq.~\eqref{ha} present one topological amplitude and the coefficient is calculated by the product of $(D_\gamma)_i(H)^{jk}_{l}(P_\alpha)^n_m(P_\beta)^s_r$.
In the case of $P_\alpha =P_\beta$, the decay amplitude $\mathcal{A}^{\rm TDA}_{(D_\gamma \to P_\alpha P_\alpha)}$ has to time $1/\sqrt{2}$ due to the symmetry factor of 1/2 appearing in the decay rate.
As we have mentioned above, it is the Wigner-Eckhart theorem \cite{Eckart30,Wigner59} ensures that topological amplitude is independent of initial and final states.
In Eq.~\eqref{ha}, we do not write the order of perturbation of four-quark operators explicitly. But notice that the same contractions with different $p$ present different topological amplitudes.
For example, $p=0$ denotes the diagrams induced by tree operators in the SM, while $p=1$ denotes the diagrams induced by penguin operators.
In fact, perturbation order $p$ provides a natural way to classify the topologies.
We will discuss this question in detail in Sec.~\ref{ppv}.

The topological diagrams contributing to the $D\to PP$ decays are showed in Fig.~\ref{top1}.
The first four diagrams, $T$, $C$, $E$ and $A$, have been analyzed in plenty of literature.
$T^{ES}$ and $T^{AS}$ are the singlet contributions which requires multi-gluon exchanges.
The last eight diagrams are quark-loop contributions.
We call them $T^{X}$ because that tree operators can be inserted into their effective vertexes.
In principle, all the topological diagrams listed in Fig.~\ref{top1} should contribute to the $D\to PP$ decays.
But some diagrams always disappear when some operators are inserted into their effective vertexes.
If the operator with three same indices (for instance $(\overline uu)(\overline uc)$) is inserted, all the 14 diagrams in Fig.~\ref{top1} contribute to the $D\to PP$ decays.
If the operator with two same indices (for instance $(\overline ud)(\overline dc)$) is inserted, only the first ten diagrams in Fig.~\ref{top1} contribute.
If the operator without same indices (for instance $(\overline ud)(\overline sc)$) is inserted, only the first six diagrams contribute.
That is why topologies $T^{QP}$, $T^{QC}$, $T^{QA}$ and $T^{QS}$ always disappear when tree operators in the SM are inserted.
But $T^{QP}$, $T^{QC}$, $T^{QA}$ and $T^{QS}$ are necessary to derive the correct relation of the topological diagram amplitudes and the $SU(3)$ irreducible amplitudes, see \ref{ira}.
In \cite{He:2018php,He:2018joe}, the last four topological diagrams in Fig.~\ref{top1} are overlooked, which results in an incorrect relation.

\subsection{$SU(3)$ irreducible amplitude}\label{ira}

Operator $O_{ij}^k$ defined in Eq.~\eqref{a5} can be regarded as a $(2,1)$-rank tensor representation of $SU(N)$ group.
There are two covariant (lower) indices and one contravariant (upper) index in $O_{ij}^k$. Indices $i$ and $j$ transform according to the foundational representation $N$ of $SU(N)$ group, and index $k$ transforms according to the complex conjugate representation $\overline N$ \cite{Georgi:1999wka}.
Now let us discuss how to decompose the general $(2,1)$-rank tensor $T_{ij}^k$ into the irreducible representations of $SU(N)$ group.
Firstly, we study a simple tensor, $T_{ij}$ with two covariant indices $i$ and $j$. $T_{ij}$ is decomposed as $N \otimes N$.
In group theory, the foundational representation of $SU(N)$ group can be expressed as one square $\square$ in young's tableaux. The decomposition of $N \otimes N$ is
\begin{align}\label{xx}
\square \,\otimes \,\square =\square\hspace{-0.45mm}\square \,\oplus \begin{array}{c}
   \square \\[-4.6mm]
  \square
 \end{array}\,,
\end{align}
in which $\square\hspace{-0.45mm}\square$ presents symmetrization of indices $i,j$ and $\begin{array}{c}\square \\[-4.6mm]\square\end{array}$ presents anti-symmetrization of indices $i,j$. The number of possible combination of antisymmetric $i,j$ is $C_N^2 = N(N-1)/2$.
And the number of possible combination of symmetric $i,j$ is $N^2-C_N^2 = N^2-N(N-1)/2 = N(N+1)/2$.
Thereby, the first term in Eq.~\eqref{xx} presents a $N(N+1)/2$ representation and the second term presents a $\overline {N(N-1)/2}$ representation of $SU(N)$ group.
If $N=3$, we have $3\otimes 3 = 6\oplus \overline 3$.
Secondly, we analyze another simple tensor, $T_i^j$ with one covariant index $i$ and one contravariant index $j$.
In group theory, a mixed tensor with contraction of one covariant index and one contravariant index is known as trace tensor.
For any mixed tensor, it can be decomposed into trace tensor and traceless tensor.
Both the subspaces composed by trace tensor and traceless tensor are invariant subspaces of $SU(N)$ group.
So $T_i^j$ can be decomposed as
\begin{align}
T_i^j = \Big\{ T_i^j - \delta_i^j\Big(\frac{1}{N}\sum_lT_l^l \Big) \Big\} + \delta_i^j\Big(\frac{1}{N}\sum_lT_l^l \Big).
\end{align}
The trace tensor is a trivial representation and the traceless tensor is $(N^2-1)$-dimensional representation of $SU(N)$ group. If $N=3$, we have $3\otimes \overline 3 = 8\oplus 1$. Finally, we discuss the $SU(N)$ decomposition of $T_{ij}^k$. There are two steps to decompose $T_{ij}^k$ into the direct sum of irreducible representations: extracting the trace tensors and symmetrizing/anti-symmetrizing indices of the remaining traceless tensor.
The result is
\begin{align}\label{q3}
T_{ij}^{k} = T_{\{ij\}}^k +T_{[ij]}^k + \frac{1}{N^2-1}\Big\{\delta_j^k\sum_l\big(N\,T_{il}^l - T_{li}^l\big)+ \delta_i^k\sum_l\big(N\,T_{lj}^l - T_{jl}^l\big)\Big\},
\end{align}
in which the two trace tensors are dimension-$N$ since only one free index left.
$T_{\{ij\}}^k$ and $T_{[ij]}^k$ are traceless with the dimensions of $(N^2-C_N^2)\times N-N =N^2(N+1)/2-N $ and $C_N^2\times N-N =N^2(N-1)/2-N $.
That is,
\begin{align}\label{c4}
N \otimes\overline N \otimes N = N^2(N+1)/2-N\oplus\overline{N^2(N-1)/2-N}\oplus N \oplus N.
\end{align}

If operator $O_{ij}^k$ is the representation of $SU(3)$ group,
 it can be decomposed as $3 \otimes\overline 3 \otimes 3 =  3_p\oplus3_t\oplus \overline 6 \oplus 15$. The explicit decomposition is \cite{Grossman:2012ry}
\begin{align}\label{hd}
  O^k_{ij}= \frac{1}{8}\,O(15)^k_{ij}+\frac{1}{4}\,\epsilon_{ijl}O(\overline 6)^{lk}+\delta_j^k\Big(\frac{3}{8}O( 3_t)_i-\frac{1}{8}O(3_p)_i\Big)+
  \delta_i^k\Big(\frac{3}{8}O( 3_p)_j-\frac{1}{8}O( 3_t)_j\Big),
\end{align}
which is consistent with Eq.~\eqref{q3}.
The coefficients $1/4$ and $1/8$ before $6$- and $15$-dimensional representations are used to match most literature.
All irreducible presentations are listed following.\\
$ 3_p$ presentation:
\begin{align}\label{3p}
  O( 3_p)_1 & = (\bar u u)(\bar u c) + (\bar dd)(\bar u c) + (\bar ss)(\bar u c),\qquad
 O(3_p)_2 = (\bar u u)(\bar d c) + (\bar dd)(\bar d c) + (\bar ss)(\bar d c),\nonumber \\
  O(3_p)_3 & = (\bar u u)(\bar s c) + (\bar dd)(\bar s c) + (\bar ss)(\bar s c).
\end{align}
$ 3_t$ presentation:
\begin{align}\label{3t}
 O(3_t)_1 & = (\bar u u)(\bar u c) + (\bar ud)(\bar d c) + (\bar us)(\bar s c),\qquad
  O(3_t)_2 = (\bar d u)(\bar u c) + (\bar dd)(\bar d c) + (\bar ds)(\bar s c),\nonumber \\
 O(3_t)_3 & = (\bar s u)(\bar u c) + (\bar sd)(\bar d c) + (\bar ss)(\bar s c).
\end{align}
$\overline 6$ presentation:
\begin{align}
    O(\overline 6)^{11} & = 2 [(\bar d u)(\bar s c) - (\bar su)(\bar d c)],\quad O(\overline 6)^{22} = 2[(\bar s d)(\bar u c) - (\bar ud)(\bar s c)],\nonumber \\
    O(\overline 6)^{33} &= 2 [(\bar u s)(\bar d c) - (\bar ds)(\bar u c)],\nonumber \\
     O(\overline 6)^{12} & = [(\bar s u)(\bar u c) - (\bar uu)(\bar s c) + (\bar dd)(\bar s c) - (\bar sd)(\bar d c)],\nonumber \\
    O(\overline 6)^{23}& = [(\bar u d)(\bar d c) - (\bar dd)(\bar u c) + (\bar ss)(\bar u c) -(\bar us)(\bar s c)],\nonumber \\
 O(\overline 6)^{31} & = [(\bar d s)(\bar s c) - (\bar ss)(\bar d c) + (\bar uu)(\bar d c) -(\bar du)(\bar u c)].
\end{align}
$ {\underline{15}}$ presentation:
\begin{align}\label{15}
    O({15})^{1}_{11} & = 4(\bar u u)(\bar u c)-2 [(\bar us)(\bar s c) + (\bar ud)(\bar dc)+(\bar dd)(\bar u c) + (\bar ss)(\bar uc)],\nonumber \\
     O({15})^{2}_{22} & = 4(\bar dd)(\bar d c)-2 [(\bar du)(\bar u c) + (\bar ds)(\bar sc)+(\bar uu)(\bar d c) + (\bar ss)(\bar dc)],\nonumber \\
   O({15})^{3}_{33} & = 4(\bar ss)(\bar s c)-2 [(\bar su)(\bar u c) + (\bar sd)(\bar dc)+(\bar uu)(\bar s c) + (\bar dd)(\bar sc)],\nonumber \\
  O({15})^{1}_{23} & = 4 [(\bar d u)(\bar s c) + (\bar su)(\bar d c)],\qquad      O({15})^{2}_{13} = 4[(\bar s d)(\bar u c) + (\bar ud)(\bar s c)],  \nonumber \\
  O({15})^{3}_{12} & = 4 [(\bar u s)(\bar d c) + (\bar ds)(\bar u c)],\nonumber \\
 O({15})^{2}_{11} & = 8(\bar u d)(\bar u c), \qquad  O({15})^{3}_{11}  = 8(\bar u s)(\bar u c), \qquad  O({15})^{1}_{22} = 8(\bar du)(\bar d c),\nonumber \\
O({15})^{3}_{22} & = 8(\bar d s)(\bar d c), \qquad O({15})^{1}_{33} = 8(\bar su)(\bar s c), \qquad
O({15})^{2}_{33}  = 8(\bar s d)(\bar s c),\nonumber \\
O({15})^{1}_{21} & = 3[(\bar u u)(\bar d c)+(\bar d u)(\bar u c)]-2(\bar d d)(\bar d c) - [(\bar d s)(\bar s c)+(\bar ss)(\bar d c)],\nonumber \\
O({15})^{2}_{12} & = 3[(\bar u d)(\bar d c)+(\bar d d)(\bar u c)]-2(\bar u u)(\bar u c) - [(\bar u s)(\bar s c)+(\bar ss)(\bar u c)],\nonumber \\
O({15})^{1}_{31} & = 3[(\bar u u)(\bar s c)+(\bar su)(\bar u c)]-2(\bar s s)(\bar s c) - [(\bar s d)(\bar d c)+(\bar dd)(\bar s c)],\nonumber \\
O({15})^{3}_{13} & = 3[(\bar u s)(\bar s c)+(\bar ss)(\bar u c)]-2(\bar u u)(\bar u c) - [(\bar u d)(\bar d c)+(\bar dd)(\bar u c)],\nonumber \\
O({15})^{2}_{32} & = 3[(\bar d d)(\bar s c)+(\bar sd)(\bar d c)]-2(\bar s s)(\bar s c) - [(\bar s u)(\bar u c)+(\bar uu)(\bar s c)],\nonumber \\
O({15})^{3}_{23} & = 3[(\bar ds)(\bar s c)+(\bar s s)(\bar d c)]-2(\bar d d)(\bar dc) - [(\bar d u)(\bar u c)+(\bar uu)(\bar d c)].
\end{align}
There are nine operators in irreducible representation $\overline 6$, but only six of them are independent because $O(\overline 6)^{ij}$ is symmetric in the interchange of its two upper indices, $O(\overline 6)^{ij} = O(\overline 6)^{ji}$.
$O(\overline 6)^{ij}$ can be written as $O(\overline 6)_{ij}^k$ by contracting
with the Levi-Civita tensor $O(\overline 6)_{ij}^k=\epsilon_{ijl}O(\overline 6)^{lk}$.
There are twenty-seven operators in irreducible representation $\emph{15}$, but only fifteen of them are independent because $O({15})_{ij}^k$ is symmetric in the interchange  of its two subscripts, $O({15})^k_{ij} = O({15})_{ji}^k$ and the following equations
\begin{align}
  O({15})^1_{11} &= -[O({15})^2_{12} +O({15})^3_{13}], \qquad O({15})^2_{22} = -[O({15})^1_{21} +O({15})^3_{23}],\nonumber \\
  O({15})^3_{33}&= -[O({15})^1_{31} +O({15})^2_{32}].
\end{align}

If operator $O_{ij}^k$ is decomposed into irreducible representations, the CKM component $(H)^{ij}_k$ should be decomposed correspondingly:
\begin{align}\label{c8}
  (H)_k^{ij}= &\frac{1}{8}(H(15))_k^{ij}+\frac{1}{4}\epsilon^{ijl}(H(\overline 6))_{lk}+\delta^j_k\Big(\frac{3}{8}(H( 3_t))^i-\frac{1}{8}(H(3_p))^i\Big)\nonumber\\&~~~~~~+
  \delta^i_k\Big(\frac{3}{8}(H( 3_p))^j-\frac{1}{8}(H( 3_t))^j\Big).
\end{align}
To obtain the $SU(3)$ irreducible amplitude of the $D \to PP$ decay, one can contract all indices in the following manner
\begin{align}\label{nonisodecomp}
{\cal A}^{\rm IRA}_{D_\gamma \to P_\alpha P_\beta} =&a_6(D_\gamma)_i (H(\overline 6))^{ij}_k(P_\alpha)_j^l(P_\beta)_l^k + b_6(D_\gamma)_i (H(\overline 6))^{ij}_k(P_\alpha)_j^k(P_\beta)^l_l \nonumber\\&+ c_6(D_\gamma)_i (H(\overline 6))^{jl}_k(P_\alpha)_j^i(P_\beta)_l^k
+a_{15}(D_\gamma)_i (H({15}))^{ij}_k(P_\alpha)_j^l(P_\beta)_l^k \nonumber \\&+ b_{15}(D_\gamma)_i (H({15}))^{ij}_k(P_\alpha)_j^k(P_\beta)^l_l + c_{15}(D_\gamma)_i (H({15}))^{jl}_k(P_\alpha)_j^i(P_\beta)_l^k\nonumber \\&+a_3^p (D_\gamma)_i (H(3_p))^i (P_\alpha)^k_j(P_\beta)^j_k +b_3^p (D_\gamma)_i (H(3_p))^i (P_\alpha)_k^k(P_\beta)_j^j\nonumber\\
  &+c_3^p (D_\gamma)_i (H(3_p))^k (P_\alpha)^i_k(P_\beta)_j^j+d_3^p (D_\gamma)_i (H(3_p))^k (P_\alpha)^i_j(P_\beta)^j_k\nonumber\\&+a_3^t (D_\gamma)_i (H(3_t))^i (P_\alpha)^k_j(P_\beta)^j_k +b_3^t (D_\gamma)_i (H(3_t))^i (P_\alpha)_k^k(P_\beta)_j^j\nonumber\\&+c_3^t (D_\gamma)_i (H(3_t))^k (P_\alpha)^i_k(P_\beta)_j^j+d_3^t (D_\gamma)_i (H(3_t))^k (P_\alpha)^i_j(P_\beta)^j_k+\alpha\leftrightarrow \beta.
\end{align}
Similar to Eq.~\eqref{ha}, there are 14 possible index-contractions in Eq.~\eqref{nonisodecomp}.
By substituting Eq.~\eqref{c8} into the amplitudes of $T$, $C$, $E$..., i.e., Eq.~\eqref{ha},
the relations between topological diagrams and the $SU(3)$ irreducible amplitudes are derived to be\footnote{Taking $T$ diagram as an example,
\begin{align}
T\times (D_\gamma)_i  (H)^{lj}_k(P_\alpha)^{i}_j  (P_\beta)^k_l = &T\times (D_\gamma)_i (P_\alpha)^{i}_j  (P_\beta)^k_l \times\Big[\delta^j_k\big(\frac{3}{8}(H( 3_t))^l-\frac{1}{8}(H( 3_p))^l\big)+\nonumber\\&~~
  \delta^l_k\big(\frac{3}{8}(H( 3_p))^j-\frac{1}{8}(H( 3_t))^j\big)+\frac{1}{4}\epsilon^{ljm}(H(\overline 6))_{mk}+\frac{1}{8}(H({15}))_k^{lj}\Big],
\end{align}
contributing to
\begin{align}
&c_{15} = \frac{1}{8}T+..., \quad c_{6} = \frac{1}{4}T+...,\quad c_{3}^t = -\frac{1}{8}T+...,\quad c_{3}^p = \frac{3}{8}T+...,\quad  d_{3}^t = \frac{3}{8}T+...,\quad d_{3}^p = -\frac{1}{8}T+...\,\,.
\end{align}}
\begin{align}\label{sol}
 a_6&  =\frac{E-A}{4},  \qquad b_6 = \frac{T^{ES}-T^{AS}}{4},  \qquad c_6 = \frac{-T+C}{4}, \nonumber\\
  a_{15}&  =\frac{E+A}{8},  \qquad b_{15} = \frac{T^{ES}+T^{AS}}{8},  \qquad c_{15} = \frac{T+C}{8},\nonumber\\
 a^t_3& = \frac{3}{8}E-\frac{1}{8}A+T^{LA},\qquad
 a^p_3 = -\frac{1}{8}E+\frac{3}{8}A+ T^{QA},\nonumber\\
  b^t_3 & = \frac{3}{8}T^{ES}-\frac{1}{8}T^{AS}+T^{LS},\qquad
 b^p_3  = -\frac{1}{8}T^{ES}+\frac{3}{8}T^{AS} + T^{QS},\nonumber\\
  c^t_3 & = -\frac{1}{8}T+\frac{3}{8}C-\frac{1}{8}T^{ES} + \frac{3}{8}T^{AS}+T^{LC},\qquad
  c^p_3  = \frac{3}{8}T-\frac{1}{8}C+\frac{3}{8}T^{ES} -\frac{1}{8}T^{AS}+T^{QC},\nonumber\\
   d^t_3 & = \frac{3}{8}T-\frac{1}{8}C-\frac{1}{8}E + \frac{3}{8}A+T^{LP},\qquad
   d^p_3  = -\frac{1}{8}T+\frac{3}{8}C+\frac{3}{8}E - \frac{1}{8}A+T^{QP}.
\end{align}
Eq.~\eqref{sol} declares the equivalence between the topological amplitudes and $SU(3)$ irreducible amplitudes.

From above discussions, one can find the sole difference between the TDA and IRA approaches is whether the four-quark operators (or equivalent, $(H)^{ij}_k$) are decomposed into the $SU(3)$ irreducible representations or not.
In Sec.~\ref{ppv}, the topological and $SU(3)$ irreducible amplitudes of the $D\to PP$ decays in the SM will be presented to verify Eq.~\eqref{sol}.
The equivalence of TDA and IRA approaches is also verified in the $D\to PV$ decays, see Appendix \ref{pv} for details.

The $SU(3)$ decomposition of tensor operator $O_{ij}^k$ can be generalized to non-leptonic $b$ decays.
Since the sole difference of $O_{ij}^k$ in $b$ decay and charm decay is the heavy quark, the decomposition should be the same.
The decomposition of $b$ decay is discussed in Appendix \ref{b}, in which some mistakes in Refs.~\cite{He:2018php,He:2018joe} are cleared.

\subsection{Linear correlation of topologies}\label{ind}
In this subsection, we discuss the linear correlation of topologies.
Because of $(H(\overline 6))^{ij}_k=\epsilon^{ijl}(H(\overline 6))_{lk}$ and the symmetric lower indices in $(H(\overline 6))_{lk}$, the terms constructed by $(H(\overline 6))^{ij}_k$ in Eq.~\eqref{nonisodecomp} can be written as
\begin{align}\label{c1}
a_6(D_\gamma)_i (H(\overline 6))^{ij}_k(P_\alpha)_j^l(P_\beta)_l^k &= a_6(D_\gamma)_i \epsilon^{ijm}(H(\overline 6))_{km}(P_\alpha)_j^l(P_\beta)_l^k \nonumber\\&~~~~= a_6(D_\gamma)^{[jm]} (H(\overline 6))_{km}(P_\alpha)_j^l(P_\beta)_l^k,
\end{align}
\begin{align}\label{c2}
b_6(D_\gamma)_i (H(\overline 6))^{ij}_k(P_\alpha)_j^k(P_\beta)^l_l &= b_6(D_\gamma)_i \epsilon^{ijm}(H(\overline 6))_{km}(P_\alpha)_j^k(P_\beta)^l_l \nonumber\\&~~~~= b_6(D_\gamma)^{[jm]} (H(\overline 6))_{km}(P_\alpha)_j^k(P_\beta)^l_l,
\end{align}
\begin{align}\label{c3}
 c_6(D_\gamma)_i (H(\overline 6))^{jl}_k(P_\alpha)_j^i(P_\beta)_l^k &=  \frac{1}{2}c_6\, \epsilon_{pqi}\epsilon^{jlm}(D_\gamma)^{[pq]} (H(\overline 6))_{km}(P_\alpha)_j^i(P_\beta)_l^k\nonumber\\&=c_6\big[-(D_\gamma)^{[jm]} (H(\overline 6))_{km}(P_\alpha)_j^l(P_\beta)_l^k+(D_\gamma)^{[jm]} (H(\overline 6))_{km}(P_\alpha)_j^k(P_\beta)^l_l\nonumber\\&~~~~~~~~~~~~~~~~+(D_\gamma)^{[jl]} (H(\overline 6))_{ki}(P_\alpha)_j^i(P_\beta)_l^k\big],
\end{align}
in which the equations
\begin{equation}
T_i=\epsilon_{ijk}T^{[jk]}/2\qquad {\rm and}\qquad
  \epsilon_{ijk}\epsilon^{lmn} =
\left|\begin{array}{cccc}
    \delta_i^l &    \delta_i^m    & \delta_i^n \\
    \delta_j^l &   \delta_j^m   & \delta_j^n\\
    \delta_k^l &\delta_k^m & \delta_k^n
\end{array}\right|
\end{equation}
are used.
The last term in Eq.~\eqref{c3}, $c_6(D_\gamma)^{[jl]} (H(\overline 6))_{ki}(P_\alpha)_j^i(P_\beta)_l^k$, cancels with its $\alpha$-$\beta$ interchanging term $c_6(D_\gamma)^{[jl]} (H(\overline 6))_{ki}(P_\beta)_j^i(P_\alpha)_l^k$
because the indices $j,l$ are antisymmetric and indices $k,i$ are symmetric.
Thereby, there are only two $SU(3)$ irreducible amplitudes associated with $\overline 6$ representation contributing to the $D\to PP$ decays.
According to Eqs.~\eqref{c1}$\sim$\eqref{c3}, parameter $c_6$ can be absorbed into $a_6$ and $b_6$ by following redefinition:
\begin{align}\label{x1}
  a_6^\prime = a_6 - c_6,\qquad b_6^\prime = b_6 + c_6.
\end{align}
This redefinition is not sole.
One can also get rid of $a_6$ or $b_6$ via the redefinition of
\begin{align}
  c^{\prime\prime}_6=-a^\prime_6 = c_6 - a_6,\qquad b^{\prime\prime}_6=a^\prime_6 +b^{\prime}_6 = b_6 + a_6,
\end{align}
or
\begin{align}
  c^{\prime\prime\prime}_6=b^\prime_6 = c_6 + b_6,\qquad a^{\prime\prime\prime}_6=a^\prime_6 +b^{\prime}_6 = a_6 + b_6.
\end{align}
Since the topological amplitudes are equivalent to the $SU(3)$ irreducible amplitudes, one of the topological amplitudes in the $D\to PP$ decays is not independent.

From above analysis, it is found the fact that one of the topological diagrams is not independent is only associated with $\overline 6$ representation.
According to Eq.~\eqref{sol}, the $SU(3)$ irreducible amplitudes of $\overline 6$ representation have nothing to do with the topologies involving quark-loop.
It means one of the topologies $T$, $C$, $E$, $A$, $T^{ES}$  $T^{AS}$ is not independent under the $SU(3)_F$ symmetry, no matter the diagrams with quark loop are ignored or not.
Besides, as pointed out in Ref.~\cite{Muller:2015lua}, if we drop the diagrams $T^{ES}$ and $T^{AS}$ but include those channels involved in $\eta_1$ in the phenomenological analysis, all the diagrams $T$, $C$, $E$ and $A$ are independent.
And if the channels with $\eta_1$ are not included, only three of $T$, $C$, $E$ and $A$ are independent.
This conclusion can also be understood in Eqs.~\eqref{c1}$\sim$ \eqref{c3}.
According to Eq.~\eqref{sol}, dropping $T^{ES}$ and $T^{AS}$ is equivalent to set $b_{15}=b_6=0$.
Eq.~\eqref{x1} shows that if $b_6$ is zero, $b_6^\prime$ is still non-zero since $b^\prime_6=b_6+c_6$.
Thereby, there are four independent parameters corresponding to the irreducible representations $\underline{15}$ and $\overline 6$.
But if $\eta_1$ is not included in the analysis, all the terms involved $(P_{\alpha,\beta})^l_l$ vanish in Eqs.~\eqref{c1}$\sim$ \eqref{c3}.
Then Eqs.~\eqref{c1}$\sim$ \eqref{c3} are simplified to be
\begin{align}
a_6(D_\gamma)_i (H(\overline 6))^{ij}_k(P_\alpha)_j^l(P_\beta)_l^k &= a_6(D_\gamma)^{[jm]} (H(\overline 6))_{km}(P_\alpha)_j^l(P_\beta)_l^k, \\
 c_6(D_\gamma)_i (H(\overline 6))^{jl}_k(P_\alpha)_j^i(P_\beta)_l^k &=-c_6(D_\gamma)^{[jm]} (H(\overline 6))_{km}(P_\alpha)_j^l(P_\beta)_l^k,
\end{align}
and hence $c_6$ can be absorbed into $a_6$ via $a_6^\prime = a_6 - c_6$. Thereby, there are three independent parameters corresponding to the irreducible representations $\underline{15}$ and $\overline 6$ if $\eta_1$ is not included.

Let us look at the prerequisites of Eqs.~\eqref{c1}$\sim$\eqref{c3}.
In the $SU(3)$ irreducible amplitudes, indices of $(H)^{ij}_k$ transform according to $SU(3)$ group.
The decomposition of $3\otimes3\otimes \overline 3$ is written as
\begin{align}\label{dec}
3\otimes3\otimes \overline 3 = (6\oplus \overline 3)\otimes \overline 3 = (6 \otimes\overline 3)\oplus (\overline 3\otimes\overline 3)=(15\oplus3)\oplus(\overline 6 \oplus 3).
\end{align}
As mentioned above,
the first step of Eq.~\eqref{dec}, $3\otimes 3=6\oplus \overline3$, is symmetrization and anti-symmetrization of $i,j$, $(H)^{ij} = (H)^{\{ij\}}+(H)^{[ij]}$.
The dimension of the antisymmetric tensor $(H)^{[ij]}$ is $C_3^2=3$ which equals to the dimension of fundamental/conjugate representation.
Because of this special character, the two antisymmetric upper indices in $(H)^{[ij]}$ can be written as one lower index via the Levi-Civita tensor: $(H)^{[ij]} = \epsilon^{ijk}(H)_k$.
In the last step of Eq.~\eqref{dec}, $\overline 6$ is gotten by $\overline 3\otimes\overline 3 = \overline 6 \oplus 3$. The two lower indices of $(H)_{kl}$ are symmetrized and anti-symmetrized again and the two lower indices in $(H(\overline 6))_{kl}$ are symmetric.
From Eq.~\eqref{c3}, one can find $(H)^{[ij]} = \epsilon^{ijk}(H)_k$ and the symmetric lower indices in $(H(\overline 6))_{kl}$ are crucial in explaining the linear correlation of the topologies.
Without these specialities of $SU(3)$ group, the Eqs.~\eqref{c1}$\sim$\eqref{c3} cannot be derived and then parameter $c_6$ cannot be absorbed into other parameters by redefinition.
For $B$ meson decays with $SU(4)_F$ symmetry, the topological amplitudes are the same with the one in the $D\to PP$ decays with $SU(3)$ symmetry.
And the $SU(4)$ irreducible amplitudes can be constructed by replacing $3_p$, $3_t$, $\overline 6$ and $\underline{15}$ in Eq.~\eqref{nonisodecomp} with $4_p$, $4_t$, $\overline{20}$ and $\underline{36}$, respectively.
But because the indices of $(H)^{ij}_k$ transform according to $SU(4)$ group, the two anti-symmetric upper index indices in $(H)^{[ij]}_k$ cannot be written as one lower index since $C_N^2=N(N-1)/2 >N$ if $N\geq 4$.
Then the equations similar to Eqs.~\eqref{c1}$\sim$~\eqref{c3} cannot be derived.
As a consequence, the linear correlation of topologies under the $SU(4)_F$ symmetry is different from the case of $SU(3)_F$ symmetry.

The topological amplitude of $K\to \pi\pi$ decay with Isospin symmetry is the same with the one in Eq.~\eqref{ha} (except for those diagrams involving singlet),
\begin{align}
{\cal A}^{\rm TDA}_{K_\gamma\to \pi_\alpha\pi_\beta} &=  T  (K_\gamma)_i  (H)^{lj}_k(\pi_\alpha)^{i}_j  (\pi_\beta)^k_l + C (K_\gamma)_i  (H)^{jl}_k  (\pi_\alpha)^{i}_j(\pi_\beta)^k_l+E  (K_\gamma)_i  (H)^{il}_j (\pi_\alpha)^j_k (\pi_\beta)^{k}_l \nonumber\\&+ A  (K_\gamma)_i (H)^{li}_j (\pi_\alpha)^j_k (\pi_\beta)^{k}_l  +T^{LP} (K_\gamma)_i (H)^{kl}_{l}(\pi_\alpha)^{i}_j (\pi_\beta)^j_k + T^{LA} (K_\gamma)_i (H)^{il}_{l}(\pi_\alpha)^j_k (P)^{k}_j
\nonumber\\&+T^{QP}(K_\gamma)_i(H)^{lk}_{l}(\pi_\alpha)^{i}_j(\pi_\beta)^j_k+ T^{QA} (K_\gamma)_i  (H)^{li}_{l}(\pi_\alpha)^j_k(\pi_\beta)^{k}_j + \alpha \leftrightarrow \beta.
\end{align}
The explicit decomposition of $O_{ij}^k$ in $2 \otimes\overline 2 \otimes 2 =  2_p\oplus2_t \oplus 4$ is
\begin{align}\label{kdec}
  O^k_{ij}= \frac{1}{3}O(4)^k_{ij}+\delta_j^k\Big(\frac{2}{3}O( 2_t)_i-\frac{1}{3}O(2_p)_i\Big)+
  \delta_i^k\Big(\frac{2}{3}O( 2_p)_j-\frac{1}{3}O( 2_t)_j\Big).
\end{align}
Notice that there is no irreducible representation of $(H)^{ij}_k$ in the decomposition of $2\otimes 2\otimes 2 = 2\oplus2\oplus4$ corresponding to the $\overline 6$ representation in the decomposition of $3\otimes \overline 3\otimes 3 = 3\oplus3\oplus\overline{6}\oplus15$ since $N^2(N-1)/2-N=0$ in $N=2$.
The $SU(2)$  irreducible amplitude of $K\to \pi\pi$ decay can be constructed by replacing $3_p$, $3_t$ and $\underline{15}$ in Eq.~\eqref{nonisodecomp} with $2_p$, $2_t$ and $\underline{4}$ respectively,
\begin{align}\label{non}
{\cal A}^{\rm IRA}_{K_\gamma\to \pi_\alpha\pi_\beta} =&a_{4}(K_\gamma)_i (H(4))^{ij}_k(\pi_\alpha)_j^l(\pi_\beta)_l^k
 + c_{4}(K_\gamma)_i (H({4}))^{jl}_k(\pi_\alpha)_j^i(\pi_\beta)_l^k
 \nonumber\\& +a_2^t (K_\gamma)_i (H(2_t))^i (\pi_\alpha)^k_j(\pi_\beta)^j_k
  +d_2^t (K_\gamma)_i (H(2_t))^k (\pi_\alpha)^i_j(\pi_\beta)^j_k \nonumber\\&+a_2^p (K_\gamma)_i (H(2_p))^i (\pi_\alpha)^k_j(\pi_\beta)^j_k +d_2^p (K_\gamma)_i (H(2_p))^k (\pi_\alpha)^i_j(\pi_\beta)^j_k + \alpha \leftrightarrow \beta.
\end{align}
Notice that there are only six $SU(2)$ irreducible amplitudes in Eq.~\eqref{non}.
Thereby, two of the topologies are not independent in the $K\to \pi\pi$ decays.
According to Eq.~\eqref{kdec}, the relations between topological diagrams and the irreducible amplitudes in the $K\to \pi\pi$ decays are
\begin{align}
  a_{4}&  =\frac{E+A}{3},  \qquad c_{4} = \frac{T+C}{3},\nonumber\\
 a^t_2& = \frac{2}{3}E-\frac{1}{3}A+T^{LA},\qquad
 a^p_2 = -\frac{1}{3}E+\frac{2}{3}A+ T^{QA},\nonumber\\
   d^t_2 & = \frac{2}{3}T-\frac{1}{3}C-\frac{1}{3}E + \frac{2}{3}A+T^{LP},\qquad
   d^p_2  = -\frac{1}{3}T+\frac{2}{3}C+\frac{2}{3}E - \frac{1}{3}A+T^{QP}.
\end{align}

The covariant index and contravariant indices of the tensor representation of $SU(N)$ group can transform to each other via the completely antisymmetric tensor $\epsilon_{i_1i_2...i_N}$ and $\epsilon^{i_1i_2...i_N}$.
The $(2,1)$-rank mixed tensor can be written as a tensor only containing upper indices via $\epsilon^{i_1i_2...i_N}$.
For examples, $(H)^{ijl} = \epsilon^{kl}(H)^{ij}_k$, $(H)^{ijlm} = \epsilon^{klm}(H)^{ij}_k$, $(H)^{ijlmn} = \epsilon^{klmn}(H)^{ij}_k$, etc.
To intuitively understand the difference of the $SU(2)$, $SU(3)$ and $SU(4)$ groups, we compare the Young's tableaux of decomposition \eqref{c4} in the cases of $N=2$, $3$ and $4$:
\begin{align}\label{c7}
N=2: & \qquad \square \,\otimes \,\square \,\otimes \,\square =\,\square\hspace{-0.45mm}\square\hspace{-0.45mm}\square \,\oplus \,\square \,\oplus\, \square, \nonumber\\
N=3: & \qquad\square\, \otimes \begin{array}{c}\square \\[-4.6mm]\square\end{array} \otimes \,\square = \begin{array}{l}\square\hspace{-0.45mm}\square\hspace{-0.45mm}\square \\ [-4.6mm]\square\end{array}  \oplus \begin{array}{l}\square\hspace{-0.45mm}\square \\[-4.6mm]\square\hspace{-0.45mm}\square \end{array}  \oplus\, \square\,\oplus\, \square, \nonumber\\
N=4: & \qquad\square \,\otimes \begin{array}{c}\square \\[-4.6mm]\square\\[-4.6mm]\square\end{array} \otimes \,\square = \begin{array}{l}\square\hspace{-0.45mm}\square\hspace{-0.45mm}\square \\ [-4.6mm]\square\\ [-4.6mm]\square\end{array}  \,\oplus \begin{array}{l}\square\hspace{-0.45mm}\square \\[-4.6mm]\square\hspace{-0.45mm}\square \\ [-4.6mm]\square\end{array}  \,\oplus \,\square \,\oplus\, \square,
\end{align}
in which each square presents one upper index of tensor $(H)^{i_1i_2...i_N}$.
For example, representation $\underline{15}$ of $SU(3)$ group needs four indices if it is written as $(H)^{i_1i_2...i_N}$. Compared to the $(2,1)$-rank mixed tensor with three indices, there is no reduction in the number of index in $\underline{15}$.
One can find only the representation $N$ reduces the number of index from three to one if all lower indices are transformed into upper indices.
Representation $N$ is associated with quark-loop diagram.
The corresponding $(H)^{ij}_k$ without $SU(N)$ decomposition always appears as $(H)^{ij}_i$ or $(H)^{ij}_j$.
The number of free index of $(H)^{ij}_i$ and $(H)^{ij}_j$ is the same as representation $N$'s.
So representation $N$ does not reduce the number of independent index.

On the other hand, one can write $(H)^{ij}_k$ and its irreducible representations as tensors only containing lower indices via $\epsilon_{i_1i_2...i_N}$.
And the young's tableaux in Eq.~\eqref{c7} are transformed to be
\begin{align}\label{c9}
N=2: & \qquad \,\blacksquare \,\otimes \,\blacksquare\,\otimes \,\blacksquare =\blacksquare\hspace{-0.25mm}\blacksquare \hspace{-0.25mm}\blacksquare\,\oplus \, \blacksquare  \,\oplus \, \blacksquare, \nonumber\\
N=3: & \qquad\begin{array}{c}\blacksquare \\[-4.4mm]\blacksquare\end{array} \otimes \,\blacksquare\,\otimes \begin{array}{c}\blacksquare \\[-4.4mm]\blacksquare\end{array}  = \begin{array}{l}\blacksquare\hspace{-0.25mm}\blacksquare\hspace{-0.25mm}\blacksquare \\ [-4.4mm]\blacksquare\hspace{-0.25mm}\blacksquare\end{array}  \oplus \, \blacksquare\hspace{-0.25mm}\blacksquare \,\oplus \begin{array}{c}\blacksquare \\[-4.4mm]\blacksquare\end{array}\oplus \begin{array}{c}\blacksquare \\[-4.4mm]\blacksquare\end{array}, \nonumber\\
N=4: & \qquad\begin{array}{c}\blacksquare \\[-4.4mm]\blacksquare\\[-4.4mm]\blacksquare\end{array}  \otimes \,\blacksquare \,\otimes \begin{array}{c}\blacksquare \\[-4.4mm]\blacksquare\\[-4.4mm]\blacksquare\end{array} = \begin{array}{l}\blacksquare\hspace{-0.25mm}\blacksquare\hspace{-0.25mm}\blacksquare \\ [-4.4mm]\blacksquare\hspace{-0.25mm}\blacksquare\\ [-4.4mm]\blacksquare\hspace{-0.25mm}\blacksquare\end{array} \oplus \begin{array}{l}\blacksquare\hspace{-0.25mm}\blacksquare \\[-4.4mm]\blacksquare\end{array}  \oplus \begin{array}{c}\blacksquare \\[-4.4mm]\blacksquare\\[-4.4mm]\blacksquare\end{array}\oplus \begin{array}{c}\blacksquare \\[-4.4mm]\blacksquare\\[-4.4mm]\blacksquare\end{array},
\end{align}
in which one $\blacksquare$ presents one lower index.
From Eq.~\eqref{c9}, one can find if all upper indices transform into lower indices, the $\overline 6$ representation of $SU(3)$ group has two indices, smaller than three.
And $\blacksquare\hspace{-0.25mm}\blacksquare$ presents the two symmetric lower indices.
For other representations showed in Eq.~\eqref{c9}, writing all indices into lower indices does not reduce the number of indices compared to $(2,1)$-rank mixed tensor.

In summary, the linear correlation of topologies depends on the symmetry of the physical system.
For different symmetry group, the linear correlation of topologies is different, and can be explained in group theory.

\section{Topologies of $D\to PP$ decays in the SM}\label{ppv}

In this section, we present the amplitude decompositions of the $D\to PP$ decays in the Standard Model and discuss the applications.

\subsection{Topologies in the SM: classification and linear correlation}\label{def}
According to Eq.~\eqref{hsm} and  Eq.~\eqref{CKM}, the CKM
components in the SM, $(H^{(p)})^{ij}_k$, can be obtained from the map $(\bar uq_1)(\bar q_2c)\rightarrow V^*_{cq_2}V_{uq_1}$ in current-current operators and $(\bar qq)(\bar uc)\rightarrow -V^*_{cb}V_{ub}$ in penguin operators and the others are set to be zero.
The non-zero CKM components induced by tree operators in the topological amplitude include
\begin{align}\label{ckm1}
 &(H^{(0)})^{13}_2 = V_{cs}^*V_{ud},  \qquad (H^{(0)})_{2}^{12}=V_{cd}^*V_{ud},\qquad (H^{(0)})_{3}^{13}= V_{cs}^*V_{us}, \qquad (H^{(0)})_{3}^{12}=V_{cd}^*V_{us}.
\end{align}
 The non-zero CKM components induced by penguin operators in the topological amplitude include
\begin{align}\label{ckm2}
 &(H^{(1)})^{11}_1 = -V_{cb}^*V_{ub}, \qquad (H^{(1)})^{21}_2=-V_{cb}^*V_{ub}, \qquad (H^{(1)})^{31}_3=-V_{cb}^*V_{ub}.
\end{align}
The superscripts (0) and (1) differentiate tree and penguin contributions.
The non-zero CKM components induced by the tree operators in the $SU(3)$ irreducible representations are
\begin{align}\label{ckm3}
 &  (H^{(0)}( \overline6))_{22}=-2\,V_{cs}^*V_{ud},\qquad (H^{(0)}( \overline 6))_{23}=(V_{cd}^*V_{ud}-V_{cs}^*V_{us}),  \qquad (H^{(0)}( \overline 6))_{33}=  2\,V_{cd}^*V_{us},\nonumber \\
   &  (H^{(0)}(15))_{1}^{11}=-2\,(V_{cd}^*V_{ud}+V_{cs}^*V_{us}), \qquad (H^{(0)}(15))_{2}^{13}= 4\,V_{cs}^*V_{ud},  \qquad  (H^{(0)}(15))_{3}^{12}=4\,V_{cd}^*V_{us},\nonumber \\
 &  (H^{(0)}(15))_{2}^{12}= 3\,V_{cd}^*V_{ud}-V_{cs}^*V_{us},\qquad (H^{(0)}(15))_{3}^{13}=3\,V_{cs}^*V_{us}-V_{cd}^*V_{ud}, \nonumber \\ & (H^{(0)}( 3_t))^1=V_{cd}^*V_{ud}+V_{cs}^*V_{us}.
\end{align}
The non-zero CKM components induced by the penguin operators in the $SU(3)$ irreducible representations are
\begin{align}\label{ckm4}
 (H^{(1)}( 3_t))^1=-V_{cb}^*V_{ub}, \qquad (H^{(1)}( 3_p))^1=-3V_{cb}^*V_{ub}.
\end{align}

In general, the topologies in the SM are classified into two types: tree diagram and penguin diagram.
The quark-loop contributions induced by tree operators are absorbed into the Wilson coefficients of penguin operators \cite{Beneke:2001ev,Beneke:2003zv,Beneke:2000ry,Beneke:1999br},
\begin{align}
  C_{3,5}(\mu) & \rightarrow  C_{3,5}(\mu) - \frac{\alpha_s(\mu)}{8\pi N_c} \sum_{q=d,s} \frac{\lambda_q}{\lambda_b}C^{(q)}(\mu,\langle l^2\rangle), \nonumber\\
    C_{4,6}(\mu) & \rightarrow  C_{4,6}(\mu) - \frac{\alpha_s(\mu)}{8\pi} \sum_{q=d,s} \frac{\lambda_q}{\lambda_b}C^{(q)}(\mu,\langle l^2\rangle),
\end{align}
with the averaged invariant mass squared of the virtual gluon emitted from the quark loop $\langle l^2\rangle$ and the function
\begin{align}
  C^{(q)}(\mu,\langle l^2\rangle) = \Big[ -4\int_0^1\,dx\,x(1-x)ln\frac{m^2_q-x(1-x)\langle l^2\rangle}{\mu^2}  - \frac{2}{3}\Big]C_2(\mu).
\end{align}
The penguin operator induced quark loop contributions are negligible.
Thereby, the penguin diagrams include those diagram induced by penguin operators without quark-loop and quark-loop diagrams induced by tree operators.
But this classification is not convenient to analyze the topologies in the tensor.

We suggest to put the traditional tree diagrams and the quark-loop diagrams induced by tree operators ($O_{1,2}$) together,  named after "tree-operator-induced diagrams", and put the diagrams induced by penguin operators ($O_{3-6}$) together, named after "penguin-operator-induced diagrams".
The new classification of topologies is according to which operators ($O_{1,2}$ or $O_{3-6}$) being inserted into the diagrams, indifferent to the diagrams with quark-loop or not.
Because the Wilson coefficients $C_{1,2}$ are larger than $C_{3-6}$, the new classification is also based on magnitude of the Wilson coefficients, or in other word, the perturbation order $p$ introduced in Eq. \eqref{hg}.
The advantage of the new classification is that it is convenient to build the one-to-one mapping between topology and invariant tensor, just like we have done in Sec.~\ref{ga}.
And then some mathematical techniques, such as group theory, can be induced to study the topological amplitudes.

Revisit the topological diagrams listed in Fig.~\ref{top1}.
There are ten tree-operator-induced diagrams contributing to the $D\to PP$ decays in the SM.
The diagrams $T^{QP}$, $T^{QC}$, $T^{QA}$, $T^{QS}$ vanish because there is no tree level FCNC transition in the SM.
But for penguin-operator-induced diagrams, all the fourteen topologies contribute to the $D\to PP$ decays.
In the rest of paper, notations $PT$, $PC$, $PE$, $PA$, $P^{ES}$, $P^{AS}$, $P^{LP}$, $P^{LC}$, $P^{LA}$, $P^{LS}$, $P^{QP}$, $P^{QC}$, $P^{QA}$ and $P^{QS}$ are used to label the penguin-operator-induced amplitudes corresponding the topologies in Fig.~\eqref{top1} orderly.
And all the $SU(3)$ irreducible amplitudes induced by penguin operators are added $P$ before their notations to differentiate the amplitudes induced by tree operators.

In order to compare with other literature, we give the relation between our classification of the topologies and the traditional one (see Ref.~\cite{Cheng:2011qh} for example).
The tree-operator-induced diagrams without quark loop defined in this work (left) and the tree diagrams defined in Ref.~\cite{Cheng:2011qh} (right) have one-to-one correspondence:
\begin{equation}
  \begin{split}
  T_{\rm this\,\,work} &= T_{\rm Ref.[100]},\qquad C_{\rm this\,\,work} = C_{\rm Ref.[100]},\qquad E_{\rm this\,\,work}=E_{\rm Ref.[100]},\\ A_{\rm this\,\,work}&=A_{\rm Ref.[100]},\qquad T^{ES}_{\rm this\,\,work} = SE_{\rm Ref.[100]},\qquad T^{AS}_{\rm this\,\,work} = SA_{\rm Ref.[100]}.
  \end{split}
\end{equation}
The tree-operator-induced diagrams with quark-loop in this work are the quark-loop contributions of the penguin diagrams in Ref.~\cite{Cheng:2011qh}. The relations between them are
\begin{align}\label{re}
 T^{LP}_{\rm this\,\,work} &= (P_{\rm loop}+PE_{\rm loop})_{\rm Ref.[100]}, \qquad T^{LC}_{\rm this\,\,work} = (S_{\rm loop}+SPE_{\rm loop})_{\rm Ref.[100]},\nonumber\\ T^{LA}_{\rm this\,\,work}&=(PA_{\rm loop})_{\rm Ref.[100]},\qquad T^{LS}_{\rm this\,\,work} = (SPA_{\rm loop})_{\rm Ref.[100]},
\end{align}
in which the subscript "loop" is used to distinguish them from the contributions induced by penguin operators.
The penguin-operator-induced diagrams without quark loop in this work are the penguin diagrams proportional to $C_{3-6}$ in Ref.~\cite{Cheng:2011qh}.
The relations between them are
\begin{align}
 & PC_{\rm this\,\,work} = (P_{\rm pen})_{\rm Ref.[100]},\qquad PE_{\rm this\,\,work}= (PE_{\rm pen})_{\rm Ref.[100]},\nonumber\\ &PA_{\rm this\,\,work}= (PA_{\rm pen})_{\rm Ref.[100]}, \qquad PT_{\rm this\,\,work}= (S_{\rm pen})_{\rm Ref.[100]},\nonumber\\ &P^{ES}_{\rm this\,\,work} = (SPE_{\rm pen})_{\rm Ref.[100]},\qquad P^{AS}_{\rm this\,\,work} = (SPA_{\rm pen})_{\rm Ref.[100]},
\end{align}
in which the subscript "pen" represent the $O_{3-6}$ contributions.
 The topologies defined in Ref.~\cite{Cheng:2011qh}, $P$ and $PE$, as well as $S$ and $SPE$, always appear as $P+PE$ and $S+SPE$.
It is easy to understand since our definition of topology only consider quark line flowing into and out of hadrons but does not care the gluon exchanges.
The only difference between $P$ and $PE$ (as well as $S$ and $SPE$), actually, is the different gluon exchanges.

\begin{table}[t!]
\caption{Decay amplitudes of the Cabibblo-allowed and doubly Cabibbo-suppressed $D\to PP$ decays.}\label{tab:1}
\begin{ruledtabular}
\scriptsize
\begin{tabular}{|c|c|c|}
{Channel} & {TDA}& {IRA}   \\\hline
$D^0\to \pi^+   K^-  $  &  $V^*_{cs}V_{ud}(T + E)$ & $2V^*_{cs}V_{ud}(2a_{15}+a_6+2c_{15}-c_6)$\\\hline
$D^0\to \pi^0   \overline K^0  $ & $\frac{1}{\sqrt{2}}V^*_{cs}V_{ud}(C-E)$ &  $\sqrt{2}V^*_{cs}V_{ud}(-2a_{15}-a_6+2c_{15}+c_6)$ \\\hline
$D^0\to \overline K^0   \eta_8  $ & $\frac{1}{\sqrt{6}}V^*_{cs}V_{ud}(C-E)$ & $\frac{2}{\sqrt{6}}V^*_{cs}V_{ud}(-2a_{15}-a_6+2c_{15}+c_6)$ \\\hline
$D^0\to \overline K^0   \eta_1  $ & ~~~~~~~~~~ $\frac{1}{\sqrt{3}}V^*_{cs}V_{ud}(C+2E+3T^{ES})$~~~~~~~~~ & ~~~~~~~~~~ $\frac{2}{\sqrt{3}}V^*_{cs}V_{ud}(4a_{15}+2a_6+6b_{15}+3b_6+2b_{15}+c_6)$~~~~~~~~~~   \\\hline
$D^+\to \pi^+   \overline K^0  $ & $V^*_{cs}V_{ud}(C+T)$ & $8V^*_{cs}V_{ud}c_{15}$  \\\hline
$D^+_s\to \pi^+\pi^0$ &$/$& $/$\\\hline
$D^+_s\to \pi^+   \eta_8  $ & $\frac{2}{\sqrt{6}}V^*_{cs}V_{ud}(A-T)$ & $\frac{4}{\sqrt{6}}V^*_{cs}V_{ud}(2a_{15}-a_6-2c_{15}+c_6)$ \\\hline
$D^+_s\to \pi^+   \eta_1  $ & $\frac{1}{\sqrt{3}}V^*_{cs}V_{ud}(T+2A+3T^{AS})$ & $\frac{2}{\sqrt{3}}V^*_{cs}V_{ud}(4a_{15}-2a_6+6b_{15}-3b_6+2c_{15}-c_6)$  \\\hline
$D^+_s\to K^+   \overline K^0  $ & $V^*_{cs}V_{ud}(C+A)$ & $2V^*_{cs}V_{ud}(2a_{15}-a_6+2c_{15}+c_6)$\\\hline\hline
$D^0\to \pi^0   K^0  $ & $\frac{1}{\sqrt{2}}V^*_{cd}V_{us}(C-E)$ & $\sqrt{2}V^*_{cd}V_{us}(-2a_{15}-a_6+2c_{15}+c_6)$  \\\hline
$D^0\to \pi^-   K^+  $ & $V^*_{cd}V_{us}(E+T)$ & $2V^*_{cd}V_{us}(2a_{15}+a_6+2c_{15}-c_6)$  \\\hline
$D^0\to K^0   \eta_8  $ & $\frac{1}{\sqrt{6}}V^*_{cd}V_{us}(C-E)$ & $\frac{2}{\sqrt{6}}V^*_{cd}V_{us}(-2a_{15}-a_6+2c_{15}+c_6)$  \\\hline
$D^0\to K^0   \eta_1  $ & $\frac{1}{\sqrt{3}}V^*_{cd}V_{us}(C+2E+3T^{ES})$ & $\frac{2}{\sqrt{3}}V^*_{cd}V_{us}(4a_{15}+2a_6+6b_{15}+3b_{6}+2c_{15}+c_6)$ \\\hline
$D^+\to \pi^+   K^0  $ & $V^*_{cd}V_{us}(C+A)$ & $2V^*_{cd}V_{us}(2a_{15}-a_6+2c_{15}+c_6)$ \\\hline
$D^+\to \pi^0   K^+  $ & $\frac{1}{\sqrt{2}}V^*_{cd}V_{us}(A-T)$ & $\sqrt{2}V^*_{cd}V_{us}(2a_{15}-a_6-2c_{15}+c_6)$  \\\hline
$D^+\to K^+   \eta_8  $ & $\frac{1}{\sqrt{6}}V^*_{cd}V_{us}(T-A)$ & $\frac{2}{\sqrt{6}}V^*_{cd}V_{us}(-2a_{15}+a_6+2c_{15}-c_6)$ \\\hline
$D^+\to K^+   \eta_1  $ & $\frac{1}{\sqrt{3}}V^*_{cd}V_{us}(T+2A+3T^{AS})$ & $\frac{2}{\sqrt{3}}V^*_{cd}V_{us}(4a_{15}-2a_6+6b_{15}-3b_6+2c_{15}-c_6)$  \\\hline
$D^+_s\to K^+   K^0  $ & $V^*_{cd}V_{us}(C+T)$ & $8V^*_{cd}V_{us}c_{15}$  \\
\end{tabular}
\end{ruledtabular}
\end{table}
\begin{table}[thp!]
\caption{Decay amplitudes of the singly Cabibblo-suppressed $D\to PP$ decays.
The CKM matrix elements are labeled as $\lambda_d=V_{cd}^*V_{ud}$, $\lambda_s=V_{cs}^*V_{us}$, $\lambda_b=V_{cb}^*V_{ub}$ and $\lambda_b = -(\lambda_d+\lambda_s)$.
The redefinitions \eqref{c5} and \eqref{c11} are used for simplification.
}\label{tab:2}
\begin{ruledtabular}
\scriptsize
\begin{tabular}{|c|c|c|}
{channel}& {TDA}  &{IRA}    \\\hline
$D^0\to \pi^+   \pi^-  $ & $\lambda_d(T+E) -\lambda_b(A^{LP}+2A^{LA})$ & $(3\lambda_d-\lambda_s)(a_{15}+c_{15})+(\lambda_d-\lambda_s)(a_6-c_6)$\\ & & $-\lambda_b(2a_3+d_3-2a_{15})$\\ \hline
$D^0\to \pi^0   \pi^0  $ & $\frac{1}{\sqrt{2}}\lambda_d(E-C)-\frac{1}{\sqrt{2}}\lambda_b(A^{LP}+2A^{LA})$ &$\frac{1}{\sqrt{2}}(3\lambda_d-\lambda_s)(a_{15}-c_{15})+\frac{1}{\sqrt{2}}(\lambda_d-\lambda_s)(a_6-c_6)$  \\&&
$-\frac{1}{\sqrt{2}}\lambda_b(2a_3+d_3-2a_{15}-2c_{15})$ \\\hline
$D^0\to \pi^0   \eta_8  $ & $-\frac{1}{\sqrt{3}}(\lambda_dE+\lambda_sC)-\frac{1}{\sqrt{3}}\lambda_bA^{LP}$ & $-\frac{1}{\sqrt{3}}(3\lambda_d-\lambda_s)a_{15}-\frac{1}{\sqrt{3}}(3\lambda_s-\lambda_d)c_{15}
$\\
&&$-\frac{1}{\sqrt{3}}(\lambda_d-\lambda_s)(a_6-c_6)
-\frac{1}{\sqrt{3}}\lambda_b(d_3-2a_{15}-2c_{15})$ \\\hline
$D^0\to \pi^0   \eta_1  $ & $-\frac{1}{\sqrt{6}}\lambda_d(2E+3T^{ES})+\frac{1}{\sqrt{6}}\lambda_sC$& $-\frac{1}{\sqrt{6}}(3\lambda_d-\lambda_s)(2a_{15}+3b_{15})+\frac{1}{\sqrt{6}}(3\lambda_s-\lambda_d)c_{15}$ \\&$-\frac{1}{\sqrt{6}}\lambda_b(2A^{LP}+3A^{LC})$ & $
-\frac{1}{\sqrt{6}}(\lambda_d-\lambda_s)(2a_6+3b_6+c_6)$  \\& & $-\frac{1}{\sqrt{6}}\lambda_b(3c_3+2d_3-2(2a_{15}+3b_{15}+2c_{15})$ \\\hline
$D^0\to K^+   K^-  $ & $\lambda_s(T+E)-\lambda_b(A^{LP}+2A^{LA})$ & $(3\lambda_s-\lambda_d)(a_{15}+c_{15})-(\lambda_d-\lambda_s)(a_6-c_6)$\\&  & $-\lambda_b(2a_3+d_3-2a_{15})$\\ \hline
$D^0\to K^0  \overline K^0  $ &$-\lambda_b(E+2A^{LA})$ & $-2\lambda_b(a_3+a_{15})$\\\hline
$D^0\to \eta_8   \eta_8  $ & $\frac{1}{3\sqrt{2}}\lambda_d(C+E)-\frac{2}{3\sqrt{2}}\lambda_s(C-2E)$ & $\frac{\sqrt{2}}{3}(3\lambda_s-\lambda_d)(2a_{15}-c_{15})+\frac{1}{3\sqrt{2}}(3\lambda_d-\lambda_s)(a_{15}+c_{15})$\\ & $-\frac{1}{3\sqrt{2}}\lambda_b(A^{LP}+6A^{LA})$& $-\frac{1}{\sqrt{2}}(\lambda_d-\lambda_s)(a_6-c_6)-\frac{1}{3\sqrt{2}}\lambda_b(6a_3+d_3-2(a_{15}+c_{15})$\\\hline
$D^0\to \eta_8   \eta_1  $ & $\frac{1}{3\sqrt{2}}\lambda_d(2C+2E+3T^{ES})-\frac{1}{3\sqrt{2}}\lambda_s(C$& $\frac{1}{3\sqrt{2}}(3\lambda_d-\lambda_s)(2a_{15}+3b_{15}+2c_{15})-\frac{1}{3\sqrt{2}}(3\lambda_s-\lambda_d)$\\&
$+4E+6T^{ES})-\frac{1}{3\sqrt{2}}\lambda_b(2A^{LP}+3A^{LC})
$ & $(4a_{15}+6b_{15}+b_{15})+
\frac{1}{\sqrt{2}}(\lambda_d-\lambda_s)(2a_6+3b_6+c_6)$ \\ & &$
-\frac{1}{3\sqrt{2}}\lambda_b(3c_3+2d_3-2(2a_{15}+3b_{15}+2c_{15}))$\\\hline
$D^0\to \eta_1   \eta_1  $ & $-\frac{\sqrt{2}}{3}\lambda_b(C+E+3T^{ES}+A^{LP}$ & $-\frac{\sqrt{2}}{3}\lambda_b(3a_3+9b_3+3c_3+d_3)$\\ &$+3A^{LA}+3A^{LC}+9A^{LS})$ & \\ \hline
$D^+\to \pi^+   \pi^0  $& $-\frac{1}{\sqrt{2}}\lambda_d(T+C)$ & $-2\sqrt{2}\lambda_dc_{15}$ \\\hline
$D^+\to \pi^+   \eta_8  $ & ~~ $\frac{1}{\sqrt{6}}\lambda_d(T+C+2A)-\frac{2}{\sqrt{6}}\lambda_sC-\frac{2}{\sqrt{6}}\lambda_bA^{LP}$~~ & $\frac{2}{\sqrt{6}}(3\lambda_d-\lambda_s)(a_{15}+c_{15})-\frac{2}{\sqrt{6}}(3\lambda_s-\lambda_d)c_{15}
$ \\ & & $-\frac{2}{\sqrt{6}}(\lambda_d-\lambda_s)(a_6-c_6)-\frac{1}{\sqrt{6}}\lambda_b(2d_3-2c_{15})$\\\hline
$D^+\to \pi^+   \eta_1  $ & $\frac{1}{\sqrt{3}}\lambda_d(T+C+2A+3T^{AS})+\frac{1}{\sqrt{3}}\lambda_sC$ & $\frac{1}{\sqrt{3}}(3\lambda_d-\lambda_s)(2a_{15}+3b_{15}+2c_{15})+\frac{1}{\sqrt{3}}(3\lambda_s-\lambda_d)c_{15}$ \\  & $-\frac{1}{\sqrt{3}}\lambda_b(2A^{LP}+3A^{LC})$  & $-\frac{1}{\sqrt{3}}(\lambda_d-\lambda_s)(2a_6+3b_6+c_6)-\frac{1}{\sqrt{3}}\lambda_b(3c_3+2d_3-2c_{15})$\\\hline
$D^+\to K^+   \overline K^0  $ & $\lambda_dA+\lambda_sT-\lambda_bA^{LP}$ & $(3\lambda_d-\lambda_s)a_{15}+(3\lambda_s-\lambda_d)c_{15}-(\lambda_d-\lambda_s)(a_6-c_6)-\lambda_bd_3$  \\\hline
$D^+_s\to \pi^+   K^0  $ & $\lambda_dT+\lambda_sA-\lambda_bA^{LP}$& $(3\lambda_d-\lambda_s)c_{15}+(3\lambda_s-\lambda_d)a_{15}+(\lambda_d-\lambda_s)(a_6-c_6)-\lambda_bd_3$  \\\hline
$D^+_s\to \pi^0   K^+  $ & $-\frac{1}{\sqrt{2}}(\lambda_dC-\lambda_sA)-\frac{1}{\sqrt{2}}\lambda_bA^{LP}$ & $-\frac{1}{\sqrt{2}}(3\lambda_d-\lambda_s)c_{15}
+\frac{1}{\sqrt{2}}(3\lambda_s-\lambda_d)a_{15}
$ \\& & $+\frac{1}{\sqrt{2}}(\lambda_d-\lambda_s)(a_6-c_6)-\frac{1}{\sqrt{2}}\lambda_b(d_3-2c_{15})$  \\\hline
$D^+_s\to K^+   \eta_8  $ & $\frac{1}{\sqrt{6}}\lambda_dC-\frac{1}{\sqrt{6}}\lambda_s(2T+2C+A)+\frac{1}{\sqrt{6}}\lambda_bA^{LP}$ & $-\frac{1}{\sqrt{6}}(\lambda_d-\lambda_s)(a_6-c_6)+\frac{1}{\sqrt{6}}(3\lambda_d-\lambda_s)c_{15}
$ \\ && $-\frac{1}{\sqrt{6}}(3\lambda_s-\lambda_d)(a_{15}+4c_{15})+
\frac{1}{\sqrt{6}}\lambda_b(d_3+2c_{15})$  \\\hline
$D^+_s\to K^+   \eta_1  $ & $\frac{1}{\sqrt{3}}\lambda_dC+\frac{1}{\sqrt{3}}\lambda_s(T+C+2A+3T^{AS})$ & $\frac{1}{\sqrt{3}}(3\lambda_d-\lambda_s)c_{15}+\frac{1}{\sqrt{3}}(3\lambda_s-\lambda_d)(2a_{15}+3b_{15}+2c_{15})$ \\ &$-\frac{1}{\sqrt{3}}\lambda_b(2A^{LP}+3A^{LC})$ & $+\frac{1}{\sqrt{3}}(\lambda_d-\lambda_s)(2a_6+3b_6+c_6)-\frac{1}{\sqrt{3}}\lambda_b(3c_3+2d_3-2c_{15})$
\end{tabular}
\end{ruledtabular}
\end{table}

The linear correlation of topological diagrams in the SM is beyond the model-independent analysis in \ref{ind} because some characters of the effective Hamiltonian of the SM.
In the IRA approach, Eq.~\eqref{ckm4} shows that there are no penguin-operator-induced amplitudes in the $15$- and $6$-dimensional irreducible representations.
From Eqs.~\eqref{ckm3} and \eqref{ckm4}, it is found the non-zero CKM components of $3$-dimensional representations, including $(H^{(0)}(3_t))^1$,  $(H^{(1)}(3_t))^1$ and $(H^{(1)}(3_p))^1$, only contain the first components.
Because of the unitarity of the CKM matrix, we have $(H^{(0)}(3_t))^1 = (V_{cd}^*V_{ud}+V_{cs}^*V_{us}) = -V_{cb}^*V_{ub}$, and then
$(H^{(0)}(3_t)):(H^{(1)}(3_t)):(H^{(1)}(3_p)) = 1:1:3$.
So the amplitudes induced by the $3$-dimensional presentations always appear simultaneously and can be absorbed into four parameters with following redefinition:
\begin{align}\label{c5}
 & a_3  = a_3^t+Pa_3^t + 3 Pa_3^p, \qquad   b_3 =b_3^t+Pb_3^t + 3 Pb_3^p, \qquad   c_3 = c_3^t+Pc_3^t + 3 Pc_3^p, \nonumber\\ &   d_3 = d_3^t+Pd_3^t + 3 Pd_3^p,
\end{align}
Notice that Eq.~\eqref{c5} only holds in the Standard Model but not a general conclusion.
According to Eq.~\eqref{c5} and Eq.~\eqref{sol}, the tree-operator-induced diagrams with quark loop and all the penguin-operator-induced diagrams can be absorbed into four parameters with following redefinition:
\begin{align}\label{c11}
 & A^{LA}  = T^{LA}+PA+P^{LA}+3P^{QA}, \qquad   A^{LS}  = T^{LS}+P^{AS}+P^{LS}+3P^{QS}, \nonumber\\ &  A^{LC}  = T^{LC}+PT+P^{ES}+P^{LC}+3P^{QC}, \qquad  A^{LP}  = T^{LP}+PC+PE+P^{LP}+3P^{QP}.
\end{align}
All the penguin-operator-induced amplitudes are determined if the tree-operator-induced amplitudes with quark loop are known.
There is no degree of freedom of the penguin-operator-induced diagrams in the SM.
It is understandable since a penguin operator can be regard as a "quark-loop" induced by "tree operator" in high energy scale.
According to \ref{ind}, one of the ten tree-operator-induced topological diagrams of the $D\to PP$ decays is not independent.
Thereby, there are only nine degrees of freedom for all the tree- and penguin-operator-induced amplitudes in the $D\to PP$ decays in the SM.

The tree- and penguin-operator-induced amplitudes of the $D\to PP$ decays in the TDA and IRA approaches are listed in Tables.~\ref{tab:1} and \ref{tab:2}.
One can check the topological amplitudes and the $SU(3)$ irreducible amplitudes follow Eq.~\eqref{sol}.
In Ref.~\cite{He:2018joe}, the penguin-operator-induced amplitudes are not listed in tables because the authors think the penguin-operator-induced amplitudes can be obtained from tree-operator-induced amplitudes by a simply replacement $T\to PT$, $C\to PC$, ... .
It should be pointed out that the penguin-operator-induced amplitudes are determined by the tree-operator-induced amplitudes through Eq.~\eqref{c5} and Eq.~\eqref{c11} rather than a simply replacement.

\subsection{$CP$ violation in charm}\label{cp}

In 2019, LHCb Collaboration observed the direct $CP$ violation in charm at $5.3\,\sigma$ \cite{Aaij:2019kcg}.
The new world average of $\Delta a_{CP}^{\rm dir}$ given by the Heavy Flavor Averaging Group (HFLAV) is \cite{Amhis:2016xyh}
\begin{align}
\Delta a_{CP}^{\rm dir} = (-1.64\pm0.28)\times 10^{-3}.
\end{align}
According to this result, Ref.~\cite{Grossman:2019xcj} proposed a $\Delta U =0$ rule in the charm physics: the ratio of $\Delta U =0$ over $\Delta U =1$ amplitudes is
\begin{align}\label{u0}
|\tilde{p}_0|\sin(\delta_{\rm strong}) = 0.65\pm 0.12.
\end{align}
For the $\Delta U =0$ rule, there are two different arguments:
it arises from new physics \cite{Chala:2019fdb,Dery:2019ysp}, or
non-perturbative QCD enhancement \cite{Buccella:2019kpn,Soni:2019xko,Li:2019hho,Cheng:2019ggx}.
On the other hand, a long-standing puzzle in charm decays is the very different $D^0\to K^+K^-$ and $D^0\to \pi^+\pi^-$ decay rates. In general, the $SU(3)$ breaking is expected to be around $30\%$.
For example, amplitude $T$ of the $D$ decaying into $KK$ and $\pi\pi$ in the factorization approach has the expressions as
\begin{align}
  T_{KK} & = \frac{G_F}{\sqrt{2}}a_1(KK)f_K(m^2_D-m^2_K)F_0^{D\to K}(m_K^2),  \nonumber\\
  T_{\pi\pi} & = \frac{G_F}{\sqrt{2}}a_1(\pi\pi)f_\pi(m^2_D-m^2_\pi)F_0^{D\to \pi}(m_\pi^2),
\end{align}
and $T_{KK}/T_{\pi\pi}\approx1.3$ \cite{Cheng:2012xb}.
Such a $SU(3)$ breaking is not enough to explain the branching fractions of $D^0\to K^+K^-$ and $D^0\to \pi^+\pi^-$ since \cite{Tanabashi:2018oca}
\begin{align}
\mathcal{B}r(D^0\to K^+K^-)= (4.08\pm 0.06)\times 10^{-3},\qquad \mathcal{B}r(D^0\to \pi^+\pi^-) = (1.445\pm 0.024)\times 10^{-3},
\end{align}
and
\begin{align}
\frac{|\mathcal{A}(D^0\to K^+K^-)|^2}{|\mathcal{A}(D^0\to \pi^+\pi^-)|^2}\simeq\frac{\mathcal{B}r(D^0\to K^+K^-)}{\mathcal{B}r(D^0\to \pi^+\pi^-)}
 = 2.80\pm 0.02.
\end{align}
In the following, we will show that assuming a large quark-loop diagram $T^{LP}$ could be a better choice to solve the puzzles of large $\Delta a^{\rm dir}_{CP}$ and the very different branching fractions in the $D^0\to K^+K^-$ and $D^0\to \pi^+\pi^-$ decays simultaneously.

In the $SU(3)_F$ limit, the amplitudes of $D^0\to K^+K^-$ and $D^0\to \pi^+\pi^-$ decays are
\begin{align}
  \mathcal{A}(D^0\to K^+K^-) & = \lambda_s(T+E)+ (\lambda_d+\lambda_s)(T^{LP}+2T^{LA})-\lambda_b(PC+PE+2PA), \\
  \mathcal{A}(D^0\to \pi^+\pi^-)& = \lambda_d(T+E)+ (\lambda_d+\lambda_s)(T^{LP}+2T^{LA})-\lambda_b(PC+PE+2PA),
\end{align}
in which the penguin-operator-induced amplitudes with quark-loop are neglected.
Considering the $U$-spin breaking, the amplitude of $D^0\to K^+K^-$ decay can be written as
\begin{align}\label{kk}
  \mathcal{A}(D^0 & \to  K^+K^-) = \lambda_s(T_{KK}+E_{KK})+ \lambda_d(T^{LP}_d+2T^{LA}_d)+\lambda_s(T^{LP}_s+2T^{LA}_s)-\lambda_b\,Pen \nonumber\\&= \lambda_s(T_{KK}+E_{KK})+\lambda_d(T^{LP}_d+2T^{LA}_d)+\lambda_s(T^{LP}_d+2T^{LA}_d +T^{LP}_{\rm break}+2T^{LA}_{\rm break})-\lambda_b\,Pen \nonumber\\&~~=\lambda_s(T_{KK}+E_{KK})+ (\lambda_d+\lambda_s)(T^{LP}_d+2T^{LA}_d) +\lambda_s(T^{LP}_{\rm break} +2T^{LA}_{\rm break})-\lambda_b\,Pen\nonumber\\&~~~~= \lambda_s(T_{KK}+E_{KK})+\lambda_s(T^{LP}_{\rm break}+2T^{LA}_{\rm break})-\lambda_b(T^{LP}_d+2T^{LA}_d+Pen),
\end{align}
where $T^{LP,LA}_{\rm break}= T^{LP,LA}_s-T^{LP,LA}_d$ and $Pen=PC+PE+2PA$.
Similarly, the amplitude of $D^0\to \pi^+\pi^-$ decay can be written as
\begin{align}\label{mq}
  \mathcal{A}(D^0\to \pi^+\pi^-) = \lambda_d(T_{\pi\pi}+E_{\pi\pi})+\lambda_s(T^{LP}_{\rm break}+2T^{LA}_{\rm break})-\lambda_b(T^{LP}_d+2T^{LA}_d+Pen).
\end{align}
In the effective Hamiltonian \eqref{hsm}, the Wilson coefficients $C_{3-6}$ are much smaller than $C_{1,2}$ \cite{Buchalla:1995vs}.
The penguin-operator-induced amplitudes are smaller than the tree-operator-induced ones.
On the other hand, topology $T^{LA}$ is suppressed by the OZI rule \cite{OZI1,OZI2,OZI3}.
Thereby, we have following pattern about topologies in the $D^0\to K^+K^-$ and $D^0\to \pi^+\pi^-$ decays:
\begin{align}
  T^{LP}\,\,\gg\,\, T^{LA},\, PC,\,PE,\,PA, \qquad T^{LP}_{\rm break}\,\, \gg \,\, T^{LA}_{\rm break}.
\end{align}
Then the decay amplitudes of $D^0\to K^+K^-$ and $D^0\to \pi^+\pi^-$ are simplified to be
\begin{align}\label{kk1}
  \mathcal{A}(D^0 \to  K^+K^-) &\simeq \lambda_s(T_{KK}+E_{KK})+\lambda_sT^{LP}_{\rm break}-\lambda_bT^{LP},\\
  \mathcal{A}(D^0\to \pi^+\pi^-) &\simeq \lambda_d(T_{\pi\pi}+E_{\pi\pi})+\lambda_sT^{LP}_{\rm break}-\lambda_bT^{LP},
\end{align}
where subscript $d$ of $T^{LP}$ is removed for convenience.

The ratio between $D^0\to K^+K^-$ and $D^0\to \pi^+\pi^-$ branching fractions is approximated to be
\begin{align}
\frac{\mathcal{B}r(D^0\to K^+K^-)}{\mathcal{B}r(D^0\to \pi^+\pi^-)} \simeq \frac{ |T_{KK}+E_{KK}+T^{LP}_{\rm break}|^2}{ |T_{\pi\pi}+E_{\pi\pi}-T^{LP}_{\rm break}|^2}.
\end{align}
If we assume $|(T_{KK}+E_{KK})/(T_{\pi\pi}+E_{\pi\pi})|\approx 1.3$, by solving the equation $(1.3+x)^2/(1-x)^2=2.8$, we get
\begin{align}
  |T^{LP}_{\rm break}/(T_{\pi\pi}+E_{\pi\pi})| \sim \mathcal{O}(0.15).
\end{align}
Considering the strong phases, the situation will be more complicated.
But it does not affect the order estimation. If we assume a normal $U$-spin breaking in $T^{LP}$ diagram:
\begin{align}
  |T^{LP}_{\rm break}/T^{LP}| \approx 20\%\sim 30\%,
\end{align}
we get
\begin{align}
  |T^{LP}/(T_{\pi\pi}+E_{\pi\pi})| \simeq |\tilde{p}_0|\sin(\delta_{\rm strong}) \approx  0.50\sim 0.75,
\end{align}
which is consistent with the value extracted from the $CP$ violation in charm given in Eq.~\eqref{u0}.
Thereby, the $KK-\pi\pi$ puzzle and the large $CP$ violation in charm can be explained simultaneously if a large $T^{LP}$ diagram is assumed.
The similar idea was proposed in Refs.~\cite{Cheng:2012xb,Bhattacharya:2012ah,Brod:2012ud,Muller:2015lua,Muller:2015rna}. But the measured $CP$ violation in several years ago was too large \cite{Aaij:2011in,Collaboration:2012qw,Aaij:2013bra} and hence the reliability was questioned.

$D$ meson decay is dominated by topologies $T$, $C$, $E$, $A$.
In the other diagrams listed in Fig.~\ref{top1}, only $T^{LP}$ cannot be separated into two disconnected parts by removing the internal gluon lines and does not suppressed by the OZI rule \cite{OZI1,OZI2,OZI3}.
For the other diagrams, they need the hard gluon exchanges to emit a color singlet, or connect initial and final states.
It is plausible that $T^{LP}$ is enhanced by strong non-perturbative final-state interaction, such as re-scattering and resonance.
The authors of Ref.~\cite{Li:2019hho} and Ref.~\cite{Cheng:2019ggx} argued that which topology, $P$ or $PE$ ($P$ is called $PC$ in Ref.~\cite{Li:2019hho}), leads to the large $CP$ violation in charm.
But since $P$ and $PE$ always appear as $P+PE$ and $T^{LP}$ include main contributions of $P$ and $PE$, assuming a large $T^{LP}$ diagram does not conflict to both Ref.~\cite{Li:2019hho} and Ref.~\cite{Cheng:2019ggx}.
On the other hand, we cannot rule out the possibility that the large $T^{LP}$ arises from new physics.

Similarly to the $D^0\to K^+K^-$ and $D^0\to \pi^+\pi^-$ decays, $T_{\rm break}^{LP}$ can be used to explain the branching fraction differences of other $D\to PV$ modes, such as $D^0\to \pi^-\rho^+$ and $D^0\to K^-K^{*+}$, $D^0\to \pi^+\rho^-$ and $D^0\to K^+K^{*-}$, $D^+\to K^0_SK^{*+}$ and $D^+_s\to K^0_S\rho^+$. The amplitudes of the $D\to PV$ decays are listed in Appendix \ref{pv}.
In Refs.~\cite{FAT1,FAT2}, a Glauber strong phase associated with $\pi$ \cite{Li:2014haa,Li:2009wba,Li:2021req} is introduced to solve $KK-\pi\pi$ puzzle.
To test which effect, $T_{\rm break}^{LP}$ or Glauber phase, is the dominate source of $U$-spin breaking, we suggest to measure the branching fractions of $D^+\to K^0_SK^{*+}$ and $D^+_s\to K^0_S\rho^+$, since there is no $\pi$ meson in the final states.
The factorization-assistant topological amplitude approach \cite{FAT2} predicts the branching fractions of $D^+\to K^0_SK^{*+}$ and $D^+_s\to K^0_S\rho^+$ are approximately equal.
The amplitudes of $D^+\to K^0_SK^{*+}$ and $D^+_s\to K^0_S\rho^+$ can be written as
\begin{align}
 \mathcal{A}(D^+\to K^0_SK^{*+}) &= \sin\theta_C\,(T_P+A_P+T^{LP}_{P,\,\rm break}),\nonumber \\ \mathcal{A}(D^+_s\to K^0_S\rho^+) &= -\sin\theta_C\,(T_P+A_P-T^{LP}_{P,\,\rm break}).
\end{align}
Analogy to $D^0\to K^+K^-$ and $D^0\to \pi^+\pi^-$, the difference of $\mathcal{B}r(D^+\to K^0_SK^{*+})$ and $\mathcal{B}r(D^+_s\to K^0_S\rho^+)$ might be large. If the ratio of $\mathcal{B}r(D^+\to K^0_SK^{*+})$ and $\mathcal{B}r(D^+_s\to K^0_S\rho^+)$ is beyond the normal $SU(3)$ breaking, it might be an evidence of a large $T_{P,\,\rm break}^{LP}$.
The branching fraction of $D^+\to K^0_SK^{*+}$ is poorly measured so far \cite{Tanabashi:2018oca}:
\begin{align}
  \mathcal{B}r(D^+\to K^0_SK^{*+})= (1.6\pm0.7)\%.
\end{align}
And the branching fraction of $D^+_s\to K^0_S\rho^+$ has not been measured yet.
The precise measurements of $\mathcal{B}r(D^+\to K^0_SK^{*+})$ and $\mathcal{B}r(D^+_s\to K^0_S\rho^+)$ are desirable.

Above discussion can be generalized into the charmed baryon decay modes $\Lambda^+_c\to \Sigma^+K^{*0}$ and $\Xi^+_c\to p\overline K^{*0}$.
In Ref.~\cite{Wang:2019dls}, we find that if two singly Cabibbo-suppressed decay modes of charmed hadrons are associated by a complete interchange of $d$ and $s$ quarks, their decay amplitudes are connected by a complete interchange of $\lambda_d$ and $\lambda_s$ in the $U$-spin limit.
As a consequence, the tree-operator-induced amplitudes of $\Lambda^+_c\to \Sigma^+K^{*0}$ and $\Xi^+_c\to p\overline K^{*0}$ under the $U$-spin symmetry can be parameterized to be
\begin{align}
\mathcal{A}(\Lambda^+_c\to \Sigma^+K^{*0})&=\,\lambda_d T^A + \lambda_s T^B +(\lambda_d + \lambda_s)T^L,\nonumber\\ \mathcal{A}(\Xi^+_c\to p\overline K^{*0})&=\,\lambda_d T^B + \lambda_s T^A +(\lambda_d + \lambda_s)T^L,
\end{align}
in which $T^A$, $T^B$ and $T^L$ are not the specific topological amplitudes but the sum of the topological amplitudes proportional to $\lambda_d$, $\lambda_s$ and $(\lambda_d+\lambda_s)$ respectively.
Neglecting the small quark-loop contributions proportional to $\lambda_b$, we have
\begin{align}
 |\mathcal{A}(\Lambda^+_c\to \Sigma^+K^{*0})| \simeq |\mathcal{A}(\Xi^+_c\to p\overline K^{*0})|.
\end{align}
However, the experimental data of branching fractions \cite{Jia:2019zxi,Tanabashi:2018oca},
\begin{align}
\mathcal{B}r(\Lambda_c^+\to \Sigma^+K^{*0})= (3.4\pm 1.0)\times 10^{-3},\qquad \mathcal{B}r(\Xi_c^+\to p\overline{K}^{*0}) = (2.75\pm1.02)\times 10^{-3},
\end{align}
show that the ratio between the decay amplitudes $\mathcal{A}(\Lambda^+_c\to \Sigma^+K^{*0})$ and $\mathcal{A}(\Xi^+_c\to p\overline K^{*0})$ is
\begin{align}\label{ra}
|\mathcal{A}(\Lambda^+_c\to \Sigma^+K^{*0})/\mathcal{A}(\Xi^+_c\to p\overline K^{*0})| \approx 2.1\pm 0.5.
\end{align}
Such a ratio, at least its central value, is larger than $|\mathcal{A}(D^0\to K^+K^-)/\mathcal{A}(D^0\to \pi^+\pi^-)|\approx 1.67$.
Considering the $U$-spin breaking, the tree-operator-induced amplitudes of $\Lambda^+_c\to \Sigma^+K^{*0}$ and $\Xi^+_c\to p\overline K^{*0}$ are
\begin{align}
\mathcal{A}(\Lambda^+_c\to \Sigma^+K^{*0})&\simeq \,\cos\theta_C\sin\theta_C( T^B_s - T^A_d + T^L_{\rm break}),\nonumber\\ \mathcal{A}(\Xi^+_c\to p\overline K^{*0})&\simeq\,-\cos\theta_C\sin\theta_C( T^B_d - T^A_s + T^L_{\rm break}).
\end{align}
Just like the $D^0\to K^+K^-$ and $D^0\to \pi^+\pi^-$ modes, we can introduce a large $T^{L}_{\rm break}$ to explain the large ratio in Eq.~\eqref{ra}.
If so, $|T^{L}/(T^A-T^B)|$ must be $\mathcal{O}(1)$. And a large $|T^{L}/(T^A-T^B)|$ results in large $CP$ asymmetries in the $\Lambda^+_c\to \Sigma^+K^{*0}$ and $\Xi^+_c\to p\overline K^{*0}$ modes. $\overline K^{*0}$ is a primary resonance in $\Xi^+_c\to pK^-\pi^+$ decay since \cite{Tanabashi:2018oca}
\begin{align}
\mathcal{B}r(\Xi^+_c\to p\overline K^{*0})/\mathcal{B}r(\Xi^+_c\to pK^-\pi^+) = 0.54\pm 0.10.
\end{align}
So we predict that $CP$ violation in the $\Xi^+_c\to pK^-\pi^+$ mode can reach to be $\mathcal{O}(10^{-3})$.
Since all the final-state particles in the $\Xi^+_c\to pK^-\pi^+$ decay are preferable in experiments, it is a promising mode to search for $CP$ violation of the charmed baryon decays.

\section{Symmetry breaking and Splitting of topologies}\label{break}

Our framework provides a simple way to formulate the flavor symmetry breaking effects.
In this section, we will use some examples to illustrate how the flavor symmetry breaking effects are included in the tensor form of topology.

\subsection{Linear $SU(3)_F$ breaking}\label{blin}
In Ref.~\cite{Muller:2015lua}, the $D\to PP$ decays without $\eta$ and $\eta^\prime$ are analyzed in the TDA approach with the linear $SU(3)_F$ breaking.
In this method, the total Hamiltonian is written as $\mathcal{H} = \mathcal{H}_0+\mathcal{H}_1$, where $\mathcal{H}_0$ is the QCD  Hamiltonian with $m_u=m_d=m_s$.
$\mathcal{H}_1$ consists of the weak $|\Delta C|=1$ Hamiltonian $ \mathcal{H}_W$ and the $SU(3)_F$ breaking  Hamiltonian: $\mathcal{H}_{\cancel{SU(3)_F}} = (m_s-m_d)\overline ss$.
In this subsection, we express the topological amplitudes with linear $SU(3)_F$ breaking in the tensor form.

In the $D\to PP$ decays without $\eta$ and $\eta^\prime$ mesons, neglecting the penguin-operator-induced amplitudes, only six terms in Eq.~\eqref{ha} left:
\begin{align}
{\cal A}^{\rm TDA}_{D_\gamma\to P_\alpha P_\beta
} &=  T  (D_\gamma)_i  (H)^{lj}_k(P_\alpha)^{i}_j  (P_\beta)^k_l + C (D_\gamma)_i  (H)^{jl}_k  (P_\alpha)^{i}_j(P_\beta)^k_l+E  (D_\gamma)_i  (H)^{il}_j (P_\alpha)^j_k (P_\beta)^{k}_l \nonumber\\& + A  (D_\gamma)_i (H)^{li}_j   (P_\alpha)^j_k (P_\beta)^{k}_l +T^{LP} (D_\gamma)_i (H)^{kl}_{l}(P_\alpha)^{i}_j   (P_\beta)^j_k \nonumber\\& + T^{LA} (D_\gamma)_i  (H)^{il}_{l}  (P_\alpha)^j_k (P_\beta)^{k}_j +\alpha \leftrightarrow\beta.
\end{align}
Considering the first order of $\mathcal{H}_{\cancel{SU(3)_F}}$, amplitude of the $D\to PP$ decay can be obtained by summing all possible invariant tensors in which index 3 (presenting $s$ quark) is written explicitly:
 \begin{align}\label{su3f}
{\cal A}^{\rm TDA, \cancel{SU(3)_F}}_{D_\gamma \to P_\alpha P_\beta} &= T  (D_\gamma)_i  (H)^{lj}_k(P_\alpha)^{i}_j  (P_\beta)^k_l+ T_1  (D_\gamma)_i  (H)^{l3}_k(P_\alpha)^{i}_3  (P_\beta)^k_l + T_2  (D_\gamma)_i  (H)^{lj}_3(P_\alpha)^{i}_j  (P_\beta)^3_l \nonumber\\ & +T_3 (D_\gamma)_3  (H)^{lj}_k(P_\alpha)^{3}_j  (P_\beta)^k_l+ C (D_\gamma)_i  (H)^{jl}_k  (P_\alpha)^{i}_j(P_\beta)^k_l+ C_1 (D_\gamma)_i  (H)^{j3}_k  (P_\alpha)^{i}_j(P_\beta)^k_3\nonumber\\&+ C_2 (D_\gamma)_i  (H)^{jl}_3  (P_\alpha)^{i}_j(P_\beta)^3_l+ C_3 (D_\gamma)_3  (H)^{jl}_k  (P_\alpha)^{3}_j(P_\beta)^k_l+E  (D_\gamma)_i  (H)^{il}_j (P_\alpha)^j_k (P_\beta)^{k}_l\nonumber\\& +E_1  (D_\gamma)_i  (H)^{i3}_j (P_\alpha)^j_k (P_\beta)^{k}_3 +E_2(D_\gamma)_i  (H)^{il}_3 (P_\alpha)^3_k (P_\beta)^{k}_l +E_3(D_\gamma)_i  (H)^{il}_j (P_\alpha)^j_3 (P_\beta)^{3}_l\nonumber\\& +A  (D_\gamma)_i (H)^{li}_j   (P_\alpha)^j_k (P_\beta)^{k}_l +A_1(D_\gamma)_3 (H)^{l3}_j   (P_\alpha)^j_k (P_\beta)^{k}_l+A_2 (D_\gamma)_i (H)^{li}_3   (P_\alpha)^3_k (P_\beta)^{k}_l\nonumber\\& +A_3  (D_\gamma)_i (H)^{li}_j   (P_\alpha)^j_3 (P_\beta)^{3}_l +T^{LP}_{\rm break} (D_\gamma)_i (H)^{k3}_{3}(P_\alpha)^{i}_j   (P_\beta)^j_k + \alpha \leftrightarrow \beta,
\end{align}
where the flavor symmetric part and other $SU(3)_F$ breaking terms of $T^{LP}$ diagram are ignored because they are proportional to $V_{cd}^*V_{ud}+V_{cs}^*V_{us}=-V_{cb}^*V_{ub}$. Following  \cite{Muller:2015lua}, topology $T^{LA}$ is also neglected in Eq.~\eqref{su3f}.
Comparing Eq.~\eqref{su3f} with the Table.~II in Ref.~\cite{Muller:2015lua}, one can find the topological amplitudes defined in this work match to the ones defined in Ref.~\cite{Muller:2015lua} one by one:
\begin{align}
  T &  = T,\qquad  T_1 = T+T_1^{(1)},\qquad T_2 = T+T_2^{(1)},\qquad T_3 = T+T_3^{(1)},\qquad...\,.
\end{align}

The emergence of topologies $T_1$, $T_2$... is analogous to the splitting of energy levels.
In the flavor $SU(3)$ symmetry, some diagrams, for instance $T$, $T_1$, $T_2$, $T_3$, are degenerate, $T=T_1=T_2=T_3$.
When the $SU(3)_F$ symmetry breaks into its $SU(2)$ subgroup, the original $T$ diagram splits into four different diagrams.

\subsection{High order $U$-spin breaking}

In this subsection, we study the $U$-spin symmetry and its breaking, taking $D^0\to K^-\pi^+$, $D^0\to K^+K^-$,  $D^0\to \pi^+\pi^-$ and $D^0\to K^+\pi^-$ decays as examples.
There are four tree operators in the SM contributing to the charm decay: $O_{us}^d$, $O_{ud}^s$, $O_{ud}^d$ and $O_{us}^s$.
The $u$ quark, which has nothing to do with $U$-spin, always appears in the first lower index of $O_{ij}^k$. Thereby, the tree operators can be written as $O_{ui}^j$.
The two indices of $O_{ui}^j$ transform according to the representation of $SU(2)$ group and $1=d$, $2=s$.

Mesons $\pi^-$ and $K^-$ form a $U$-spin doublet, $(P)_i^u|P\rangle^i_u$.
Mesons $K^+$ and $\pi^+$ form another $U$-spin doublet, $(P)^i_u|P\rangle_i^u$.
Under the $U$-spin symmetry, the amplitude of $D^0$ decay is expressed as
\begin{align}\label{up}
 \mathcal{A}_{D^0 \to P^uP_u} = A\,(D^0)(H)_{i}^{uj}(P)_u^i(P)_j^u + A^L\,(D^0)(H)_{i}^{ui}(P)^j_u(P)_j^u,
\end{align}
Expanding Eq.~\eqref{up} in the $D^0\to K^-\pi^+$, $D^0\to K^+K^-$,  $D^0\to \pi^+\pi^-$ and $D^0\to K^+\pi^-$ decays, we have
\begin{align}
 \mathcal{A}(D^0\to K^-\pi^+)  & = V_{cs}^*V_{ud} A, \qquad  \mathcal{A}(D^0\to K^+K^-)  = V_{cs}^*V_{us} A + (V_{cd}^*V_{ud}+V_{cs}^*V_{us})A^L, \nonumber\\
   \mathcal{A}(D^0\to K^+\pi^-)  & = V_{cd}^*V_{us} A, \qquad\mathcal{A}(D^0\to \pi^+\pi^-)  = V_{cd}^*V_{ud} A + (V_{cd}^*V_{ud}+V_{cs}^*V_{us})A^L.
\end{align}
These results are consistent with the results in flavor $SU(3)$ symmetry if $A=T+E$ and $A^L=T^{LP}+2T^{LA}$.
Considering the approximation of $V_{cs}^*V_{ud}\simeq\cos^2\theta_C$, $V_{cs}^*V_{us} \simeq- V_{cd}^*V_{ud} \simeq \cos\theta_C\sin\theta_C$ and $V_{cd}^*V_{us}\simeq-\sin^2\theta_C$, our results are consistent with Eq.~(3) in Ref.~\cite{Gronau:2013xba}.
\begin{align}
 \mathcal{A}(D^0\to K^-\pi^+)  & =  \cos^2\theta_CA, \qquad  \mathcal{A}(D^0\to K^+K^-)  = \cos\theta_C\sin\theta_C  A, \nonumber\\
   \mathcal{A}(D^0\to K^+\pi^-)  & = -\sin^2\theta_C A, \qquad\mathcal{A}(D^0\to \pi^+\pi^-)  = -\cos\theta_C\sin\theta_C A.
\end{align}

In Ref.~\cite{Gronau:2013xba,Gronau:2015rda,Brod:2012ud,Gronau:2013mda,Jung:2009pb,
Kwong:1993ri,Hinchliffe:1995hz,Feldmann:2012js,Hiller:2012xm,
Pirtskhalava:2011va}, a perturbative method to analyze the $U$-spin breaking was proposed.
In this method, the arbitrary order $U$-spin breaking corrections to decay amplitude $\langle f|\mathcal{H}_{\rm eff}|D^0\rangle$ are obtained by introducing a $s-d$ spurion mass operator, $m_{s}^{s}-m^{d}_d$, into the Hamiltonian or the initial and final states.
The $s-d$ spurion mass operator is the $U_3=0$ component of $U$-spin triplet.
Using the $s-d$ spurion mass operator, the author of \cite{Gronau:2013xba} derived the first and second order $U$-spin breaking corrections to the $D^0\to K^-\pi^+$, $D^0\to K^+K^-$,  $D^0\to \pi^+\pi^-$ and $D^0\to K^+\pi^-$ decays.
Since the two indices of the $s-d$ spurion mass operator are transformed as the representations of $SU(2)$ group, we can write the $s-d$ spurion mass operator as $(m)^{i}_j\,m^j_i $, in analogy with Eq.~\eqref{h} and Eq.~\eqref{a3}.
The non-zero components of $(m)^{i}_j$ are $(m)^{1}_1=-1$ and $(m)^{2}_2=1$.
To include the $U$-spin breaking in the tensor form of topology, we should contract the indices of $(m)^{b}_a$ with $(H)^{uj}_i$, $(P)_u^k$ and $(P)^u_l$ (in the case of $D^0$ decay).
For example, the $D^0$ decay amplitude with the first order $U$-spin breaking corrections can be written as
\begin{align}\label{ub}
 \mathcal{A}_{\rm 1th}=&  A\, \varepsilon_1^{(1)}(D^0)(H)_i^{uk}(m)_k^j\,(P)^i_u(P)_j^u+  A\, \varepsilon_2^{(1)}(D^0)(H)_k^{uj}(m)_i^k\,(P)^i_u(P)_j^u\nonumber\\ & + A\, \varepsilon_3^{(1)}(D^0)(H)_i^{uk}(m)_k^i\,(P)^j_u(P)_j^u,
\end{align}
in which some terms are dropped due to $(m)_i^i = 1-1=0$ and $(H)^{ui}_i = V_{cd}^*V_{ud}+V_{cs}^*V_{us}\approx 0$.
Parameters $\varepsilon_n^{(1)}$ presents the first order $U$-spin breaking corrections to $A$.
With Eq.~\eqref{ub}, the decay amplitudes of $D^0\to K^-\pi^+$, $D^0\to K^+K^-$,  $D^0\to \pi^+\pi^-$ and $D^0\to K^+\pi^-$ read as
\begin{align}\label{x3}
 \mathcal{A}_{\rm 0,1th}(D^0\to K^-\pi^+)  & =  \cos^2\theta_CA(1+\varepsilon_1^{(1)}-\varepsilon_2^{(1)}),  \nonumber\\
   \mathcal{A}_{\rm 0,1th}(D^0\to K^+\pi^-)  & = -\sin^2\theta_C A(1-\varepsilon_1^{(1)}+\varepsilon_2^{(1)}), \nonumber\\
 \mathcal{A}_{\rm 0,1th}(D^0\to K^+K^-)  &= \cos\theta_C\sin\theta_C  A(1+\varepsilon_1^{(1)}+\varepsilon_2^{(1)}+2\varepsilon_3^{(1)}), \nonumber\\ \mathcal{A}_{\rm 0,1th}(D^0\to \pi^+\pi^-)  &= -\cos\theta_C\sin\theta_C A(1-\varepsilon_1^{(1)}-\varepsilon_2^{(1)}-2\varepsilon_3^{(1)}).
\end{align}
Notice that the first order $U$-spin breaking corrections are opposite in $\mathcal{A}(D^0\to K^-\pi^+)$ and $\mathcal{A}(D^0\to K^+\pi^-)$, and in $\mathcal{A}(D^0\to K^+K^-)$ and $\mathcal{A}(D^0\to \pi^+\pi^-)$, being consistent with Ref.~\cite{Gronau:2013xba}.
By comparing the last two equations of Eq.~\eqref{x3} with Eqs.~\eqref{kk} and \eqref{mq}, one can find $2(\varepsilon_1^{(1)}+\varepsilon_2^{(1)})A$ is the difference between $T_{KK}+E_{KK}$ and $T_{\pi\pi}+E_{\pi\pi}$, and $2\varepsilon_3^{(1)}
A$ is $(T^{LP}_{\rm break}+2T_{\rm break})$.

The first order $U$-spin breaking induced by the $s-d$ spurion mass operator and the linear $SU(3)_F$ breaking are equivalent in the $U$-spin multiplet.
If we "translate" the $m^{s}_s-m^{d}_d$ to $m^{s}_s$, the amplitudes of  $D^0\to K^-\pi^+$, $D^0\to K^+K^-$,  $D^0\to \pi^+\pi^-$ and $D^0\to K^+\pi^-$ with the 0th and first order $U$-spin breaking are
\begin{align}
 \mathcal{A}_{\rm 0,1th}(D^0\to K^-\pi^+)  & =  \cos^2\theta_CA(1+\varepsilon_1^{(1)}),  \qquad
   \mathcal{A}_{\rm 0,1th}(D^0\to K^+\pi^-)  = -\sin^2\theta_C A(1+\varepsilon_2^{(1)}), \nonumber\\
 \mathcal{A}_{\rm 0,1th}(D^0\to K^+K^-)  &= \cos\theta_C\sin\theta_C  A(1+\varepsilon_1^{(1)}+\varepsilon_2^{(1)}+ \varepsilon_3^{(1)}), \nonumber\\ \mathcal{A}_{\rm 0,1th}(D^0\to \pi^+\pi^-)  &= -\cos\theta_C\sin\theta_C A(1-\varepsilon_3^{(1)}).
\end{align}
Compared to the results given in linear $SU(3)_F$ breaking \cite{Muller:2015lua}, the relations of the two methods in the $D^0\to K^-\pi^+$, $D^0\to K^+K^-$, $D^0\to \pi^+\pi^-$ and $D^0\to K^+\pi^-$ decays are
\begin{align}
 A\varepsilon_1^{(1)} = T_1^{(1)} + E_1^{(1)},\qquad A\varepsilon_2^{(1)} = T_2^{(1)} + E_2^{(1)},\qquad A\varepsilon_3^{(1)} = P_{\rm break}.
\end{align}

The $D^0$ decay amplitude with the second order $U$-spin breaking can be constructed as
\begin{align}\label{b2}
\mathcal{A}_{\rm 2th}=&~ A\,\varepsilon_1^{(2)}(D^0)(H)_i^{uk}(m)_l^j\,(m)_k^l\,(P)^i_u(P)_j^u
+A\,\varepsilon_2^{(2)}(D^0)(H)_k^{uj}(m)_i^l\,(m)_l^k\,(P)^i_u(P)_j^u\nonumber\\ &~~~+   A\,\varepsilon_3^{(2)}(D^0)(H)_k^{ul}(m)_i^k\,(m)_l^j\,(P)^i_u(P)_j^u
+A\,\varepsilon_4^{(2)}(D^0)(H)_i^{uj}(m)_l^k\,(m)_k^l\,(P)^i_u(P)^u_j\nonumber\\ &~~~~~+   A\,\varepsilon_5^{(2)}(D^0)(H)_l^{uk}(m)_i^j\,(m)_k^l\,(P)^i_u(P)_j^u+A\,\varepsilon_6^{(2)}(D^0)(H)_i^{uk}(m)_j^i\,(m)_k^j\,(P)^l_u(P)_l^u,
\end{align}
And the corrections to the $D^0\to K^-\pi^+$, $D^0\to K^+K^-$,  $D^0\to \pi^+\pi^-$ and $D^0\to K^+\pi^-$ are
\begin{align}
 \mathcal{A}_{\rm 2th}(D^0\to K^-\pi^+)  & =  \cos^2\theta_CA(\varepsilon_1^{(2)}+\varepsilon_2^{(2)}-\varepsilon_3^{(2)}+2\varepsilon_4^{(2)}),  \nonumber\\
   \mathcal{A}_{\rm 2th}(D^0\to K^+\pi^-)  & = -\sin^2\theta_C A(\varepsilon_1^{(2)}+\varepsilon_2^{(2)}-\varepsilon_3^{(2)}+2\varepsilon_4^{(2)}), \nonumber\\
 \mathcal{A}_{\rm 2th}(D^0\to K^+K^-)  &= \cos\theta_C\sin\theta_C  A(\varepsilon_1^{(2)}+\varepsilon_2^{(2)} +\varepsilon_3^{(2)}+2\varepsilon_4^{(2)}+2 \varepsilon_5^{(2)}), \nonumber\\ \mathcal{A}_{\rm 2th}(D^0\to \pi^+\pi^-)  &= -\cos\theta_C\sin\theta_C A(\varepsilon_1^{(2)}+\varepsilon_2^{(2)}+\varepsilon_3^{(2)}+2\varepsilon_4^{(2)}
 +2\varepsilon_5^{(2)}).
\end{align}
One can find the second order $U$-spin breaking corrections are the same in $\mathcal{A}(D^0\to K^-\pi^+)$ and $\mathcal{A}(D^0\to K^+\pi^-)$, and in $\mathcal{A}(D^0\to K^+K^-)$ and $\mathcal{A}(D^0\to \pi^+\pi^-)$, being consistent with Ref.~\cite{Gronau:2013xba}.

\subsection{Strange-less charm decay $v.s.$ Charm-less bottom decay}\label{other}
From above two examples, the linear $SU(3)_F$ breaking and the the high order $U$-spin breaking, one can find the tensor form of topology provides a simple and systematic way to formulate the flavor symmetry breaking effects.
In this subsection, we study a more complected physical system, the charm-less $B$ decay.
The $SU(3)$ analysis on the charm-less $B$ decays is different from the $D$ decays because the charm-quark loop is beyond the $SU(3)$ symmetry.
To describe the charm-less $B$ decay in the topological amplitudes and the $SU(3)$ irreducible amplitudes, the charm quark loop should be treated differently from the $u$-quark loop.
Before discussing the charm-less $B$ decays, the strange-less charm decays are studied to show the basic idea.

\subsubsection{Strange-less charm decay}
For the strange-less charm decay, the flavor symmetry is the isospin symmetry. $D^0$ and $D^+$ form an isospin doublet $|D^i\rangle = (|D^0\rangle,\,\,|D^+\rangle)$ and $\pi^+$, $\pi^0$, $\pi^-$, $\eta_q$ form a quartet
\begin{eqnarray}
 |P\rangle^i_j=  \left( \begin{array}{cc}
   \frac{1}{\sqrt 2} |\pi^0\rangle,    & |\pi^+\rangle   \\
    |\pi^-\rangle, &   - \frac{1}{\sqrt 2} |\pi^0\rangle \\
  \end{array}\right) +  \frac{1}{\sqrt 2} \left( \begin{array}{ccc}
   |\eta_q\rangle,    & 0   \\
    0, &  |\eta_q\rangle    \\
  \end{array}\right).
\end{eqnarray}
To find all the topological amplitudes contributing to the strange-less charm decay, the first step is to find a appropriate assemble of the four-quark operators.
One might use $O_{ij}^k$ to describe the strange-less charm decay, just like we have done in Sec.~\ref{ga}.
But $O_{ij}^k$ is not enough. $O_{ij}^k$ means that all the indices $i$, $j$ and $k$ transform as the foundational or conjugate representations of $SU(2)$ group.
So $O_{ij}^k$ cannot contain $s$-quark loop contributions.
To give a complete description to the strange-less charm decay, $O_{sj}^s$ and $O_{js}^s$, i.e., the $s$-quark loop contributions, should be included.
In analogy with the linear $SU(3)_F$ breaking, the amplitude of the strange-less charm decay, in which index $3\, = \,s$ is written explicitly, is
\begin{align}\label{hx}
{\cal A}^{\rm TDA}_{\rm s-less} &=  T\,  (D_\gamma)_i  (H)^{lj}_k(P_\alpha)^{i}_j  (P_\beta)^k_l + C\, (D_\gamma)_i  (H)^{jl}_k  (P_\alpha)^{i}_j(P_\beta)^k_l+E\,  (D_\gamma)_i  (H)^{il}_j (P_\alpha)^j_k (P_\beta)^{k}_l  \nonumber\\&+ A\,  (D_\gamma)_i (H)^{li}_j   (P_\alpha)^j_k (P_\beta)^{k}_l+T^{ES} (D_\gamma)_i   (H)^{ij}_{l}   (P_\alpha)^{l}_j    (P_\beta)_k^k+T^{AS} (D_\gamma)_i  (H)^{ji}_{l}  (P_\alpha)^{l}_j  (P_\beta)_k^k \nonumber\\&+T^{LP} (D_\gamma)_i (H)^{kl}_{l}(P_\alpha)^{i}_j (P_\beta)^j_k + T^{LC} (D_\gamma)_i  (H)^{jl}_{l} (P_\alpha)^{i}_j(P_\beta)^k_k  + T^{LA} (D_\gamma)_i  (H)^{il}_{l}  (P_\alpha)^j_k (P_\beta)^{k}_j \nonumber\\&  + T^{LS} (D_\gamma)_i   (H)^{il}_{l} (P_\alpha)^{j}_j (P_\beta)^k_k +T^{QP}(D_\gamma)_i(H)^{lk}_{l} (P_\alpha)^{i}_j(P_\beta)^j_k  + T^{QC} (D_\gamma)_i   (H)^{lj}_{l}(P_\alpha)^i_j (P_\beta)_{k}^{k}   \nonumber\\
&  + T^{QA} (D_\gamma)_i  (H)^{li}_{l} (P_\alpha)^j_k (P_\beta)^{k}_j + T^{QS} (D_\gamma)_i  (H)^{li}_{l} (P_\alpha)_j^j (P_\beta)_{k}^{k}+T^{LP}_s (D_\gamma)_i (H)^{ks}_{s}(P_\alpha)^{i}_j   (P_\beta)^j_k \nonumber\\&+ T^{LC}_s (D_\gamma)_i  (H)^{js}_{s} (P_\alpha)^{i}_j (P_\beta)^k_k + T^{LA}_s (D_\gamma)_i  (H)^{is}_{s}  (P_\alpha)^j_k (P_\beta)^{k}_j   + T^{LS}_s (D_\gamma)_i   (H)^{is}_{s} (P_\alpha)_{j}^j (P_\beta)^k_k  \nonumber\\&+T^{QP}_s(D_\gamma)_i(H)^{sk}_{s} (P_\alpha)^{i}_j(P_\beta)^j_k  + T^{QC}_s (D_\gamma)_i   (H)^{sj}_{s}  (P_\alpha)^i_j (P_\beta)^{k}_{k}
 + T^{QA}_s (D_\gamma)_i  (H)^{si}_{s} (P_\alpha)^j_k (P_\beta)^{k}_j \nonumber\\&+ T^{QS}_s (D_\gamma)_i  (H)^{si}_{s} (P_\alpha)_j^j (P_\beta)_{k}^{k} + \alpha \leftrightarrow \beta.
\end{align}
In Eq.~\eqref{hx}, the first six terms denote the diagrams without quark loop, the middle eight terms denote the diagrams with light-quark loop diagrams, and the last eight terms denote the $s$-quark loop diagrams.

The explicit $SU(2)$ decomposition of $O_{ij}^k$ is found in Eq.~\eqref{kdec}.
All components of the $O_{ij}^k$ irreducible presentation are listed following.\\
$ 2_p$ presentation:
\begin{align}\label{2p}
  O( 2_p)_1 & = (\bar u u)(\bar u c) + (\bar dd)(\bar u c),\qquad
 O(2_p)_2 = (\bar u u)(\bar d c) + (\bar dd)(\bar d c).
\end{align}
$ 2_t$ presentation:
\begin{align}\label{2t}
 O(2_t)_1 & = (\bar u u)(\bar u c) + (\bar ud)(\bar d c),\qquad
 O(2_t)_2 = (\bar d u)(\bar u c) + (\bar dd)(\bar d c).
\end{align}
$ {\underline{4}}$ presentation:
\begin{align}\label{4}
     O({4})^{1}_{11} & = (\bar u u)(\bar u c)-[ (\bar ud)(\bar dc)+(\bar dd)(\bar u c) ],\qquad
     O({4})^{2}_{22} = (\bar dd)(\bar d c)-[(\bar du)(\bar u c) +(\bar uu)(\bar d c)],\nonumber \\
     O({4})^{1}_{21} & = [(\bar u u)(\bar d c)+(\bar d u)(\bar u c)]-(\bar d d)(\bar d c)],\qquad
      O({4})^{2}_{12} = [(\bar u d)(\bar d c)+(\bar d d)(\bar u c)]-(\bar u u)(\bar u c)],\nonumber \\
     O({4})^{2}_{11} & = 3\,(\bar u d)(\bar u c), \qquad  O({4})^{1}_{22} = 3\,(\bar du)(\bar d c).
\end{align}
There are only four independent operators in Eq.~\eqref{4} since
\begin{align}
  O({4})^1_{11} &= -O({4})^2_{12}, \qquad O({4})^2_{22} = -O({4})^1_{21}.
\end{align}
The operators $O_{js}^s$ and $O_{sj}^s$ are the $SU(2)$ irreducible representations themselves, labeled by $2^\prime$ and $2^{\prime\prime}$, respectively.
The $SU(2)$ irreducible amplitude of the strange-less charm decay is expressed as
\begin{align}\label{n}
{\cal A}^{\rm IRA}_{\rm s-less} &=a_{4}(D_\gamma)_i (H({4}))^{ij}_k(P_\alpha)_j^l(P_\beta)_l^k
+ b_{4}(D_\gamma)_i (H({4}))^{ij}_k(P_\alpha)_j^k(P_\beta)^l_l+ c_{4}(D_\gamma)_i (H({4}))^{jl}_k(P_\alpha)_j^i(P_\beta)_l^k\nonumber\\&
+a_2^p (D_\gamma)_i (H(2_p))^i (P_\alpha)^k_j(P_\beta)^j_k +b_2^p (D_\gamma)_i (H(2_p))^i (P_\alpha)_k^k(P_\beta)_j^j+c_2^p (D_\gamma)_i (H(2_p))^k (P_\alpha)^i_k(P_\beta)_j^j\nonumber\\&+d_2^p (D_\gamma)_i (H(2_p))^k (P_\alpha)^i_j(P_\beta)^j_k
  +a_2^t (D_\gamma)_i (H(2_t))^i (P_\alpha)^k_j(P_\beta)^j_k +b_2^t (D_\gamma)_i (H(2_t))^i (P_\alpha)_k^k(P_\beta)_j^j\nonumber\\
  & +c_2^t (D_\gamma)_i (H(2_t))^k (P_\alpha)^i_k(P_\beta)_j^j+d_2^t (D_\gamma)_i (H(2_t))^k (P_\alpha)^i_j(P_\beta)^j_k +a_2^\prime (D_\gamma)_i (H(2^\prime))^i (P_\alpha)^k_j(P_\beta)^j_k \nonumber\\&+b_2^\prime (D_\gamma)_i (H(2^\prime))^i (P_\alpha)_k^k(P_\beta)_j^j+c_2^\prime (D_\gamma)_i (H(2^\prime))^k (P_\alpha)^i_k(P_\beta)_j^j+d_2^\prime (D_\gamma)_i (H(2^\prime))^k (P_\alpha)^i_j(P_\beta)^j_k\nonumber\\
  &+a_2^{\prime\prime} (D_\gamma)_i (H(2^{\prime\prime}))^i (P_\alpha)^k_j(P_\beta)^j_k +b_2^{\prime\prime} (D_\gamma)_i (H(2^{\prime\prime}))^i(P_\alpha)_k^k(P_\beta)_j^j+c_2^{\prime\prime} (D_\gamma)_i (H(2^{\prime\prime}))^k (P_\alpha)^i_k(P_\beta)_j^j\nonumber\\&+d_2^{\prime\prime} (D_\gamma)_i (H(2^{\prime\prime}))^k (P_\alpha)^i_j(P_\beta)^j_k +\alpha\leftrightarrow\beta.
\end{align}
By substituting Eq.~\eqref{kdec} into the amplitudes of $T$, $C$, $E$..., the relations between topological diagrams and the irreducible amplitudes in the strange-less charm decay are derived to be
\begin{align}\label{sol3}
  a_{4}&  =\frac{E+A}{3},  \qquad b_{4} = \frac{T^{ES}+T^{AS}}{3},  \qquad c_{4} = \frac{T+C}{3},\nonumber\\
 a^t_2& = \frac{2}{3}E-\frac{1}{3}A+T^{LA},\qquad
 a^p_2 = -\frac{1}{3}E+\frac{2}{3}A+ T^{QA},\nonumber\\
  b^t_2 & = \frac{2}{3}T^{ES}-\frac{1}{3}T^{AS}+T^{LS},\qquad
 b^p_2  = -\frac{1}{3}T^{ES}+\frac{2}{3}T^{AS} + T^{QS},\nonumber\\
  c^t_2 & = -\frac{1}{3}T+\frac{2}{3}C-\frac{1}{3}T^{ES} + \frac{2}{3}T^{AS}+T^{LC},\qquad
  c^p_2  = \frac{2}{3}T-\frac{1}{3}C+\frac{2}{3}T^{ES} -\frac{1}{3}T^{AS}+T^{QC},\nonumber\\
   d^t_2 & = \frac{2}{3}T-\frac{1}{3}C-\frac{1}{3}E + \frac{2}{3}A+T^{LP},\qquad
   d^p_2  = -\frac{1}{3}T+\frac{2}{3}C+\frac{2}{3}E - \frac{1}{3}A+T^{QP},\nonumber\\
   a_2^\prime & = T^{LA}_s,\qquad  b_2^\prime = T^{LS}_s, \qquad  c_2^\prime = T^{LC}_s, \qquad d_2^\prime = T^{LP}_s,\nonumber\\
   a_2^{\prime\prime} & = T^{QA}_s,\qquad  b_2^{\prime\prime} = T^{QS}_s, \qquad  c_2^{\prime\prime} = T^{QC}_s, \qquad d_2^{\prime\prime} = T^{QP}_s.
\end{align}

According to Eq.~\eqref{hsm}, the non-zero CKM components induced by the tree operators in the SM are
\begin{align}
 & (H^{(0)})_{2}^{12}=V_{cd}^*V_{ud},\qquad (H^{(0)})_{s}^{1s}= V_{cs}^*V_{us}.
\end{align}
The non-zero components induced by the penguin operators are
\begin{align}
 &(H^{(1)})^{11}_1 = -V_{cb}^*V_{ub}, \qquad (H^{(1)})^{21}_2=-V_{cb}^*V_{ub}, \qquad (H^{(1)})^{s1}_s=-V_{cb}^*V_{ub}.
\end{align}
In the $SU(2)$ irreducible amplitudes, the non-zero CKM components induced by the tree operators are
\begin{align}
  (H^{(0)}(2_t))^1 &= V^*_{cd}V_{ud}, \qquad (H^{(0)}(2^\prime))^1 = V^*_{cs}V_{us}, \nonumber\\
  (H^{(0)}(4))^{11}_{1} &= -V^*_{cd}V_{ud}, \qquad (H^{(0)}(4))^{12}_2 = V^*_{cd}V_{ud}.
\end{align}
The non-zero CKM components induced by the penguin operators are
\begin{align}
  (H^{(1)}(2_t))^1 &= -V^*_{cb}V_{ub}, \qquad (H^{(1)}(2_p))^1 = -2V^*_{cb}V_{ub}, \qquad (H^{(1)}(2^{\prime\prime}))^1 = -V^*_{cb}V_{ub}.
\end{align}
In the SM, not all topologies in Eq.~\eqref{hx} contribute to the strange-less charm decays. If the tree operators are inserted, all the diagrams with superscript $QX$ vanish because there is no tree level FCNC transition in the SM.
If the penguin operators are inserted, $P^{LP}_s$, $P^{LC}_s$, $P^{LA}_s$ and $P^{LS}_s$ vanish.
Similar to Eq.~\eqref{c5}, the amplitudes associated with the $2$-dimensional presentations in Eq.~\eqref{n} can be absorbed into four parameters with following redefinition:
\begin{align}
 & a_2  = -\frac{\lambda_d}{\lambda_b}a_2^t-\frac{\lambda_s}{\lambda_b}a^\prime_2+Pa_2^t + 2 Pa_2^p+Pa^{\prime\prime}_2, \qquad   b_2  = -\frac{\lambda_d}{\lambda_b}b_2^t-\frac{\lambda_s}{\lambda_b}b^\prime_2+Pb_2^t + 2 Pb_2^p+Pb^{\prime\prime}_2, \nonumber\\ &   c_2  = -\frac{\lambda_d}{\lambda_b}c_2^t-\frac{\lambda_s}{\lambda_b}c^\prime_2+Pc_2^t + 2 Pc_2^p+Pc^{\prime\prime}_2, \qquad d_2  = -\frac{\lambda_d}{\lambda_b}d_2^t-\frac{\lambda_s}{\lambda_b}d^\prime_2+Pd_2^t + 2 Pd_2^p+Pd^{\prime\prime}_2.
\end{align}
So there are only seven independent parameters in the strange-less charm decay in the SM.
The tree-operator-induced amplitudes with quark loop and all penguin-operator-induced amplitudes in Eq.~\eqref{hx} can be absorbed into four parameters with following redefinition:
\begin{align}
 & A^{LA}  = -\frac{\lambda_d}{\lambda_b}T^{LA}-\frac{\lambda_s}{\lambda_b}T^{LA}_s+PA+P^{LA}+2P^{QA}+P^{QA}_s, \nonumber\\ &   A^{LS}  =  -\frac{\lambda_d}{\lambda_b}T^{LS}-\frac{\lambda_s}{\lambda_b}T^{LS}_s+P^{AS}+P^{LS}+2P^{QS} + +P^{QS}_s, \nonumber\\ &  A^{LC}  = -\frac{\lambda_d}{\lambda_b}T^{LC}-\frac{\lambda_s}{\lambda_b}T^{LC}_s+PT+P^{ES}+P^{LC}+2P^{QC}+P^{QC}_s, \nonumber\\ &  A^{LP}  =  -\frac{\lambda_d}{\lambda_b}T^{LP}-\frac{\lambda_s}{\lambda_b}T^{LP}_s+PC+PE+P^{LP}+2P^{QP}+P^{QP}_s.
\end{align}

As an example of the strange-less charm decays, we write down the decay amplitude of $D^0\to \pi^+\pi^-$.
The $SU(2)$ irreducible amplitude of $D^0\to \pi^+\pi^-$ is
\begin{align}
\mathcal{A}(D^0\to \pi^+\pi^-)= \lambda_dc_4-\lambda_b(2a_2+d_2).
\end{align}
The topological amplitude of $D^0\to \pi^+\pi^-$ reads as
\begin{align}\label{aq}
\mathcal{A}(D^0\to \pi^+\pi^-)&= \lambda_d(T+E) -\lambda_b(A^{LP}+2A^{LA})\nonumber\\&=\lambda_d(T+E) + \lambda_d(T^{LP}+2T^{LA})+ \lambda_s(T^{LP}_s+2T^{LA}_s)\nonumber\\&~-\lambda_b(PC+PE +2PA+P^{LP}+2P^{LA}+2P^{QP}+4P^{QA}+P^{QP}_s+2P^{QA}_s).
\end{align}
If the difference between the $s$-quark loop and $u/d$-quark loop is neglected, Eq.~\eqref{aq} returns to the result in the $SU(3)_F$ symmetry:
\begin{align}
\mathcal{A}(D^0\to \pi^+\pi^-)&=\lambda_d(T+E) + (\lambda_d+\lambda_s)(T^{LP}+2T^{LA})\nonumber\\&~~~-\lambda_b(PC+PE +2PA+P^{LP}+2P^{LA}+3P^{QP}+6P^{QA}).
\end{align}

\subsubsection{Charm-less bottom decay}

In the charm-less $B$ decay, the $SU(4)_F$ symmetry breaks into the $SU(3)_F$ symmetry.
Analogy to the strange-less charm decay, the index $4\, = \,c$ is written explicitly in the amplitude.
The $\overline B$ meson triplet is $|\overline B^i\rangle = (|\overline B^+\rangle,\,\,|\overline B^0\rangle,\,\,|\overline B^0_s\rangle)$.
The amplitude of the charm-less $\overline B$ decay is constructed by
\begin{align}\label{hxx}
{\cal A}^{\rm TDA}_{\rm c-less}  &= T\,  (\overline B_\gamma)_i  (H)^{jl}_k(P_\alpha)^{i}_j  (P_\beta)^k_l + C \,(\overline B_\gamma)_i  (H)^{lj}_k  (P_\alpha)^{i}_j(P_\beta)^k_l+E\,  (\overline B_\gamma)_i  (H)^{li}_j (P_\alpha)^j_k (P_\beta)^{k}_l \nonumber\\&+ A \, (\overline B_\gamma)_i (H)^{il}_j   (P_\alpha)^j_k (P_\beta)^{k}_l +T^{ES} (\overline B_\gamma)_i   (H)^{ji}_{l}   (P_\alpha)^{l}_j    (P_\beta)_k^k+T^{AS} (\overline B_\gamma)_i  (H)^{ij}_{l}  (P_\alpha)^{l}_j  (P_\beta)_k^k \nonumber\\& +T^{LP} (\overline B_\gamma)_i (H)^{lk}_{l}(P_\alpha)^{i}_j(P_\beta)^j_k + T^{LC} (\overline B_\gamma)_i  (H)^{lj}_{l} (P_\alpha)^{i}_j (P_\beta)^k_k + T^{LA} (\overline B_\gamma)_i  (H)^{li}_{l} (P_\alpha)^j_k (P_\beta)^{k}_j \nonumber\\&  + T^{LS} (\overline B_\gamma)_i   (H)^{li}_{l} (P_\alpha)^{j}_j (P_\beta)^k_k +T^{QP}(\overline B_\gamma)_i(H)^{kl}_{l}(P_\alpha)^{i}_j(P_\beta)^j_k  + T^{QC} (\overline B_\gamma)_i   (H)^{jl}_{l}(P_\alpha)^i_j (P_\beta)_{k}^{k}   \nonumber\\
& + T^{QA} (\overline B_\gamma)_i  (H)^{il}_{l} (P_\alpha)^j_k (P_\beta)^{k}_j + T^{QS} (\overline B_\gamma)_i  (H)^{il}_{l}(P_\alpha)_j^j (P_\beta)_{k}^{k}+T^{LP}_c (\overline B_\gamma)_i (H)^{ck}_{c}(P_\alpha)^{i}_j (P_\beta)^j_k \nonumber\\& + T^{LC}_c (\overline B_\gamma)_i  (H)^{cj}_{c} (P_\alpha)^{i}_j(P_\beta)^k_k+ T^{LA}_c (\overline B_\gamma)_i  (H)^{ci}_{c} (P_\alpha)^j_k (P_\beta)^{k}_j   + T^{LS}_c (\overline B_\gamma)_i   (H)^{ci}_{c} (P_\alpha)^{j}_j (P_\beta)^k_k    \nonumber\\
& +T^{QP}_c(\overline B_\gamma)_i(H)^{kc}_{c} (P_\alpha)^{i}_j(P_\beta)^j_k  + T^{QC}_c (\overline B_\gamma)_i   (H)^{jc}_{c}(P_\alpha)^i_j (P_\beta)_{k}^{k} + T^{QA}_c (\overline B_\gamma)_i  (H)^{ic}_{c} (P_\alpha)^j_k (P_\beta)^{k}_j \nonumber\\
&+ T^{QS}_c (\overline B_\gamma)_i  (H)^{ic}_{c} (P_\alpha)_j^j (P_\beta)_{k}^{k} + \alpha \leftrightarrow \beta.
\end{align}
The $SU(3)$ decomposition of $O_{ij}^k$ in the $B$ decay is presented in Appendix~\ref{b}.
The $SU(3)$ irreducible amplitude of the charm-less $\overline B$ decay is expressed to be
\begin{align}\label{nx}
{\cal A}^{\rm IRA}_{\rm c-less} &=a_{15}(\overline B_\gamma)_i (H({15}))^{ij}_k(P_\alpha)_j^l(P_\beta)_l^k+ b_{15}(\overline B_\gamma)_i (H({15}))^{ij}_k(P_\alpha)_j^k(P_\beta)^l_l + c_{15}(\overline B_\gamma)_i (H({15}))^{jl}_k(P_\alpha)_j^i(P_\beta)_l^k\nonumber\\
&+a_{6}(\overline B_\gamma)_i (H({\overline{6}}))^{ji}_k(P_\alpha)_j^l(P_\beta)_l^k
+ b_{6}(\overline B_\gamma)_i (H({\overline{6}}))^{ji}_k(P_\alpha)_j^k(P_\beta)^l_l + c_{6}(\overline B_\gamma)_i(H({\overline{6}}))^{lj}_k(P_\alpha)_j^i(P_\beta)_l^k\nonumber\\
&+a_3 (\overline B_\gamma)_i (H(3))^i (P_\alpha)^k_j(P_\beta)^j_k +b_3 (\overline B_\gamma)_i (H(3))^i (P_\alpha)_k^k(P_\beta)_j^j+c_3 (\overline B_\gamma)_i (H(3))^k (P_\alpha)^i_k(P_\beta)_j^j\nonumber\\
&+d_3 (\overline B_\gamma)_i (H(3))^k (P_\alpha)^i_j(P_\beta)^j_k+a_3^\prime (\overline B_\gamma)_i (H(3^\prime))^i (P_\alpha)^k_j(P_\beta)^j_k +b_3^\prime (\overline B_\gamma)_i (H(3^\prime))^i (P_\alpha)_k^k(P_\beta)_j^j\nonumber\\
  &+c_3^\prime (\overline B_\gamma)_i (H(3^\prime))^k (P_\alpha)^i_k(P_\beta)_j^j+d_3^\prime (\overline B_\gamma)_i (H(3^\prime))^k (P_\alpha)^i_j(P_\beta)^j_k +
 +a_3^{\prime\prime} (\overline B_\gamma)_i (H(3^{\prime\prime}))^i (P_\alpha)^k_j(P_\beta)^j_k\nonumber\\
  & +b_3^{\prime\prime} (\overline B_\gamma)_i (H(3^{\prime\prime}))^i (P_\alpha)_k^k(P_\beta)_j^j+c_3^{\prime\prime} (\overline B_\gamma)_i (H(3^{\prime\prime}))^k (P_\alpha)^i_k(P_\beta)_j^j+d_3^{\prime\prime} (\overline B_\gamma)_i (H(3^{\prime\prime}))^k (P_\alpha)^i_j(P_\beta)^j_k \nonumber\\&+a_3^{\prime\prime\prime} (\overline B_\gamma)_i (H(3^{\prime\prime\prime}))^i (P_\alpha)^k_j(P_\beta)^j_k +b_3^{\prime\prime\prime} (\overline B_\gamma)_i (H(3^{\prime\prime\prime}))^i (P_\alpha)_k^k(P_\beta)_j^j+c_3^{\prime\prime\prime} (\overline B_\gamma)_i (H(3^{\prime\prime\prime}))^k (P_\alpha)^i_k(P_\beta)_j^j \nonumber\\&+d_3^{\prime\prime\prime} (\overline B_\gamma)_i (H(3^{\prime\prime\prime}))^k (P_\alpha)^i_j(P_\beta)^j_k+\alpha \leftrightarrow \beta.
\end{align}
By substituting Eq.~\eqref{xb} into the amplitudes of $T$, $C$, $E$..., the relations between topological diagrams and the $SU(3)$ irreducible amplitudes in the charm-less $B$ decay are derived to be
\begin{align}\label{sol4}
 a_6&  =\frac{E-A}{4},  \qquad b_6 = \frac{T^{ES}-T^{AS}}{4},  \qquad c_6 = \frac{-T+C}{4}, \nonumber\\
  a_{15}&  =\frac{E+A}{8},  \qquad b_{15} = \frac{T^{ES}+T^{AS}}{8},  \qquad c_{15} = \frac{T+C}{8},\nonumber\\
 a^\prime_3& = \frac{3}{8}E-\frac{1}{8}A+T^{LA},\qquad
 a_3 = -\frac{1}{8}E+\frac{3}{8}A+ T^{QA},\nonumber\\
  b^\prime_3 & = \frac{3}{8}T^{ES}-\frac{1}{8}T^{AS}+T^{LS},\qquad
 b_3  = -\frac{1}{8}T^{ES}+\frac{3}{8}T^{AS} + T^{QS},\nonumber\\
  c^\prime_3 & = -\frac{1}{8}T+\frac{3}{8}C-\frac{1}{8}T^{ES} + \frac{3}{8}T^{AS}+T^{LC},\qquad
  c_3  = \frac{3}{8}T-\frac{1}{8}C+\frac{3}{8}T^{ES} -\frac{1}{8}T^{AS}+T^{QC},\nonumber\\
   d^\prime_3 & = \frac{3}{8}T-\frac{1}{8}C-\frac{1}{8}E + \frac{3}{8}A+T^{LP},\qquad
   d_3  = -\frac{1}{8}T+\frac{3}{8}C+\frac{3}{8}E - \frac{1}{8}A+T^{QP},\nonumber\\
  a_3^{\prime\prime} & = T^{LA}_c,\qquad  b_3^{\prime\prime} = T^{LS}_c, \qquad  c_3^{\prime\prime} = T^{LC}_c, \qquad d_3^{\prime\prime} = T^{LP}_c,\nonumber\\
   a_3^{\prime\prime\prime} & = T^{QA}_c,\qquad  b_3^{\prime\prime\prime} = T^{QS}_c, \qquad  c_3^{\prime\prime\prime} = T^{QC}_c, \qquad d_3^{\prime\prime\prime} = T^{QP}_c.
\end{align}

The non-zero CKM components $(H^{(0,1)})^{ij}_k$, $(H^{(0,1)})^{ci}_c$, $(H^{(0,1)})^{ic}_c$ in the SM and their $SU(3)$ decompositions are listed in Appendix~\ref{b}.
Similar to the strange-less charm decay, if the tree operators are inserted into the diagrams in Eq.~\eqref{nx}, all the diagrams with superscript $QX$ vanish.
If the penguin operators are inserted into the diagrams, $P^{LP}_c$, $P^{LC}_c$, $P^{LA}_c$ and $P^{LS}_c$ vanish.
The $SU(3)$ irreducible amplitudes contributing to $\Delta S =0$ transition associated with $3$-dimensional presentations can be absorbed into four parameters with following redefinition:
\begin{align}\label{c14}
 & a_3  = -\frac{\lambda_u}{\lambda_t}a_3^\prime-\frac{\lambda_c}{\lambda_t}a_3^{\prime\prime}+Pa_3^\prime + 3Pa_3+Pa^{\prime\prime\prime}_3, \qquad  b_3  = -\frac{\lambda_u}{\lambda_t}b_3^\prime-\frac{\lambda_c}{\lambda_t}b_3^{\prime\prime}+Pb_3^\prime + 3Pb_3+Pb^{\prime\prime\prime}_3, \nonumber\\ &   c_3  = -\frac{\lambda_u}{\lambda_t}c_3^\prime-\frac{\lambda_c}{\lambda_t}c_3^{\prime\prime}+Pc_3^\prime + 3Pc_3+Pc^{\prime\prime\prime}_3, \qquad d_3  = -\frac{\lambda_u}{\lambda_t}d_3^\prime-\frac{\lambda_c}{\lambda_t}d_3^{\prime\prime}+Pd_3^\prime + 3Pd_3+Pd^{\prime\prime\prime}_3,
\end{align}
where $\lambda_u = V_{ub}V^*_{ud}$, $\lambda_c = V_{cb}V^*_{cd}$, $\lambda_t = V_{tb}V^*_{td}$.
The tree-operator-induced topological amplitudes with quark loop and all the penguin-operator-induced topological amplitudes contributing to $\Delta S =0$ transition can be absorbed into four parameters with following redefinition:
\begin{align}\label{c15}
 & A^{LA}  = -\frac{\lambda_u}{\lambda_t}T^{LA}-\frac{\lambda_c}{\lambda_t}T^{LA}_c+PA+P^{LA}+3P^{QA}+P^{QA}_c, \nonumber\\ &   A^{LS}  =  -\frac{\lambda_u}{\lambda_t}T^{LS}-\frac{\lambda_c}{\lambda_t}T^{LS}_c+P^{AS}+P^{LS}+3P^{QS} + +P^{QS}_c, \nonumber\\ &  A^{LC}  = -\frac{\lambda_u}{\lambda_t}T^{LC}-\frac{\lambda_c}{\lambda_t}T^{LC}_c+PT+P^{ES}+P^{LC}+3P^{QC}+P^{QC}_c, \nonumber\\ &  A^{LP}  =  -\frac{\lambda_u}{\lambda_t}T^{LP}-\frac{\lambda_c}{\lambda_t}T^{LP}_c+PC+PE+P^{LP}+3P^{QP}+P^{QP}_c.
\end{align}
For $\Delta S =-1$ transition, $\lambda_{u,c,t}$ in Eqs.~\eqref{c14} and \eqref{c15} are replaced by  $\lambda^\prime_{u,c,t}$ and $\lambda^\prime_u = V_{ub}V^*_{us}$, $\lambda^\prime_c = V_{cb}V^*_{cs}$, $\lambda^\prime_t = V_{tb}V^*_{ts}$.
After the redefinitions, there are ten parameters left in the charm-less $B$ decay in the $SU(3)_F$ symmetry. Considering the analysis in subsection \ref{ind},
there are only nine independent parameters in the charm-less $B$ decays in the SM.
The tree- and penguin-operator-induced amplitudes of the $\overline B\to PP$ decays in the TDA and IRA approaches are listed in Tables.~\ref{tab:bppd} and \ref{tab:bpps}.

\begin{table}[t!]
\caption{Decay amplitudes of the $\overline B\to PP$ decays induced by the $b-d$ transition.}\label{tab:bppd}
\begin{ruledtabular}
\scriptsize
\begin{tabular}{|c|c|c|}
{Channel} & {TDA}& {IRA}   \\\hline
$B^-\to \pi^- \pi^0  $  &  $\frac{1}{\sqrt{2}}\lambda_u(T + C)$ & $4\sqrt{2} c_{15}\lambda_u$\\\hline
$B^-\to \pi^- \eta_8  $  &  $\frac{1}{\sqrt{6}}\lambda_u(T + C+2A)-\frac{2}{\sqrt{6}}\lambda_t A^{LP}$ & $\sqrt{\frac{2}{3}}(-a_6+3a_{15}+c_6+3c_{15})\lambda_u-\sqrt{\frac{2}{3}} d_3\lambda_t$\\\hline
$B^-\to \pi^- \eta_1  $  &  $\frac{1}{\sqrt{3}}\lambda_u(T + C+2A+3T^{AS})$ & $-\frac{1}{\sqrt{3}}(2 a_6-6 a_{15}+3 b_6-9 b_{15}+c_6-3 c_{15})\lambda_u$\\ & $-\frac{1}{\sqrt{3}}\lambda_t(2A^{LP}+3A^{LC})$& $-\frac{1}{\sqrt{3}}(3 c_3+2 d_3)\lambda_t$\\\hline
$B^-\to K^- K^0  $  &  $\lambda_u A-\lambda_t A^{LP}$ &$-\left(a_6-3 a_{15}-c_6+c_{15}\right) \lambda _u-d_3 \lambda _t$ \\\hline
$\overline B^0\to \pi^+ \pi^-  $  &  $\lambda_u(T + E)-\lambda_t(A^{LP}+2A^{LA})$ & $(a_6+a_{15}-c_6+3 c_{15}) \lambda _u-(2 a_3+d_3) \lambda _t$\\\hline
$\overline B^0\to \pi^0 \pi^0  $  &  $\frac{1}{\sqrt{2}}\lambda_u(-C + E)-\frac{1}{\sqrt{2}}\lambda_t(A^{LP}+2A^{LA})$ & $\frac{1}{\sqrt{2}}(a_6+a_{15}-c_6-5 c_{15})\lambda _u-\frac{1}{\sqrt{2}}(2 a_3+d_3)\lambda _t$ \\\hline
$\overline B^0\to \pi^0 \eta_8  $  &  $\frac{1}{\sqrt{3}}\lambda_u E+\frac{1}{\sqrt{3}}\lambda_tA^{LP}$ & $\frac{1}{\sqrt{3}}(a_6+5 a_{15}-c_6+c_{15})\lambda_u+\frac{1}{\sqrt{3}} d_3\lambda_t$\\\hline
$\overline B^0\to \pi^0 \eta_1  $  &  $\frac{1}{\sqrt{6}}\lambda_u (2E+3T^{ES})+\frac{1}{\sqrt{6}}\lambda_t(2A^{LP}+3A^{LC})$ & $\frac{1}{\sqrt{6}}(2 a_6+10 a_{15}+3 b_6+15 b_{15}+c_6+5 c_{15})\lambda _u$ \\ &&$+\frac{1}{\sqrt{6}}(3 c_3+2 d_3) \lambda _t$\\\hline
$\overline B^0\to K^+ K^-  $  &  $\lambda_u E-2\lambda_tA^{LA}$ & $2 a_{15} \lambda _u-2 a_3 \lambda _t$\\\hline
$\overline B^0\to K^0 \overline K^0  $  &  $-\lambda_t(A^{LP}+2A^{LA})$ & $-(a_6+3 a_{15}-c_6+c_{15}) \lambda _u-(2 a_3+d_3) \lambda _t$ \\\hline
$\overline B^0\to \eta_8 \eta_8  $  &  $\frac{1}{3\sqrt{2}}\lambda_u(C+E)$ & $\frac{1}{\sqrt{2}}(-a_6-a_{15}+c_6+c_{15}) \lambda _u$\\ &$-\frac{1}{3\sqrt{2}}\lambda_t(A^{LP}+6A^{LA})$& $-\frac{1}{3\sqrt{2}}(6 a_3+d_3) \lambda _t$\\\hline
$\overline B^0\to \eta_8 \eta_1  $  &
$\frac{1}{3\sqrt{2}}\lambda_u(2C+2E+3T^{ES})
$ & $\frac{1}{\sqrt{2}}(2 a_6+2 a_{15}+3 b_6+3 b_{15}+c_6+c_{15})\lambda _u$\\ &$-\frac{1}{3\sqrt{2}}\lambda_t(2A^{LP}+3A^{LC})$&-$\frac{1}{3\sqrt{2}}(3 c_3+2 d_3) \lambda _t$\\\hline
$\overline B^0\to \eta_1 \eta_1  $  &
$\frac{\sqrt{2}}{3}\lambda_u(C+E+3T^{ES})
$ & $-\frac{\sqrt{2}}{3} (3(a_3+3 b_3+c_3)+d_3)\lambda _t$\\ &$-\frac{\sqrt{2}}{3}\lambda_t(A^{LP}+3A^{LC}+3A^{LA}+9A^{LS})$& \\\hline
$\overline B^0_s\to \pi^0 K^0  $  &$\frac{1}{\sqrt{2}}\lambda_u C+\frac{1}{\sqrt{2}}\lambda_t A^{LP}$
&$\frac{1}{\sqrt{2}}(-a_6+a_{15}+c_6+5 c_{15}) \lambda _u+\frac{1}{\sqrt{2}}d_3 \lambda _t$ \\\hline
$\overline B^0_s\to \pi^- K^+  $  &$\lambda_u T-\lambda_t A^{LP}$
&$(a_6-a_{15}-c_6+3 c_{15}) \lambda _u-d_3 \lambda _t$ \\\hline
$\overline B^0_s\to  K^0\eta_8  $  &$\frac{1}{\sqrt{6}}\lambda_u C+\frac{1}{\sqrt{6}}\lambda_t A^{LP}$
&$\frac{1}{\sqrt{6}}(-a_6+a_{15}+c_6+5 c_{15}) \lambda _u+\frac{1}{\sqrt{6}}d_3 \lambda _t$ \\\hline
$\overline B^0_s\to  K^0\eta_1  $  &$\frac{1}{\sqrt{3}}\lambda_u C-\frac{1}{\sqrt{3}}\lambda_t (2A^{LP}+3A^{LC})$
& $\frac{1}{\sqrt{3}} (2 a_6-2 a_{15}+3 b_6-3 b_{15}+c_6-c_{15})\lambda _u-\frac{1}{\sqrt{3}}(3 c_3+2 d_3) \lambda _t$~ \\
\end{tabular}
\end{ruledtabular}
\end{table}
\begin{table}[t!]
\caption{Decay amplitudes of the $\overline B\to PP$ decays induced by the $b-s$ transition.}\label{tab:bpps}
\begin{ruledtabular}
\scriptsize
\begin{tabular}{|c|c|c|}
{Channel} & {TDA}& {IRA}   \\\hline
$B^-\to \pi^0 K^-  $  &  $\frac{1}{\sqrt{2}}\lambda^\prime_u(T + C+A)-\frac{1}{\sqrt{2}}\lambda^\prime_tA^{LP}$ & $\frac{1}{\sqrt{2}}(-a_6+3 a_{15}+c_6+7 c_{15}) \lambda _u'-\frac{1}{\sqrt{2}}d_3 \lambda _t'$ \\\hline
$B^-\to \pi^- \overline K^0  $  &  $\lambda^\prime_uA-\lambda^\prime_tA^{LP}$ &$(-a_6+3 a_{15}+c_6-c_{15}) \lambda _u'-d_3 \lambda _t'$ \\\hline
$B^-\to K^- \eta_8  $  &  $\frac{1}{\sqrt{6}}\lambda^\prime_u(T+C-A)+\frac{1}{\sqrt{6}}\lambda^\prime_tA^{LP}$ & $\frac{1}{\sqrt{6}}(a_6-3 a_{15}-c_6+9 c_{15}) \lambda _u'+\frac{1}{\sqrt{6}}d_3 \lambda _t'$\\\hline
$B^-\to K^- \eta_1  $  &  $\frac{1}{\sqrt{3}}\lambda^\prime_u(T+C+2A+3T^{AS})
$ & $\frac{1}{\sqrt{3}}(-2 a_6+6 a_{15}-3 b_6+9 b_{15}-c_6+3 c_{15}) \lambda _u'$\\ &$-\frac{1}{\sqrt{3}}\lambda^\prime_t(2A^{LP}+3A^{LC})$&$-\frac{1}{\sqrt{3}}(3 c_3+2 d_3) \lambda _t'$ \\\hline
$\overline B^0\to \pi^+ K^-  $  &  $\lambda^\prime_uT-\lambda^\prime_tA^{LP}$ & $(a_6-a_{15}-c_6+3 c_{15}) \lambda _u'-d_3 \lambda _t'$\\\hline
$\overline B^0\to \pi^0 \overline K^0  $  &  $\frac{1}{\sqrt{2}}\lambda^\prime_uC+\frac{1}{\sqrt{2}}\lambda^\prime_tA^{LP}$ & $\frac{1}{\sqrt{2}}(-a_6+a_{15}+c_6+5 c_{15}) \lambda _u'+\frac{1}{\sqrt{2}}d_3 \lambda _t'$\\\hline
$\overline B^0\to \overline K^0 \eta_8 $  &  $\frac{1}{\sqrt{6}}\lambda^\prime_uC+\frac{1}{\sqrt{6}}\lambda^\prime_tA^{LP}$ & $\frac{1}{\sqrt{6}}(-a_6+a_{15}+c_6+5 c_{15}) \lambda _u'+\frac{1}{\sqrt{6}}d_3 \lambda _t'$\\\hline
$\overline B^0\to \overline K^0 \eta_1 $  &  $\frac{1}{\sqrt{3}}\lambda^\prime_uC
-\frac{1}{\sqrt{3}}\lambda^\prime_t(2A^{LP}+3A^{LC})$ &$\frac{1}{\sqrt{3}}(2 a_6-2 a_{15}+3 b_6-3 b_{15}+c_6-c_{15}) \lambda _u'-\frac{1}{\sqrt{3}}(3 c_3+2 d_3) \lambda _t'$~~ \\\hline
$\overline B^0_s\to \pi^+ \pi^- $  &  $\lambda^\prime_uE
-2\lambda^\prime_tA^{LA}$ & $-2 a_3 \lambda _t'+2 a_{15} \lambda _u'$\\\hline
$\overline B^0_s\to \pi^0 \pi^0 $  &  $\frac{1}{\sqrt{2}}\lambda^\prime_uE-\sqrt{2}\lambda^\prime_tA^{LA}$ & $\sqrt{2}(-a_3 \lambda _t'+a_{15} \lambda _u')$\\\hline
$\overline B^0_s\to \pi^0 \eta_8 $  &  $\frac{1}{\sqrt{3}}\lambda^\prime_u(-C+E)$ & $\frac{2}{\sqrt{3}}(a_6+2 a_{15}-c_6-2 c_{15}) \lambda _u'$\\\hline
$\overline B^0_s\to \pi^0 \eta_1 $  &  $\frac{1}{\sqrt{6}}\lambda^\prime_u(C+2E+3T^{ES})$ &$\sqrt{\frac{2}{3}}(2 a_6+4 a_{15}+3 b_6+6 b_{15}+c_6+2 c_{15}) \lambda _u'$ \\\hline
$\overline B^0_s\to K^+ K^- $  &  $\lambda^\prime_u(T+E)-\lambda^\prime_t(A^{LP}+2A^{LA})$ &$-(2 a_3+d_3) \lambda _t'+(a_6+a_{15}-c_6+3 c_{15}) \lambda _u'$ \\\hline
$\overline B^0_s\to K^0\overline K^0 $  &  $-\lambda^\prime_t(A^{LP}+2A^{LA})$ &$-(2 a_3+d_3) \lambda _t'-(a_6+3 a_{15}-c_6+c_{15}) \lambda _u'$ \\\hline
$\overline B^0_s\to \eta_8\eta_8 $  &  $\frac{1}{3\sqrt{2}}\lambda^\prime_u(-2C+E)
-\frac{\sqrt{2}}{3}\lambda^\prime_t(2A^{LP}+3A^{LA})$ & $-\frac{\sqrt{2}}{3}  ((3 a_3+2 d_3) \lambda _t'+3(a_{15}+2 c_{15}) \lambda _u')$\\\hline
$\overline B^0_s\to \eta_8\eta_1 $  &  $\frac{1}{3\sqrt{2}}\lambda^\prime_u(-C+2E+3T^{ES})
$ &$\frac{\sqrt{2}}{3}  ((3 c_3+2 d_3) \lambda _t'+3(2 a_{15}+3 b_{15}+c_{15}) \lambda _u')$ \\ &$+\frac{\sqrt{2}}{3}\lambda^\prime_t(2A^{LP}+3A^{LC})$&\\\hline
$\overline B^0_s\to \eta_1\eta_1 $  &  $\frac{\sqrt{2}}{3}\lambda^\prime_u(C+E+3T^{ES})
$ & $-\frac{\sqrt{2}}{3}  (3(a_3+3 b_3+c_3)+d_3) \lambda _t'$\\ &$-\frac{\sqrt{2}}{3}\lambda^\prime_t(A^{LP}+3A^{LA}+3A^{LC}+9A^{LS})$& \\
\end{tabular}
\end{ruledtabular}
\end{table}

As an example of the charm-less $B$ decays, we write down the amplitude of $\overline B^0\to \pi^+\pi^-$ decay.
The $SU(3)$ irreducible amplitude of $\overline B^0\to \pi^+\pi^-$ decay is
\begin{align}
&\mathcal{A}(\overline{B}^0\to \pi^+\pi^-)=\lambda_u(a_{15}+3c_{15}+a_6-c_6)-\lambda_t(2a_3+d_3).
\end{align}
The topological amplitude of $\overline B^0\to \pi^+\pi^-$ decay is
\begin{align}\label{aqx}
\mathcal{A}(\overline{B}^0\to \pi^+\pi^-)& = \lambda_u(T+E) - \lambda_t(A^{LP}+2A^{LA}) \nonumber\\& =\lambda_u(T+E) + \lambda_u(T^{LP}+2T^{LA})+ \lambda_c(T^{LP}_c+2T^{LA}_c)\nonumber\\&-\lambda_t(PC+PE +2PA+P^{LP}+2P^{LA}+3P^{QP}+6P^{QA}+P^{QP}_c+2P^{QA}_c).
\end{align}
In above formula, the charm-quark loop amplitudes are written explicitly.
If the difference between the $c$-quark loop and $u/d/s$-quark loop is neglected, Eq.~\eqref{aqx} is simplified to the result under the flavor $SU(4)$ symmetry:
\begin{align}
\mathcal{A}(\overline{B}^0\to \pi^+\pi^-)=&\lambda_u(T+E) + (\lambda_u+\lambda_c)(T^{LP}+2T^{LA})\nonumber\\&-\lambda_t(PC+PE +2PA+P^{LP}+2P^{LA}+4P^{QP}+8P^{QA}).
\end{align}

In \cite{He:2018joe}, the topological diagrams are classified according to CKM matrix element:
\begin{align}
\mathcal{A} = V_{ub}V^*_{uq}\mathcal{A}_u + V_{tb}V^*_{tq}\mathcal{A}_t.
\end{align}
The $c$-quark loop is absorbed into $\mathcal{A}_u$ and $\mathcal{A}_t$ with unitarity of the CKM matrix $V_{ub}V^*_{uq}+V_{cb}V^*_{cq}+V_{tb}V^*_{tq}=0$.
$\mathcal{A}_u$ is called as "tree" amplitude and $\mathcal{A}_t$ is "penguin" amplitude since $\mathcal{A}_u$ is dominated by tree contributions but $\mathcal{A}_t$ is not.
This classification of topologies is ambiguous.
In our scheme, the topologies is classified according to which operators, tree or penguin, being inserted into the diagrams, do not care the diagram with quark-loop or not. This classification is definite and convenient.

The strange-less charm decay and the charm-less bottom decay are two examples of the degeneracy and splitting of topologies.
In the strange-less charm decay, the $u,d$-quark loops and $s$-quark loop are degenerate in the flavor $SU(3)$ symmetry.
When the $SU(3)_F$ symmetry breaks into isospin symmetry, the identical $u,d,s$-quark loops split into the unequal $u,d$-quark loops and $s$-quark loop.
The similar situation also exist in the charm-less $B$ decay and the sole difference is that the $SU(4)$ group breaking into $SU(3)$ group.
The operator $O_{ij}^k$ with its indices transforming according to one symmetry group might not include all the contributions.
Operators beyond the given flavor symmetry should be included to get a complete description of topological amplitudes.
And then the corresponding irreducible amplitudes should be modified to match the topological amplitudes.
If not, the topological amplitudes and the irreducible amplitudes are not equivalent.
Refs.~\cite{He:2018php,He:2018joe} did not introduce $O_{ci}^c$ and $O_{ic}^c$ to describe the charm-less $B$ decay, which leads to some confusion.

\section{Conclusion}\label{sum}

In this work, we proposed a systematic theoretical framework for the topological and $SU(N)$ irreducible amplitudes in the two-body non-leptonic heavy meson decays.
Some model-independent conclusions are listed following.
\begin{enumerate}
  \item The topological diagrams can be formalized as invariant tensors constituted by the four-fermion operators and the initial and final states.
\item  The number of possible topologies contributing to one type of decay can be counted in permutations and combinations.
\item The Wigner-Eckhart theorem ensures that the topological amplitudes under flavor symmetry are independent of the initial/final states.
\item The difference between topological amplitudes and $SU(N)$ irreducible amplitudes is whether the four-quark operators are decomposed into irreducible representations of $SU(N)$ group or not, so they are equivalent.
\item  The fact that one of the topologies in the $D$ and $B$ decays is not independent under the $SU(3)_F$ limit can be explained in the group theory.
\item The linear correlation of topologies depends on the symmetry of the physical system.
\end{enumerate}

Applying our framework to the $D\to PP$ decays in the Standard Model, we drew some useful conclusions following.
\begin{enumerate}
  \item The topological diagrams can be classified into tree- and penguin-operator-induced diagrams according to which operators, tree or penguin, being inserted into the effective vertexes, no matter whether the diagrams involving quark-loop or not.
 \item  Once the tree-operator-induced amplitudes in one decay channel are determined, the penguin-operator-induced amplitudes are determined too.
  \item There are ten tree-operator-induced and fourteen penguin-operator-induced diagrams contributing to the $D\to PP$ decays in the SM, but only nine of the twenty-four diagrams are independent in the $SU(3)_F$ limit.
\item  Assuming a large quark loop diagram $T^{LP}$ could explain the large $CP$ violation in charm and the very different branching fractions of the $D^0\to K^+K^-$ and $D^0\to \pi^+\pi^-$ decays with a normal $U$-spin breaking.
  \item  $\Xi^+_c\to pK^-\pi^+$ might be a promising mode to search for $CP$ violation in the charmed baryon decays.
\end{enumerate}

Our framework can include the flavor $SU(N)$ breaking effects naturally. Some conclusions are listed following.
\begin{enumerate}
\item The linear $SU(3)_F$ breaking and the high-order $U$-spin breaking in the charm decays can be reformulated as tensor form of topology, being consistent with literature.
\item The degeneracy/splitting of topologies in the heavy quark decays is similar to the degeneracy/splitting of energy levels.
\item The $SU(3)_F$ analysis for the charm-less $B$ meson decays is different from the $D$ meson decays because the charm-quark loop is beyond the $SU(3)$ symmetry and should be analyzed in the symmetry breaking of $SU(4)\to SU(3)$.
\end{enumerate}

Our theoretical framework can be generalized into other decay modes, which we leave for future work.

\begin{acknowledgements}
We are grateful to Hai-Yang Cheng, Wei Wang, Yu-Ming Wang and  Alexander Lenz for useful discussions.
This work was supported in part by the National Natural Science Foundation of China under
Grants No.U1732101 and 11975112.
\end{acknowledgements}

\begin{appendix}

%%%%%%%%%%%%%%%%%%%%%%%%%%%%%%%
\section{Topologies in $D\to PV$ decays}\label{pv}
\begin{figure}
  \centering
  \includegraphics[width=14cm]{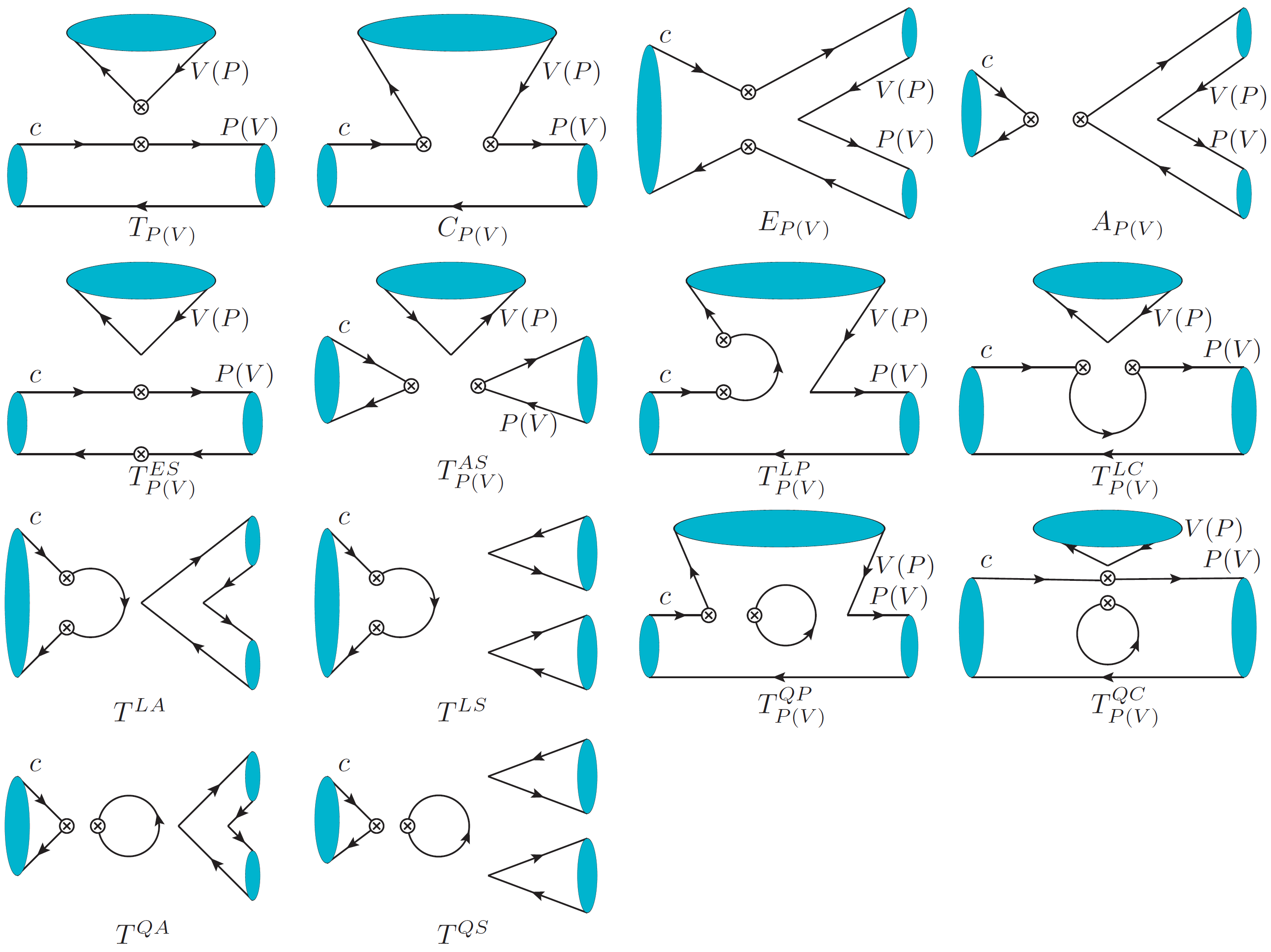}
  \caption{Topological diagrams in the $D\to PV$ decays.}\label{top2}
\end{figure}

\begin{table}[t!]
\caption{Decay amplitudes of the Cabibblo-allowed and doubly Cabibbo-suppressed $D\to PV$ decays.}\label{tab:pv1}
\begin{ruledtabular}
\scriptsize
\begin{tabular}{|c|c|c|}
{channel} & {TDA} & {IRA}   \\\hline
$D^0\to \pi^+   K^{*-}  $  &  $V^*_{cs}V_{ud}(T_V + E_P)$ & $2V^*_{cs}V_{ud}(2d_{15}+d_6+2f_{15}-f_6)$\\\hline
$D^0\to    K^{-} \rho^+ $  &  $V^*_{cs}V_{ud}(T_P + E_V)$ & $2V^*_{cs}V_{ud}(2a_{15}+a_6+2c_{15}-c_6)$\\\hline
$D^0\to \pi^0   \overline K^{*0}  $ & $\frac{1}{\sqrt{2}}V^*_{cs}V_{ud}(C_P-E_P)$ &  $\sqrt{2}V^*_{cs}V_{ud}(-2d_{15}-d_6+2c_{15}+c_6)$ \\\hline
$D^0\to \overline K^{0}\rho^0     $ & $\frac{1}{\sqrt{2}}V^*_{cs}V_{ud}(C_V-E_V)$ &  $\sqrt{2}V^*_{cs}V_{ud}(-2a_{15}-a_6+2f_{15}+f_6)$ \\\hline
$D^0\to \overline K^0   \omega_8  $ & $\frac{1}{\sqrt{6}}V^*_{cs}V_{ud}(C_V+E_V-2E_P)$ & $\frac{2}{\sqrt{6}}V^*_{cs}V_{ud}(2a_{15}+a_6-4d_{15}-2d_{6}+2f_{15}+f_6)$ \\\hline
$D^0\to \overline K^0   \omega_1  $ & ~~$\frac{1}{\sqrt{3}}V^*_{cs}V_{ud}(C_V+E_P+E_V+3T^{ES}_P)$~~ & ~~$\frac{2}{\sqrt{3}}V^*_{cs}V_{ud}(2a_{15}+a_6+2d_{15}+d_6+6b_{15}+3b_6+2f_{15}+f_6)$~~  \\\hline
$D^0\to \eta_8 \overline K^{*0}     $ & $\frac{1}{\sqrt{6}}V^*_{cs}V_{ud}(C_P+E_P-2E_V)$ & $\frac{2}{\sqrt{6}}V^*_{cs}V_{ud}(2d_{15}+d_6-4a_{15}-2a_{6}+2c_{15}+c_6)$ \\\hline
$D^0\to  \eta_1\overline K^{*0}    $ & $\frac{1}{\sqrt{3}}V^*_{cs}V_{ud}(C_P+E_P+E_V+3T^{ES}_V)$ & $\frac{2}{\sqrt{3}}V^*_{cs}V_{ud}(2a_{15}+a_6+2d_{15}+d_6+6e_{15}+3e_6+2c_{15}+c_6)$  \\\hline
$D^+\to \pi^+   \overline K^{*0}  $ & $V^*_{cs}V_{ud}(C_P+T_V)$ & $2V^*_{cs}V_{ud}(2c_{15}+c_6+2f_{15}-f_6)$  \\\hline
$D^+\to \overline K^{0}\rho^+     $ & $V^*_{cs}V_{ud}(C_V+T_P)$ & $2V^*_{cs}V_{ud}(2c_{15}-c_6+2f_{15}+f_6)$  \\\hline
$D^+_s\to \pi^+\rho^0$ &$\frac{1}{\sqrt{2}}V^*_{cs}V_{ud}(A_P-A_V)$&$\sqrt{2}V^*_{cs}V_{ud}(2d_{15}-d_6-2a_{15}+a_6)$\\\hline
$D^+_s\to \pi^0\rho^+$ &$\frac{1}{\sqrt{2}}V^*_{cs}V_{ud}(A_V-A_P)$&$\sqrt{2}V^*_{cs}V_{ud}(2a_{15}-a_6-2d_{15}+d_6)$\\\hline
$D^+_s\to \pi^+   \omega_8  $ & $\frac{1}{\sqrt{6}}V^*_{cs}V_{ud}(A_P+A_V-2T_V)$ & $\frac{2}{\sqrt{6}}V^*_{cs}V_{ud}(2a_{15}-a_6+2d_{15}-d_6-4f_{15}+2f_6)$ \\\hline
$D^+_s\to \pi^+   \omega_1  $ & $\frac{1}{\sqrt{3}}V^*_{cs}V_{ud}(T_V+A_P+A_V+3T^{AS}_P)$ & $\frac{2}{\sqrt{3}}V^*_{cs}V_{ud}(2a_{15}-a_6+2d_{15}-d_6+6b_{15}-3b_6+2f_{15}-f_6)$  \\\hline
$D^+_s\to  \eta_8\rho^+    $ & $\frac{1}{\sqrt{6}}V^*_{cs}V_{ud}(A_P+A_V-2T_P)$ & $\frac{2}{\sqrt{6}}V^*_{cs}V_{ud}(2a_{15}-a_6+2d_{15}-d_6-4c_{15}+2c_6)$ \\\hline
$D^+_s\to \eta_1\rho^+     $ & $\frac{1}{\sqrt{3}}V^*_{cs}V_{ud}(T_P+A_P+A_V+3T^{AS}_V)$ & $\frac{2}{\sqrt{3}}V^*_{cs}V_{ud}(2a_{15}-a_6+2d_{15}-d_6+6e_{15}-3e_6+2c_{15}-c_6)$  \\\hline
$D^+_s\to K^+   \overline K^{*0}  $ & $V^*_{cs}V_{ud}(C_P+A_V)$ & $2V^*_{cs}V_{ud}(2a_{15}-a_6+2c_{15}+c_6)$\\\hline
$D^+_s\to \overline K^0    K^{*+}  $ & $V^*_{cs}V_{ud}(C_V+A_P)$ & $2V^*_{cs}V_{ud}(2d_{15}-d_6+2f_{15}+f_6)$\\\hline
$D^0\to \pi^0   K^{*0}  $ & $\frac{1}{\sqrt{2}}V^*_{cd}V_{us}(C_P-E_V)$ & $\sqrt{2}V^*_{cd}V_{us}(-2a_{15}-a_6+2c_{15}+c_6)$  \\\hline
$D^0\to K^0   \rho^0  $ & $\frac{1}{\sqrt{2}}V^*_{cd}V_{us}(C_V-E_P)$ & $\sqrt{2}V^*_{cd}V_{us}(-2d_{15}-d_6+2f_{15}+f_6)$  \\\hline
$D^0\to \pi^-   K^{*+}  $ & $V^*_{cd}V_{us}(E_V+T_P)$ & $2V^*_{cd}V_{us}(2a_{15}+a_6+2c_{15}-c_6)$  \\\hline
$D^0\to K^+   \rho^{-}  $ & $V^*_{cd}V_{us}(E_P+T_V)$ & $2V^*_{cd}V_{us}(2d_{15}+d_6+2f_{15}-f_6)$  \\\hline
$D^0\to K^0   \omega_8  $ & $\frac{1}{\sqrt{6}}V^*_{cd}V_{us}(C_V+E_P-2E_V)$ & $\frac{2}{\sqrt{6}}V^*_{cd}V_{us}(-2a_{6}+d_6+f_{6}-4a_{15}+2d_{15}+2f_{15})$  \\\hline
$D^0\to K^0   \omega_1  $ & $\frac{1}{\sqrt{3}}V^*_{cd}V_{us}(C_V+E_P+E_V+3T^{ES}_P)$ & $\frac{2}{\sqrt{3}}V^*_{cd}V_{us}(2a_{15}+a_6+2d_{15}+d_6+6b_{15}+3b_{6}+2f_{15}+f_6)$ \\\hline
$D^0\to \eta_8K^{*0}     $ & $\frac{1}{\sqrt{6}}V^*_{cd}V_{us}(C_P+E_V-2E_P)$ & $\frac{2}{\sqrt{6}}V^*_{cd}V_{us}(2c_{15}+c_6+2a_{15}+a_6-4d_{15}-2d_6)$  \\\hline
$D^0\to \eta_1K^{*0}     $ & $\frac{1}{\sqrt{3}}V^*_{cd}V_{us}(C_P+E_P+E_V+3T^{ES}_V)$ & $\frac{2}{\sqrt{3}}V^*_{cd}V_{us}(2a_{15}+a_6+2d_{15}+d_6+6e_{15}+3e_{6}+2c_{15}+c_6)$ \\\hline
$D^+\to \pi^+   K^{*0}  $ & $V^*_{cd}V_{us}(C_P+A_V)$ & $2V^*_{cd}V_{us}(2a_{15}-a_6+2c_{15}+c_6)$ \\\hline
$D^+\to K^0   \rho^+  $ & $V^*_{cd}V_{us}(C_V+A_P)$ & $2V^*_{cd}V_{us}(2d_{15}-d_6+2f_{15}+f_6)$ \\\hline
$D^+\to \pi^0   K^{*+}  $ & $\frac{1}{\sqrt{2}}V^*_{cd}V_{us}(A_V-T_P)$ & $\sqrt{2}V^*_{cd}V_{us}(2a_{15}-a_6-2c_{15}+c_6)$  \\\hline
$D^+\to K^+   \rho^{0}  $ & $\frac{1}{\sqrt{2}}V^*_{cd}V_{us}(A_P-T_V)$ & $\sqrt{2}V^*_{cd}V_{us}(2d_{15}-d_6-2f_{15}+f_6)$  \\\hline
$D^+\to K^+   \omega_8  $ & $\frac{1}{\sqrt{6}}V^*_{cd}V_{us}(T_V+A_P-2A_V)$ & $\frac{2}{\sqrt{6}}V^*_{cd}V_{us}(-4a_{15}+2a_6+2d_{15}-d_6+2f_{15}-f_6)$ \\\hline
$D^+\to K^+   \omega_1  $ & $\frac{1}{\sqrt{3}}V^*_{cd}V_{us}(T_V+A_P+A_V+3T^{AS}_P)$ & $\frac{2}{\sqrt{3}}V^*_{cd}V_{us}(2a_{15}-a_6+2d_{15}-d_6+6b_{15}-3b_6+2f_{15}-f_6)$  \\\hline
$D^+\to  \eta_8 K^{*+}   $ & $\frac{1}{\sqrt{6}}V^*_{cd}V_{us}(T_P-2A_P+A_V)$ & $\frac{2}{\sqrt{6}}V^*_{cd}V_{us}(2a_{15}-a_6-4d_{15}+2d_6+2c_{15}-c_6)$ \\\hline
$D^+\to \eta_1 K^{*+}    $ & $\frac{1}{\sqrt{3}}V^*_{cd}V_{us}(T_P+A_P+A_V+3T^{AS}_V)$ & $\frac{2}{\sqrt{3}}V^*_{cd}V_{us}(2a_{15}-a_6+2d_{15}-d_6+6e_{15}-3e_6+2c_{15}-c_6)$  \\\hline
$D^+_s\to K^+   K^{*0}  $ & $V^*_{cd}V_{us}(C_P+T_V)$ & $2V^*_{cd}V_{us}(2c_{15}+c_6+2f_{15}-f_6)$  \\\hline
$D^+_s\to K^0   K^{*+}  $ & $V^*_{cd}V_{us}(C_V+T_P)$ & $2V^*_{cd}V_{us}(2c_{15}-c_6+2f_{15}+f_6)$
\end{tabular}
\end{ruledtabular}
\end{table}

\begin{table}[thp!]
\caption{Decay amplitudes of the Singly Cabibblo-suppressed $D^0\to PV$ decays. }\label{tab:pv2}
\begin{ruledtabular}
\scriptsize
\begin{tabular}{|c|c|c|}
{channel} & {TDA} & {IRA}   \\\hline
$D^0\to \pi^+   \rho^-  $ & $\lambda_d(T_V+E_P)-\lambda_b(A^{LP}_V+A^{LA})$ & $(3\lambda_d-\lambda_s)(d_{15}+f_{15})+(\lambda_d-\lambda_s)(d_6-f_6)$\\ & & $-\lambda_b(a_3+f_3-2a_{15})$\\\hline
$D^0\to \pi^-   \rho^+  $ & $\lambda_d(T_P+E_V)-\lambda_b(A^{LP}_P+A^{LA})$ & $(3\lambda_d-\lambda_s)(a_{15}+c_{15})+(\lambda_d-\lambda_s) (a_6-c_6)$\\ & & $-\lambda_b(a_3+d_3-2d_{15})$\\\hline
$D^0\to \pi^0   \rho^0  $ & $\frac{1}{2}\lambda_d(E_P+E_V-C_P-C_V)$ &  $\frac{1}{2}(3\lambda_d-\lambda_s)(a_{15}+d_{15}-c_{15}-f_{15})$ \\&$-\frac{1}{2}\lambda_b(A^{LP}_P+A^{LP}_V+2A^{LA})$&$+(\lambda_d-\lambda_s)(a_6+d_6-c_6-f_6)$ \\& & $-\frac{1}{2}\lambda_b(2a_3+d_3+f_3-2(a_{15}+c_{15}+d_{15}+f_{15}))$
\\\hline
$D^0\to \pi^0   \omega_8  $ & $-\frac{1}{2\sqrt{3}}\lambda_d(E_P+E_V-C_P+C_V)$ & $-\frac{1}{2\sqrt{3}}\lambda_b(d_3+f_3-2(a_{15}+d_{15}+c_{15}+f_{15}))$  \\  & $-\frac{1}{\sqrt{3}}\lambda_sC_P-\frac{1}{2\sqrt{3}}\lambda_b(A^{LP}_P+A^{LP}_V)$ & $-\frac{1}{2\sqrt{3}}(3\lambda_d-\lambda_s) (a_{15}+d_{15}-c_{15}+f_{15})$ \\ &&$-\frac{1}{\sqrt{3}}(3\lambda_s-\lambda_d)c_{15}-\frac{1}{\sqrt{3}}(\lambda_d-\lambda_s) (a_6+d_6-3c_6+f_6)$
 \\\hline
$D^0\to \pi^0   \omega_1  $ & $\frac{1}{\sqrt{6}}\lambda_d(C_P-C_V-E_P-E_V-3T^{ES}_P)$& $-\frac{1}{\sqrt{6}}\lambda_b(3c_3+d_3+f_3-2(a_{15}+d_{15}+3b_{15}+c_{15}+f_{15}))$ \\&$+\frac{1}{\sqrt{6}}\lambda_sC_P-\frac{1}{\sqrt{6}}\lambda_b(A^{LP}_P+A^{LP}_V+3A^{LC}_P)$ & $-\frac{1}{\sqrt{6}}(3\lambda_d-\lambda_s)(a_{15}+d_{15}-c_{15}+f_{15}+3b_{15})$\\
&&$+\frac{1}{\sqrt{6}}(3\lambda_s-\lambda_d)c_{15}-\frac{1}{\sqrt{6}}(\lambda_d-\lambda_s)(a_6+d_6+f_6+3b_6)$ \\\hline
$D^0\to  \eta_8\rho^0 $  & $-\frac{1}{2\sqrt{3}}\lambda_d(E_P+E_V+C_P-C_V)-\frac{1}{\sqrt{3}}\lambda_sC_V$ & $-\frac{1}{2\sqrt{3}}\lambda_b(d_3+f_3-2(a_{15}+d_{15}+c_{15}+f_{15}))$  \\  & $-\frac{1}{2\sqrt{3}}\lambda_b(A^{LP}_P+A^{LP}_V)$ & $-\frac{1}{2\sqrt{3}}(3\lambda_d-\lambda_s)(a_{15}+d_{15}+c_{15}-f_{15})$ \\ &&$-\frac{1}{\sqrt{3}}(3\lambda_s-\lambda_d)f_{15}-\frac{1}{\sqrt{3}}(\lambda_d-\lambda_s) (a_6+d_6+c_6-3f_6)$ \\\hline
$D^0\to  \eta_1\rho^0  $ & $\frac{1}{\sqrt{6}}\lambda_d(C_V-C_P-E_P-E_V-3T^{ES}_V)$& $-\frac{1}{\sqrt{6}}\lambda_b(3e_3+d_3+f_3-2(a_{15}+d_{15}+3e_{15}+c_{15}+f_{15}))$ \\&$+\frac{1}{\sqrt{6}}\lambda_sC_V-\frac{1}{\sqrt{6}}\lambda_b(A^{LP}_P+A^{LP}_V+3A^{LC}_V)$ & $-\frac{1}{\sqrt{6}}(3\lambda_d-\lambda_s)(a_{15}+d_{15}+c_{15}-f_{15}+3e_{15})
$\\
&&$+\frac{1}{\sqrt{6}}(3\lambda_s-\lambda_d)f_{15}-\frac{1}{\sqrt{6}}(\lambda_d-\lambda_s)(a_6+d_6+c_6+3e_6)$   \\\hline
$D^0\to K^+   K^{*-}  $ & $\lambda_s(T_V+E_P) -\lambda_b(A^{LP}_V+A^{LA})$ & $-\lambda_b(a_3+f_3-2a_{15})+(3\lambda_s-\lambda_d)(d_{15}+f_{15})$\\& & $-(\lambda_d-\lambda_s) (d_6-f_6)$ \\ \hline
$D^0\to K^-   K^{*+}  $ & $\lambda_s(T_P+E_V) -\lambda_b(A^{LP}_P+A^{LA})$ & $-\lambda_b(a_3+d_3-2d_{15})+(3\lambda_s-\lambda_d)(a_{15}+c_{15})$\\&  & $-(\lambda_d-\lambda_s) (a_6-c_6)$\\ \hline
$D^0\to K^0  \overline K^{*0}  $ &$\lambda_dE_V+\lambda_sE_P-\lambda_bA^{LA}$ & $-\lambda_ba_3+(3\lambda_d-\lambda_s) a_{15}+(3\lambda_s-\lambda_d)d_{15}$\\
& & $+(\lambda_d-\lambda_s)(a_6-d_6)$ \\\hline
$D^0\to \overline K^0  K^{*0}  $ &$\lambda_dE_P+\lambda_sE_V-\lambda_bA^{LA}$ & $-\lambda_ba_3+(3\lambda_d-\lambda_s)d_{15}+
(3\lambda_s-\lambda_d)a_{15}$
\\& & $-(\lambda_d-\lambda_s)(a_6-d_6)$\\\hline
$D^0\to \eta_8   \omega_8  $ & $\frac{1}{6}\lambda_d(C_P+C_V+E_P+E_V)-\frac{1}{3}\lambda_s(C_P+C_V$ & $-\frac{1}{6}\lambda_b(6a_3+d_3+f_3-2(a_{15}+d_{15}+c_{15}+f_{15}))$\\ &$-2E_P-2E_V)-\frac{1}{6}\lambda_b(A^{LP}_P+A^{LP}_V+6A^{LA})$ & $-\frac{1}{2}(\lambda_d-\lambda_s) (a_6+d_6-c_6-f_6)+\frac{1}{6}(3\lambda_d-\lambda_s)(a_{15}+d_{15}$\\
&&$+c_{15}+f_{15})+\frac{1}{3}(3\lambda_s-\lambda_d)(2a_{15}+2d_{15}-c_{15}-f_{15})$ \\\hline
$D^0\to \eta_8   \omega_1  $ & $\frac{1}{3\sqrt{2}}\lambda_d(C_P+C_V+E_P+E_V+3T^{ES}_P)+$& $-\frac{1}{3\sqrt{2}}\lambda_b(3c_3+d_3+f_3-2(a_{15}+d_{15}+3b_{15}+c_{15}+f_{15}))$\\
&$\frac{1}{3\sqrt{2}}\lambda_s(C_P-2C_V-2E_P-2E_V-6T^{ES}_P)$ & $+\frac{1}{\sqrt{2}}(\lambda_d-\lambda_s) (a_6+d_6+3b_6+f_6)$ \\ & $-\frac{1}{3\sqrt{2}}\lambda_b(A^{LP}_P+A^{LP}_V+3A^{LC}_P)$&
$+\frac{1}{3\sqrt{2}}(3\lambda_d-\lambda_s) (a_{15}+d_{15}+3b_{15}+c_{15}+f_{15})$\\&&$-\frac{1}{3\sqrt{2}}(3\lambda_s-\lambda_d)(2a_{15}+2d_{15}+6b_{15}-c_{15}+2f_{15})$\\\hline
$D^0\to \eta_1   \omega_8  $ & $\frac{1}{3\sqrt{2}}\lambda_d(C_P+C_V+E_P+E_V+3T^{ES}_V)+$& $-\frac{1}{3\sqrt{2}}\lambda_b(3e_3+d_3+f_3-2(a_{15}+d_{15}+3e_{15}+c_{15}+f_{15}))$\\
&$\frac{1}{3\sqrt{2}}\lambda_s(C_V-2C_P-2E_P-2E_V-6T^{ES}_V)$ & $+\frac{1}{\sqrt{2}}(\lambda_d-\lambda_s)(a_6+d_6+3e_6+c_6)$ \\ & $-\frac{1}{3\sqrt{2}}\lambda_b(A^{LP}_P+A^{LP}_V+3A^{LC}_V)$&
$+\frac{1}{3\sqrt{2}}(3\lambda_d-\lambda_s)(a_{15}+d_{15}+3e_{15}+c_{15}+f_{15})$\\
&&$-\frac{1}{3\sqrt{2}}(3\lambda_s-\lambda_d)(2a_{15}+2d_{15}+6e_{15}+2c_{15}-f_{15})$\\\hline
$D^0\to \eta_1   \omega_1  $ & $-\frac{1}{3}\lambda_b(C_P+C_V+E_P+E_V+3T^{ES}_P+3T^{ES}_V$ & $-\frac{1}{3}\lambda_b(3a_3+9b_3+3c_3+3e_3+d_3+f_3)$ \\
&$+A^{LP}_P+A^{LP}_V+3A^{LA}+3A^{LC}_P+3A^{LC}_V+9A^{LS})$&
\end{tabular}
\end{ruledtabular}
\end{table}

\begin{table}[thp!]
\caption{Decay amplitudes of the singly Cabibblo-suppressed $D^+_{(s)}\to PV$ decays. }\label{tab:pv3}
\begin{ruledtabular}
\scriptsize
\begin{tabular}{|c|c|c|}
{channel}& {TDA}  &{IRA}    \\\hline
$D^+\to \pi^+   \rho^0  $& $-\frac{1}{\sqrt{2}}\lambda_d(T_V+C_P-A_P+A_V)$ & $-\frac{1}{\sqrt{2}}(3\lambda_d-\lambda_s)(a_{15}-d_{15}+c_{15}+f_{15})$ \\
&$-\frac{1}{\sqrt{2}}\lambda_b(A^{LP}_P-A^{LP}_V)$&
$+\frac{1}{\sqrt{2}}(\lambda_d-\lambda_s) (a_6-d_{6}-c_6+f_6)-\frac{1}{\sqrt{2}}\lambda_b(d_3-f_3-2c_{15})$\\\hline
$D^+\to \pi^0   \rho^+  $& $-\frac{1}{\sqrt{2}}\lambda_d(T_P+C_V+A_P-A_V)$ & $\frac{1}{\sqrt{2}}(3\lambda_d-\lambda_s)(a_{15}-d_{15}-c_{15}-f_{15})$ \\
&$-\frac{1}{\sqrt{2}}\lambda_b(A^{LP}_V-A^{LP}_P)$&
$-\frac{1}{\sqrt{2}}(\lambda_d-\lambda_s)(a_{6}-d_{6}-c_{6}+f_{6})
-\frac{1}{\sqrt{2}}\lambda_b(f_3-d_3-2f_{15})$ \\\hline
$D^+\to \pi^+   \omega_8  $ & $\frac{1}{\sqrt{6}}\lambda_d(T_V+C_P+A_P+A_V)$ & $-\frac{1}{\sqrt{6}}\lambda_b(d_3+f_3-2c_{15})+\frac{1}{\sqrt{6}}(3\lambda_d-\lambda_s)(a_{15}+d_{15}+c_{15}+f_{15})$ \\ & $-\frac{2}{\sqrt{6}}\lambda_sC_P-\frac{1}{\sqrt{6}}\lambda_b(A^{LP}_P+A^{LP}_V)$& $-\frac{2}{\sqrt{6}}(3\lambda_s-\lambda_d)c_{15}-\frac{1}{\sqrt{6}}(\lambda_d-\lambda_s) (a_6+d_6-3c_6+f_6)$ \\\hline
$D^+\to \pi^+   \omega_1  $ & $\frac{1}{\sqrt{3}}\lambda_d(T_V+C_P+A_P+A_V+3T^{AS}_P)$ & $-\frac{1}{\sqrt{3}}\lambda_b(3c_3+d_3+f_3-2c_{15})-\frac{1}{\sqrt{3}}(\lambda_d-\lambda_s)(a_6+d_6+3b_6+f_6)$ \\  &  $+\frac{1}{\sqrt{3}}\lambda_sC_P-\frac{1}{\sqrt{3}}\lambda_b(A^{LP}_P+A^{LP}_V+3A^{LC}_P)$ & $+\frac{1}{\sqrt{3}}(3\lambda_d-\lambda_s) (a_{15}+d_{15}+3b_{15}+c_{15}+f_{15})+\frac{1}{\sqrt{3}}(3\lambda_s-\lambda_d)c_{15}$\\\hline
$D^+\to \eta_8\rho^+     $ & $\frac{1}{\sqrt{6}}\lambda_d(T_P+C_V+A_P+A_V)$ & $-\frac{1}{\sqrt{6}}\lambda_b(d_3+f_3-2f_{15})+\frac{1}{\sqrt{6}}(3\lambda_d-\lambda_s)(a_{15}+d_{15}+c_{15}+f_{15})$ \\ & $-\frac{2}{\sqrt{6}}\lambda_sC_V-\frac{1}{\sqrt{6}}\lambda_b(A^{LP}_P+A^{LP}_V)$& $-\frac{2}{\sqrt{6}}(3\lambda_s-\lambda_d)f_{15}-\frac{1}{\sqrt{6}}(\lambda_d-\lambda_s) (a_6+d_6+c_6-3f_6)$ \\\hline
$D^+\to  \eta_1\rho^+    $ & $\frac{1}{\sqrt{3}}\lambda_d(T_P+C_V+A_P+A_V+3T^{AS}_V)$ & $-\frac{1}{\sqrt{3}}\lambda_b(3e_3+d_3+f_3-2f_{15})-\frac{1}{\sqrt{6}}(\lambda_d-\lambda_s)(a_6+d_6+3e_6+c_6)$ \\  &  $+\frac{1}{\sqrt{3}}\lambda_sC_V-\frac{1}{\sqrt{3}}\lambda_b(A^{LP}_P+A^{LP}_V+3A^{LC}_V)$ & $+\frac{1}{\sqrt{3}}(3\lambda_d-\lambda_s)(a_{15}+d_{15}+3e_{15}+c_{15}+f_{15})+\frac{1}{\sqrt{3}}(3\lambda_s-\lambda_d)f_{15}$\\\hline
$D^+\to K^+   \overline K^{*0}  $ & $\lambda_dA_V+\lambda_sT_V-\lambda_bA^{LP}_V$ & $-\lambda_bf_3+(3\lambda_d-\lambda_s)a_{15}+(3\lambda_s-\lambda_d)f_{15}-(\lambda_d-\lambda_s) (a_6-f_6)$ \\\hline
$D^+\to \overline K^{0}K^{*+}     $ & $\lambda_dA_P+\lambda_sT_P-\lambda_bA^{LP}_P$ & $-\lambda_bd_3+(3\lambda_d-\lambda_s)d_{15}+(3\lambda_s-\lambda_d)c_{15}-(\lambda_d-\lambda_s)(d_6-c_6)$ \\\hline
$D^+_s\to \pi^+   K^{*0}  $ & $\lambda_dT_V+\lambda_sA_V-\lambda_bA^{LP}_V$& $-\lambda_bf_3+(3\lambda_d-\lambda_s) f_{15}+(3\lambda_s-\lambda_d)a_{15}+(\lambda_d-\lambda_s) (a_6-f_6)$  \\\hline
$D^+_s\to K^{0} \rho^+    $ & $\lambda_dT_P+\lambda_sA_P-\lambda_bA^{LP}_P$& $-\lambda_bd_3+(3\lambda_d-\lambda_s)c_{15}+(3\lambda_s-\lambda_d)d_{15}+(\lambda_d-\lambda_s)(d_6-c_6)$  \\\hline
$D^+_s\to \pi^0   K^{*+}  $ & $-\frac{1}{\sqrt{2}}(\lambda_dC_V-\lambda_sA_V+\lambda_bA^{LP}_V)$ & $-\frac{1}{\sqrt{2}}(3\lambda_d-\lambda_s) f_{15}
+\frac{1}{\sqrt{2}}(3\lambda_s-\lambda_d)a_{15}$ \\& & $+\frac{1}{\sqrt{2}}(\lambda_d-\lambda_s)(a_6-f_6)-\frac{1}{\sqrt{2}}\lambda_b(f_3-2f_{15})$ \\\hline
$D^+_s\to  K^{+}\rho^0    $ & $-\frac{1}{\sqrt{2}}(\lambda_dC_P-\lambda_sA_P+\lambda_bA^{LP}_P)$ & $-\frac{1}{\sqrt{2}}(3\lambda_d-\lambda_s)c_{15}
+\frac{1}{\sqrt{2}}(3\lambda_s-\lambda_d)d_{15}$ \\& & $+\frac{1}{\sqrt{2}}(\lambda_d-\lambda_s) (d_6-c_6)-\frac{1}{\sqrt{2}}\lambda_b(d_3-2c_{15})$  \\\hline
$D^+_s\to K^+   \omega_8  $ & $-\frac{1}{\sqrt{6}}\lambda_s(2T_V+2C_P-A_P+2A_V)$ & $-\frac{1}{\sqrt{6}}\lambda_b(d_3-2f_3-2c_{15})-\frac{1}{\sqrt{6}}(\lambda_d-\lambda_s)(2a_6-d_6-3c_6+2f_6)$ \\ &$+\frac{1}{\sqrt{6}}\lambda_dC_P-\frac{1}{\sqrt{6}}\lambda_b(A^{LP}_P-2A^{LP}_V)$& $+\frac{1}{\sqrt{6}}(3\lambda_d-\lambda_s)c_{15}-\frac{1}{\sqrt{6}}(3\lambda_s-\lambda_d)(2a_{15}-d_{15}+2c_{15}+2f_{15})$\\\hline
$D^+_s\to K^+   \omega_1  $ & $\frac{1}{\sqrt{3}}\lambda_s(T_V+C_P+A_P+A_V+3T^{AS}_P)$ & $-\frac{1}{\sqrt{3}}\lambda_b(3c_3+d_3+f_3-2c_{15})+\frac{1}{\sqrt{3}}(3\lambda_d-\lambda_s)c_{15}$ \\ &$+\frac{1}{\sqrt{3}}\lambda_dC_P-\frac{1}{\sqrt{3}}\lambda_b(A^{LP}_P+A^{LP}_V+3A^{LC}_P)$ & $+\frac{1}{\sqrt{3}}(3\lambda_s-\lambda_d)(a_{15}+d_{15}+3b_{15}+c_{15}+f_{15})$\\& & $+\frac{1}{\sqrt{3}}(\lambda_d-\lambda_s) (a_6+d_6+3b_6+f_6)$ \\\hline
$D^+_s\to  \eta_8 K^{*+}   $ & $-\frac{1}{\sqrt{6}}\lambda_s(2T_P+2C_V-A_V+2A_P)$ & $-\frac{1}{\sqrt{6}}\lambda_b(f_3-2d_3-2f_{15})-\frac{1}{\sqrt{6}}(\lambda_d-\lambda_s) (2d_6-a_6-3f_6+2c_6)$ \\ &$+\frac{1}{\sqrt{6}}\lambda_dC_V-\frac{1}{\sqrt{6}}\lambda_b(A^{LP}_V-2A^{LP}_P)$& $+\frac{1}{\sqrt{6}}(3\lambda_d-\lambda_s)f_{15}-\frac{1}{\sqrt{6}}(3\lambda_s-\lambda_d)(2d_{15}-a_{15}+2c_{15}+2f_{15})$\\\hline
$D^+_s\to \eta_1K^{*+}     $ & $\frac{1}{\sqrt{3}}\lambda_s(T_P+C_V+A_P+A_V+3T^{AS}_V)$ & $-\frac{1}{\sqrt{3}}\lambda_b(3e_3+d_3+f_3-2f_{15})+\frac{1}{\sqrt{3}}(\lambda_d-\lambda_s) (a_6+d_6+3e_6+c_6)$ \\ &$+\frac{1}{\sqrt{3}}\lambda_dC_V-\frac{1}{\sqrt{3}}\lambda_b(A^{LP}_P+A^{LP}_V+3A^{LC}_V)$ & $+\frac{1}{\sqrt{3}}(3\lambda_d-\lambda_s)f_{15}+\frac{1}{\sqrt{3}}(3\lambda_s-\lambda_d)(a_{15}+d_{15}+3e_{15}+c_{15}+f_{15})$
\end{tabular}
\end{ruledtabular}
\end{table}

In this appendix, we present the topological amplitudes of the $D\to PV$ decays.
The vector meson nonet is
\begin{eqnarray}
 |V\rangle ^i_j=  \left( \begin{array}{ccc}
   \frac{1}{\sqrt 2} |\rho^0\rangle+  \frac{1}{\sqrt 6} |\omega_8\rangle,    & |\rho^+\rangle,  & |K^{*+}\rangle \\
    |\rho^-\rangle, &   - \frac{1}{\sqrt 2} |\rho^0\rangle+ \frac{1}{\sqrt 6} |\omega_8\rangle,   & |K^{*0}\rangle \\
    |K^{*-}\rangle, & |\overline K^{*0}\rangle, & -\sqrt{2/3}|\omega_8\rangle \\
  \end{array}\right) +  \frac{1}{\sqrt 3} \left( \begin{array}{ccc}
   |\omega_1\rangle,    & 0,  & 0 \\
    0, &  |\omega_1\rangle,   & 0 \\
   0, & 0, & |\omega_1\rangle \\
  \end{array}\right).
\end{eqnarray}
In the $D\to PV$ mode, there are $N=A^4_4=24$ possible topological diagrams.
Amplitude of the $D\to PV$ decay can be written as
\begin{align}\label{hv}
{\cal A}^{\rm TDA}_{D_\gamma \to P_\alpha V_\beta}& = T_P  (D_\gamma)_i (H)^{lj}_k (P_\alpha)^{i}_j  (V_\beta)^k_l +T_V  (D_\gamma)_i (H)^{lj}_k (V_\beta)^{i}_j  (P_\alpha)^k_l + C_P (D_\gamma)_i(H)^{jl}_k  (P_\alpha)^{i}_j  (V_\beta)^k_l   \nonumber\\& + C_V (D_\gamma)_i (H)^{jl}_k(V_\beta)^{i}_j   (P_\alpha)^k_l + E_P  (D_\gamma)_i  (H)^{il}_j (P_\alpha)^j_k (V_\beta)^{k}_l + E_V  (D_\gamma)_i (H)^{il}_j (V_\beta)^j_k (P_\alpha)^{k}_l   \nonumber\\& + A_P  (D_\gamma)_i (H)^{li}_j   (P_\alpha)^j_k (V_\beta)^{k}_l + A_V  (D_\gamma)_i (H)^{li}_j   (V_\beta)^j_k (P_\alpha)^{k}_l+T^{ES}_P (D_\gamma)_i   (H)^{ij}_{l}   (P_\alpha)^{l}_j    (V_\beta)_k^k  \nonumber\\& +T^{ES}_V (D_\gamma)_i   (H)^{ij}_{l} (V_\beta)^{l}_j (P_\alpha)_k^k  +T^{AS}_P (D_\gamma)_i  (H)^{ji}_{l}  (P_\alpha)^{l}_j  (V_\beta)_k^k +T^{AS}_V (D_\gamma)_i (H)^{ji}_{l} (V_\beta)^{l}_j  (P_\alpha)_k^k \nonumber\\&+T^{LP}_P (D_\gamma)_i (H)^{kl}_{l} (P_\alpha)^{i}_j  (V_\beta)^j_k  +T^{LP}_V (D_\gamma)_i (H)^{kl}_{l} (V_\beta)^{i}_j   (P_\alpha)^j_k  + T^{LC}_P (D_\gamma)_i  (H)^{jl}_{l} (P_\alpha)^{i}_j  (V_\beta)^k_k\nonumber\\
&+ T^{LC}_V (D_\gamma)_i (H)^{jl}_{l} (V_\beta)^{i}_j  (P_\alpha)^k_k+T^{QP}_P (D_\gamma)_i  (H)^{lk}_{l}(P_\alpha)^{i}_j   (V_\beta)^j_k +T^{QP}_V (D_\gamma)_i (H)^{lk}_{l} (V_\beta)^{i}_j   (P_\alpha)^j_k   \nonumber\\
&+ T^{QC}_P (D_\gamma)_i (H)^{lj}_{l} (P_\alpha)^{i}_j  (V_\beta)^k_k + T^{QC}_V (D_\gamma)_i (H)^{lj}_{l}  (V_\beta)^{i}_j   (P_\alpha)^k_k + T^{LA} (D_\gamma)_i  (H)^{il}_{l}  (P_\alpha)^j_k (V_\beta)^{k}_j   \nonumber\\
& + T^{LS} (D_\gamma)_i  (H)^{il}_{l}  (P_\alpha)_j^j (V_\beta)_{k}^{k} + T^{QA} (D_\gamma)_i  (H)^{li}_{l}  (P_\alpha)^j_k (V_\beta)^{k}_j  \nonumber\\
&+ T^{QS} (D_\gamma)_i (H)^{li}_{l}  (P_\alpha)_j^j (V_\beta)_{k}^{k}.
\end{align}
The topological diagrams in the $D\to PV$ decays are showed in Fig.~\ref{top2}.
The $SU(3)$ irreducible amplitude of the $D \to PV$ decay is
\begin{align}
&~~~~~~~~~~{\cal A}^{\rm IRA}_{D_\gamma \to P_\alpha V_\beta} =\nonumber\\&~~~~
a_6(D_\gamma)_i (H(\overline 6))^{ij}_k(P_\alpha)_j^l(V_\beta)_l^k +d_6(D_\gamma)_i (H(\overline 6))^{ij}_k(V_\beta)_j^l(P_\alpha)_l^k+ b_6(D_\gamma)_i (H(\overline 6))^{ij}_k(P_\alpha)_j^k(V_\beta)^l_l \nonumber\\ & + e_6(D_\gamma)_i (H(\overline 6))^{ij}_k(V_\beta)_j^k(P_\alpha)^l_l +c_6(D_\gamma)_i (H(\overline 6))^{jl}_k(P_\alpha)_j^i(V_\beta)_l^k +f_6(D_\gamma)_i (H(\overline 6))^{jl}_k(V_\beta)_j^i(P_\alpha)_l^k \nonumber\\
  & + a_{15}(D_\gamma)_i (H({15}))^{ij}_k(P_\alpha)_j^l(V_\beta)_l^k+ d_{15}(D_\gamma)_i (H({15}))^{ij}_k(V_\beta)_j^l(P_\alpha)_l^k+ b_{15}(D_\gamma)_i (H({15}))^{ij}_k(P_\alpha)_j^k(V_\beta)^l_l\nonumber\\
  & +e_{15}(D_\gamma)_i (H({15}))^{ij}_k(V_\beta)_j^k(P_\alpha)^l_l  + c_{15}(D_\gamma)_i (H({15}))^{jl}_k(P_\alpha)_j^i(V_\beta)_l^k + f_{15}(D_\gamma)_i (H({15}))^{jl}_k(V_\beta)_j^i(P_\alpha)_l^k\nonumber\\ &
+a_3^p (D_\gamma)_i (H(3_p))^i (P_\alpha)^k_j(V_\beta)^j_k +b_3^p (D_\gamma)_i (H(3_p))^i (P_\alpha)_k^k(V_\beta)_j^j+c_{3}^{p} (D_\gamma)_i (H(3_p))^k (P_\alpha)^i_k(V_\beta)_j^j\nonumber\\
  &+e_{3}^{p} (D_\gamma)_i (H(3_p))^k (V_\beta)^i_k(P_\alpha)_j^j+d_3^p (D_\gamma)_i (H(3_p))^k (P_\alpha)^i_j(V_\beta)^j_k +f_3^p (D_\gamma)_i (H(3_p))^k (V_\beta)^i_j(P_\alpha)^j_k
\nonumber\\  &+a_3^t (D_\gamma)_i (H(3_t))^i (P_\alpha)^k_j(V_\beta)^j_k +b_3^t (D_\gamma)_i (H(3_t))^i (P_\alpha)_k^k(V_\beta)_j^j+c_{3}^{t} (D_\gamma)_i (H(3_t))^k (P_\alpha)^i_k(V_\beta)_j^j
\nonumber\\  & +e_{3}^{t} (D_\gamma)_i (H(3_t))^k (V_\beta)^i_k(P_\alpha)_j^j +d_3^t (D_\gamma)_i (H(3_t))^k (P_\alpha)^i_j(V_\beta)^j_k \nonumber\\  & +f_3^t (D_\gamma)_i (H(3_t))^k (V_\beta)^i_j(P_\alpha)^j_k.
\end{align}
The relations between topological amplitudes and the irreducible amplitudes are
\begin{align}\label{sol2}
 a_6&  =\frac{E_V-A_V}{4},  \qquad b_6 = \frac{T^{ES}_P-T^{AS}_P}{4},  \qquad c_6 = \frac{-T_P+C_P}{4}, \qquad
 d_6  =\frac{E_P-A_P}{4}, \nonumber\\ \ e_6 & = \frac{T^{ES}_V-T^{AS}_V}{4},  \qquad f_6 = \frac{-T_V+C_V}{4}, \qquad
  a_{15}  =\frac{E_V+A_V}{8},  \qquad b_{15} = \frac{T^{ES}_P+T^{AS}_P}{8}, \nonumber\\ c_{15}& = \frac{T_P+C_P}{8},\qquad
  d_{15}  =\frac{E_P+A_P}{8},  \qquad e_{15} = \frac{T^{ES}_V+T^{AS}_V}{8},  \qquad f_{15} = \frac{T_V+C_V}{8},\nonumber\\
 a^t_3& = \frac{3}{8}E_P-\frac{1}{8}A_P+\frac{3}{8}E_V-\frac{1}{8}A_V+T^{LA},\qquad
 a^p_3 = -\frac{1}{8}E_P+\frac{3}{8}A_P-\frac{1}{8}E_V+\frac{3}{8}A_V+ T^{QA},\nonumber\\
  b^t_3 & = \frac{3}{8}T^{ES}_P-\frac{1}{8}T^{AS}_P+\frac{3}{8}T^{ES}_V-\frac{1}{8}T^{AS}_V+T^{LS},\qquad
 b^p_3 = -\frac{1}{8}T^{ES}_P+\frac{3}{8}T^{AS}_P-\frac{1}{8}T^{ES}_V+\frac{3}{8}T^{AS}_V + T^{QS},\nonumber\\
  c^t_3 & = -\frac{1}{8}T_P+\frac{3}{8}C_P-\frac{1}{8}T^{ES}_P + \frac{3}{8}T^{AS}_P+T^{LC}_P,\qquad
  c^p_3  = \frac{3}{8}T_P-\frac{1}{8}C_P+\frac{3}{8}T^{ES}_P -\frac{1}{8}T^{AS}_P+T^{QC}_P,\nonumber\\
   d^t_3 & = \frac{3}{8}T_P-\frac{1}{8}C_P-\frac{1}{8}E_P + \frac{3}{8}A_P+T^{LP}_P,\qquad
   d^p_3  = -\frac{1}{8}T_P+\frac{3}{8}C_P+\frac{3}{8}E_P - \frac{1}{8}A_P+T^{QP}_P,\nonumber\\
   e^t_3 & = -\frac{1}{8}T_V+\frac{3}{8}C_V-\frac{1}{8}T^{ES}_V + \frac{3}{8}T_V^{AS}+T^{LC}_V,\qquad
   e^p_3  = \frac{3}{8}T_V-\frac{1}{8}C_V+\frac{3}{8}T^{ES}_V - \frac{1}{8}T_V^{AS}+T^{QC}_V,\nonumber\\
   f^t_3 & = \frac{3}{8}T_V-\frac{1}{8}C_V-\frac{1}{8}E_V + \frac{3}{8}A_V+T^{LP}_V,\qquad
    f^p_3  = -\frac{1}{8}T_V+\frac{3}{8}C_V+\frac{3}{8}E_V - \frac{1}{8}A_V+T^{QP}_V.
\end{align}
Similar to the $D\to PP$ decay, the irreducible amplitude $c_6$ in the $D\to PV$ decay can be absorbed into $a_6$, $b_6$, $d_6$, $e_6$, $f_6$ with following redefinition:
\begin{align}
  a_6^\prime = a_6 - c_6,\qquad b_6^\prime = b_6 + c_6, \qquad d_6^\prime = d_6 - c_6, \qquad e_6^\prime = e_6 + c_6, \qquad f_6^\prime = f_6 - c_6.
\end{align}

Similarly to Eqs.~\eqref{c5} and \eqref{c11}, all the penguin-operator-induced amplitudes of the $D\to PV$ modes in the SM can be absorbed into six parameters in both IRA and TDA approaches with following redefinitions.\\
IRA:
\begin{align}\label{c12}
 & a_3  = a_3^t+Pa_3^t + 3 Pa_3^p, \qquad   b_3 =b_3^t+Pb_3^t + 3 Pb_3^p, \qquad   c_3 = c_3^t+Pc_3^t + 3 Pc_3^p, \nonumber\\ &   d_3 = d_3^t+Pd_3^t + 3 Pd_3^p,  \qquad   e_3 =e_3^t+Pe_3^t + 3 Pe_3^p, \qquad   f_3 = f_3^t+Pf_3^t + 3 Pf_3^p.
\end{align}
TDA:
\begin{align}\label{c13}
 & A^{LA}  = T^{LA}+PA_P+PA_V+P^{LA}+3P^{QA}, \quad   A^{LS}  = T^{LS}+P^{AS}_P+P^{AS}_V+P^{LS}+3P^{QS}, \nonumber\\ &  A^{LC}_P  = T^{LC}_P+PT_P+P^{ES}_P+P^{LC}_P+3P^{QC}_P, \quad  A^{LP}_P  = T^{LP}_P+PC_P+PE_P+P^{LP}_P+3P^{QP}_P,\nonumber\\ & A^{LC}_V  = T^{LC}_V+PT_V+P^{ES}_V+P^{LC}_V+3P^{QC}_V, \quad  A^{LP}_V  = T^{LP}_V+PC_V+PE_V+P^{LP}_V+3P^{QP}_V.
\end{align}
The tree- and penguin-operator-induced amplitudes of all the $D\to PV$ modes are listed in Tables.~\ref{tab:pv1}, \ref{tab:pv2} and \ref{tab:pv3}.

%%%%%%%%%%%%%%%%%%%%%%%%%%%%%%%
\section{$SU(3)$ decomposition in $b$ quark decay }\label{b}
In this appendix, we discuss the $SU(3)$ decomposition of the non-leptonic $b$ decays.
The explicit $SU(3)$  decomposition of operator $O_{ij}^k$ in the charm-less $B$ decay, similar to Eq.~\eqref{hd}, is
\begin{align}\label{xb}
O_{ij}^k  = \frac{1}{8}O(15)_{ij}^k +  \frac{1}{4}O(\overline 6)_{ij}^k+\delta^k_j\Big(\frac{3}{8}O(3)_i-\frac{1}{8}O(3^\prime)_i\Big)
+\delta^k_i\Big(\frac{3}{8}O(3^\prime)_j-\frac{1}{8}O(3)_j\Big),
\end{align}
in which $O(\overline 6)_{ij}^k = \epsilon_{ijl}O(\overline 6)^{kl}$.
To compare with literature, we use the convention that index $i$ of $O_{ij}^k$ presents quark $q_i$ produces in the effective vertex connecting with $b$ quark line, and indices $j$ and $k$ present quark $q_j$ and anti-quark $\overline q^k$ produce in the other effective vertex.
Notice that this convention is different from the convention in the charm decay.
All components of the $SU(3)$ irreducible presentation are listed following.\\
$ 3$ presentation:
\begin{align}\label{3p1}
  O( 3)_1 & = (\bar u b)(\bar u u) + (\bar u b)(\bar dd) + (\bar u b)(\bar ss),\qquad
 O(3)_2 = (\bar d b)(\bar u u) + (\bar d b)(\bar dd) + (\bar d b)(\bar ss),\nonumber \\
 O(3)_3 & = (\bar s b)(\bar u u) + (\bar s b)(\bar dd) + (\bar s b)(\bar ss).
\end{align}
$ 3^\prime$ presentation:
\begin{align}\label{3t1}
 O(3^\prime)_1 & = (\bar u b)(\bar u u) + (\bar d b)(\bar ud) + (\bar s b)(\bar us),\qquad
 O(3^\prime)_2 = (\bar u b)(\bar d u) + (\bar d b)(\bar dd) + (\bar s b)(\bar ds),\nonumber \\
O(3^\prime)_3 & = (\bar u b)(\bar s u) + (\bar d b)(\bar sd) + (\bar s b)(\bar ss).
\end{align}
$\overline 6$ presentation:
\begin{align}
    O(\overline 6)^{11} & = 2[(\bar d b)(\bar su)-(\bar s b)(\bar d u) ],\qquad O(\overline 6)^{22} = 2[(\bar s b)(\bar ud)-(\bar u b)(\bar s d)],\nonumber \\
     O(\overline 6)^{33}& = 2[ (\bar u b)(\bar ds)-(\bar d b)(\bar u s)],\nonumber \\
     O(\overline 6)^{12} & = -[(\bar u b)(\bar s u) - (\bar s b)(\bar uu) + (\bar s b)(\bar dd) - (\bar d b)(\bar sd)],\nonumber \\
    O(\overline 6)^{23}& = -[(\bar d b)(\bar u d) - (\bar u b)(\bar dd) + (\bar u b)(\bar ss) -(\bar s b)(\bar us)],\nonumber \\
    O(\overline 6)^{31} & =  -[(\bar s b)(\bar d s) - (\bar d b)(\bar ss) + (\bar d b)(\bar uu) -(\bar u b)(\bar du)].
\end{align}
$ {15}$ presentation:
\begin{align}
     O({15})^{1}_{11} & = 4((\bar u b)\bar u u)-2[(\bar s b)(\bar us) + (\bar db)(\bar ud)+(\bar u b)(\bar dd) + (\bar ub)(\bar ss)],\nonumber \\
     O({15})^{2}_{22} & = 4(\bar d b)(\bar dd)- 2[(\bar u b)(\bar du) + (\bar sb)(\bar ds)+(\bar d b)(\bar uu) + (\bar db)(\bar ss)],\nonumber \\
    O({15})^{3}_{33} & = 4(\bar s b)(\bar ss)-2[(\bar u b)(\bar su) + (\bar db)(\bar sd)+(\bar s b)(\bar uu) + (\bar sb)(\bar dd)],\nonumber \\
     O({15})^{1}_{32} & = 4[(\bar s b)(\bar d u) + (\bar d b)(\bar su)],\qquad         O({15})^{2}_{31} = 4[(\bar u b)(\bar s d) + (\bar s b)(\bar ud)],  \nonumber \\
    O({15})^{3}_{21} & = 4[(\bar d b)(\bar u s) + (\bar u b)(\bar ds)],\nonumber \\
  O({15})^{2}_{11} & = 8(\bar u b)(\bar u d), \qquad O({15})^{3}_{11}  = 8(\bar u b)(\bar u s), \qquad  O({15})^{1}_{22} = 8(\bar d b)(\bar du),\nonumber \\
     O({15})^{3}_{22} & = 8(\bar d b)(\bar d s), \qquad O({15})^{1}_{33} = 8(\bar s b)(\bar su), \qquad
     O({15})^{2}_{33}  = 8(\bar s b)(\bar s d),\nonumber \\
      O({15})^{1}_{12} & = 3[(\bar d b)(\bar u u)+(\bar u b)(\bar d u)]-2(\bar d b)(\bar d d) - [(\bar s b)(\bar d s)+(\bar d b)(\bar ss)],\nonumber \\
     O({15})^{2}_{21} & = 3[(\bar d b)(\bar u d)+(\bar u b)(\bar d d)]-2(\bar u b)(\bar u u) - [(\bar s b)(\bar u s)+(\bar u b)(\bar ss)],\nonumber \\
      O({15})^{1}_{13} & = 3[(\bar s b)(\bar u u)+(\bar u b)(\bar su)]-2(\bar s b)(\bar s s) - [(\bar d b)(\bar s d)+(\bar s b)(\bar dd)],\nonumber \\
     O({15})^{3}_{31} & = 3[(\bar s b)(\bar u s)+(\bar u b)(\bar ss)]-2(\bar u b)(\bar u u) - [(\bar d b)(\bar u d)+(\bar u b)(\bar dd)],\nonumber \\
     O({15})^{2}_{23} & = 3[(\bar s b)(\bar d d)+(\bar d b)(\bar sd)]-2(\bar s b)(\bar s s) - [(\bar u b)(\bar s u)+(\bar s b)(\bar uu)],\nonumber \\
    O({15})^{3}_{32} & = 3[(\bar s b)(\bar ds)+(\bar d b)(\bar s s)]-2(\bar db)(\bar d d) - [(\bar u b)(\bar d u)+(\bar d b)(\bar uu)].
\end{align}

There are two tree operators without $c$ quark contributing to the charm-less $B$ decay in the SM: $O^{(0)1}_{12}=(\bar u b)(\bar d u)$ for $\Delta S = 0$ transition, and $O_{13}^{(0)1}=(\bar u b)(\bar s u)$ for $\Delta S = -1$ transition. According to Eq.~\eqref{xb}, $O_{12}^{(0)1}$ is decomposed to be
\begin{align}\label{q2}
O_{12}^{(0)1}  &= \frac{3}{8}O^{(0)}(3^\prime)_2-\frac{1}{8}O^{(0)}(3)_2+  \frac{1}{4}O^{(0)}(\overline 6)_{12}^{1}+\frac{1}{8}O^{(0)}(15)_{12}^{1} \nonumber\\
&= \frac{1}{8}\Big\{3[(\bar u b)(\bar d u) + (\bar d b)(\bar dd) + (\bar s b)(\bar ds)]_{3^\prime} -[(\bar d b)(\bar u u) + (\bar d b)(\bar dd) + (\bar d b)(\bar ss)]_{3} \nonumber\\ & ~~~~-[(\bar s b)(\bar d s) - (\bar d b)(\bar ss) + (\bar d b)(\bar uu) -(\bar u b)(\bar du)]_{\overline 6} \nonumber\\ &~~~~+ \big[3[(\bar d b)(\bar u u)+(\bar u b)(\bar d u)]-2(\bar d b)(\bar d d) - [(\bar s b)(\bar d s)+(\bar d b)(\bar ss)]\big]_{15}\Big\}.
\end{align}
$O^{(0)1}_{13}$  is decomposed to be
\begin{align}\label{qx}
O_{13}^{(0)1}  &=  \frac{3}{8}O^{(0)}(3^\prime)_3-\frac{1}{8}O^{(0)}(3)_3+  \frac{1}{4}O^{(0)}(\overline 6)_{13}^1+\frac{1}{8}O^{(0)}(15)_{13}^1\nonumber\\
&= \frac{1}{8}\Big\{3[(\bar u b)(\bar s u) + (\bar d b)(\bar sd) + (\bar s b)(\bar ss)]_{3^\prime} -[(\bar s b)(\bar u u) + (\bar s b)(\bar dd) + (\bar s b)(\bar ss)]_{3} \nonumber\\ & ~~~~-[(\bar s b)(\bar uu) - (\bar u b)(\bar s u) + (\bar d b)(\bar sd) - (\bar s b)(\bar dd) ]_{\overline 6} \nonumber\\ &~~~~+ \big[3[(\bar s b)(\bar u u)+(\bar u b)(\bar su)]-2(\bar s b)(\bar s s) - [(\bar d b)(\bar s d)+(\bar s b)(\bar dd)]\big]_{15}\Big\}.
\end{align}
Eqs.~\eqref{q2} and \eqref{qx} are consistent with the results in Refs.~\cite{He:2000ys,Savage:1989ub}.
In Ref.~\cite{He:2018php}, the $SU(3)$ decomposition of $O_{ij}^k$ is given by
\begin{align}\label{qx1}
O_{ij}^k  = \frac{1}{8}O(15)_{ij}^k +  \frac{1}{4}O(\overline 6)_{ij}^k-\frac{1}{8}\delta^k_jO(3)_i+\frac{3}{8}\delta^k_iO(3^\prime)_j.
\end{align}
However, this general formula is not consistent with Eq.~\eqref{q2}. If we set $i=1$, $j=2$, $k=1$ in Eq.~\eqref{qx1}, the coefficient of $O(3)$  is zero but not $-1/8$ since $\delta^1_2 = 0$. Similarly, Eq.~\eqref{qx1} is not consistent with Eq.~\eqref{qx} either. Thereby, Eq.~\eqref{qx1} is incorrect. Because of the mistake in $SU(3)$ decomposition, the relation of topological amplitudes and $SU(3)$ irreducible amplitudes in \cite{He:2018php,He:2018joe} is incorrect.

The non-zero CKM components of the $SU(3)$ irreducible representation in the $b\to d$ transition are
\begin{align}
&(H^{(0)}(3^\prime))^2 = V_{ub}V^*_{ud}, \qquad (H^{(0)}(\overline 6))_{31} = V_{ub}V^*_{ud}, \qquad (H^{(0)}(15))^{12}_1 = 3V_{ub}V^*_{ud}, \nonumber\\& (H^{(0)}(15))^{22}_2 = -2V_{ub}V^*_{ud}, \qquad (H^{(0)}(15))^{32}_3 = -V_{ub}V^*_{ud}.
\end{align}
Notice that $(H^{(0)}(3))^2$ is zero because the operator $O^{(0)}(3)_2$ does not have $(\bar ub)(\bar du)$ constituent according to Eq.~\eqref{3p}, and hence $(H^{(0)}(3))^2 \neq (H^{(0)}(3^\prime))^2$.
Similarly, the non-zero CKM components of the $SU(3)$ irreducible representation in the $b\to s$ transition are
\begin{align}
&(H^{(0)}(3^\prime))^3 =  V_{ub}V^*_{us}, \qquad (H^{(0)}(\overline 6))_{12} = -V_{ub}V^*_{us}, \qquad (H^{(0)}(15))^{13}_1 = 3V_{ub}V^*_{us}, \nonumber\\&  (H^{(0)}(15))^{33}_3 = -2V_{ub}V^*_{us}, \qquad (H^{(0)}(15))^{32}_2 = -V_{ub}V^*_{us}.
\end{align}
And again, $(H^{(0)}(3))^3 = 0$ and $(H^{(0)}(3))^3\neq (H^{(0)}(3^\prime))^3$.

There are six penguin operators without $c$ quark contributing to the charm-less $B$ decay in the SM: $O^{(1)1}_{21}=(\bar d b)(\bar u u)$, $O^{(1)2}_{22}=(\bar d b)(\bar d d)$, $O^{(1)3}_{23}=(\bar d b)(\bar s s)$ in the $\Delta S = 0$ transition, and $O_{31}^{(1)1}=(\bar s b)(\bar u u)$, $O_{32}^{(1)2}=(\bar s b)(\bar d d)$, $O_{33}^{(1)3}=(\bar s b)(\bar s s)$ in the $\Delta S = -1$ transition.  The non-zero CKM components of the $SU(3)$ irreducible representation in the $\Delta S = 0$ transition are
\begin{align}
(H^{(1)}(3))^2 = -3V_{tb}V^*_{td}, \qquad (H^{(1)}(3^\prime))^2 = -V_{tb}V^*_{td}.
\end{align}
The non-zero CKM components in the $SU(3)$ irreducible representation in the $\Delta S = -1$ transition are
\begin{align}
(H^{(1)}(3))^3 = -3V_{tb}V^*_{ts}, \qquad (H^{(1)}(3^\prime))^3 = -V_{tb}V^*_{ts}.
\end{align}

The operator $O_{ij}^k$ is not enough to describe the charm-less $B$ decay because $O_{ij}^k$ cannot include $c$-quark loop.
To give a complete description of the charm-less $B$ decay, $O_{ci}^c$ and $O_{ic}^c$ should be included.
Operators $O_{ci}^c$ and $O_{ic}^c$ are the $SU(3)$ irreducible representations themselves, $O(3^{\prime\prime})_i = O_{ci}^c$ and $O(3^{\prime\prime\prime})_i = O_{ic}^c$. There are two tree operators with $c$ quark contributing to the charm-less $B$ decay in the SM: $O_{c2}^{(0)c} = (\bar cb)(\overline dc)$ for the $\Delta S = 0$ transition and $O_{c3}^{(0)c} = (\bar cb)(\overline sc)$ for the $\Delta S = -1$ transition. The non-zero CKM components induced by $O_{c2}^{(0)c}$ and $O_{c3}^{(0)c}$ in the $SU(3)$ irreducible representation are
\begin{align}
(H^{(0)}(3^{\prime\prime}))^2 = V_{cb}V^*_{cd}, \qquad (H^{(0)}(3^{\prime\prime}))^3 = V_{cb}V^*_{cs}.
\end{align}
There are two penguin operators with $c$ quark contributing to the charm-less $B$ decay in the SM: $O_{2c}^{(1)c} = (\bar db)(\overline cc)$ for the $\Delta S = 0$ transition and $O_{3c}^{(1)c} = (\bar sb)(\overline cc)$ for the $\Delta S = -1$ transition.  The non-zero CKM components induced by $O_{2c}^{(1)c}$ and $O_{3c}^{(1)c}$ in the $SU(3)$ irreducible representation are
\begin{align}
(H^{(1)}(3^{\prime\prime\prime}))^2 = -V_{tb}V^*_{td}, \qquad (H^{(1)}(3^{\prime\prime\prime}))^3 = -V_{tb}V^*_{ts}.
\end{align}

%%%%%%%%%%%%%%%%%%%%%%%%%%%%%%%
\section{Topologies in charm-less $\overline B\to PV$ decays}\label{b-pv}

\begin{table}[t!]
\caption{Decay amplitudes of the $\overline B_{u,d}\to PV$ decays induced by the $b\to d$ transition.}\label{tab:bpvd1}
\begin{ruledtabular}
\scriptsize
\begin{tabular}{|c|c|c|}
{Channel} & {TDA}& {IRA}   \\\hline
$  B^-\to \pi^-\rho^0$  &  $\frac{1}{\sqrt{2}}\lambda_u(-A_P+A_V+C_P+T_V)
+\frac{1}{\sqrt{2}}\lambda_t(A_P^{LP}-A_V^{LP})$ & $\frac{1}{\sqrt{2}}(-a_6+3 a_{15}+c_6+5 c_{15}+d_6-3 d_{15}$\\&& $-f_6+3 f_{15})\lambda _u+\frac{1}{\sqrt{2}}(d_3-f_3)\lambda _t$\\\hline
$  B^-\to \pi^0\rho^-$  &  $\frac{1}{\sqrt{2}}\lambda_u(A_P-A_V+C_V+T_P)
-\frac{1}{\sqrt{2}}\lambda_t(A_P^{LP}-A_V^{LP})$ &$\frac{1}{\sqrt{2}}(a_6-3 a_{15}-c_6+3 c_{15}-d_6+3 d_{15}$ \\ &&$+f_6+5 f_{15})\lambda _u+\frac{1}{\sqrt{2}}(-d_3+f_3)\lambda _t$ \\\hline
$  B^-\to \eta_8\rho^-$  &  $\frac{1}{\sqrt{6}}\lambda_u(A_P+A_V+C_V+T_P)
-\frac{1}{\sqrt{6}}\lambda_t(A_P^{LP}+A_V^{LP})$ &$-\frac{1}{\sqrt{6}}(a_6-3 a_{15}+c_6-3 c_{15}+d_6$ \\ &&$-3 (d_{15}+f_6+f_{15})) \lambda _u-\frac{1}{\sqrt{6}}(d_3+f_3) \lambda _t$\\\hline
$  B^-\to \eta_1\rho^-$  &  $\frac{1}{\sqrt{3}}\lambda_u(A_P+A_V+C_V+T_P+3 T_V^{AS})$ &$-\frac{1}{\sqrt{3}}(a_6-3 a_{15}+c_6-3 c_{15}+d_6-3 d_{15}$\\ &$-\frac{1}{\sqrt{3}}\lambda_t(A_P^{LP}+A_V^{LP}+3A_V^{LC})$& $+3 e_6-9 e_{15}) \lambda _u-\frac{1}{\sqrt{3}}(d_3+3 e_3+f_3) \lambda _t$ \\\hline
$  B^-\to K^-K^{*0}$  &  $\lambda _uA_P -\lambda _tA_P^{LP} $ & $-d_3 \lambda _t+(c_6-c_{15}-d_6+3 d_{15}) \lambda _u$ \\\hline
$  B^-\to K^0K^{*-}$  &  $\lambda _uA_V -\lambda _tA_V^{LP}$ & $-f_3 \lambda _t-(a_6-3 a_{15}-f_6+f_{15}) \lambda _u$ \\\hline
$  B^-\to \pi^-\omega_8$  &  $\frac{1}{\sqrt{6}}\lambda_u(A_P+A_V+C_P+T_V)-\frac{1}{\sqrt{6}}\lambda_t(A_P^{LP}+A_V^{LP})$ & $-\frac{1}{\sqrt{6}}(a_6-3 a_{15}-3 c_6-3 c_{15}+d_6-3 d_{15}$\\ &&$+f_6-3 f_{15}) \lambda _u)-\frac{1}{\sqrt{6}}(d_3+f_3) \lambda _t$ \\ \hline
$  B^-\to \pi^-\omega_1$  &  $\frac{1}{\sqrt{3}}\lambda_u(A_P+A_V+C_P+3 T_P^{AS}+T_V)$ & $-\frac{1}{\sqrt{3}}(a_6-3 a_{15}+3 b_6-9 b_{15}+d_6-3 d_{15}$\\ & $-\frac{1}{\sqrt{3}}\lambda_t(A_P^{LP}+A_V^{LP}+3A_P^{LC})$ & $+f_6-3 f_{15})\lambda _u-\frac{1}{\sqrt{3}}(3 c_3+d_3+f_3) \lambda _t$ \\\hline
$  \overline B^0\to \pi^-\rho^+$  & $\lambda _u(E_P+T_V) -\lambda _t(A^{LA}+A_V^{LP}) $ & $(-2 a_{15}+d_6+3 d_{15}-f_6+3 f_{15}) \lambda _u-(a_3+f_3) \lambda _t~~$\\\hline
$  \overline B^0\to \pi^0\rho^0$  &  $\frac{1}{2}\lambda _u(-C_P-C_V+E_P+E_V)$ & $\frac{1}{2}(a_6+a_{15}-c_6-5 c_{15}+d_6+d_{15}$\\ & $-\frac{1}{2}\lambda _t(2 A^{LA}+A_P^{LP}+A_V^{LP})$ & $-f_6-5 f_{15}) \lambda _u-\frac{1}{2}(2 a_3+d_3+f_3) \lambda _t$ \\\hline
$  \overline B^0\to \eta_8\rho^0$  &  $\frac{1}{2\sqrt{3}}\lambda _u(C_P-C_V+E_P+E_V)+\frac{1}{2\sqrt{3}}\lambda _t(A_P^{LP}+A_V^{LP}) $ & $\frac{1}{2 \sqrt{3}}(a_6+5 a_{15}+c_6+5 c_{15}+d_6+5 d_{15}$ \\ & &$-3 (f_6+f_{15})) \lambda _u+\frac{1}{2\sqrt{3}}(d_3+f_3) \lambda _t$ \\\hline
$  \overline B^0\to \eta_1\rho^0$  &  $\frac{1}{\sqrt{6}}\lambda _u(C_P-C_V+E_P+E_V+3 T_V^{ES})$ & $\frac{1}{\sqrt{6}}(a_6+5 a_{15}+c_6+5 c_{15}+d_6+5 d_{15}$\\ & $+\frac{1}{\sqrt{6}}\lambda _t(A_P^{LP}+A_V^{LP}+3 A_V^{LC}) $ & $+3 e_6+15 e_{15}) \lambda _u+\frac{1}{\sqrt{6}}(d_3+3 e_3+f_3) \lambda _t$ \\\hline
$  \overline B^0\to \pi^+\rho^-$  &  $\lambda _u(E_V+T_P)-\lambda _t(A^{LA}+A_P^{LP}) $ & $-(a_3+d_3) \lambda _t+(a_6+3 a_{15}-c_6+3 c_{15}-2 d_{15}) \lambda _u$ \\\hline
$  \overline B^0\to K^-K^{*+}$  &
$\lambda _uE_P-\lambda _tA^{LA} $ & $-a_3 \lambda _t-(a_6+a_{15}-d_6-3 d_{15}) \lambda _u$\\\hline
$  \overline B^0\to \overline K^0K^{*0}$  &
$- \lambda _t(A^{LA}+A_P^{LP})$ & $-(a_3+d_3) \lambda _t-(a_6+a_{15}-c_6+c_{15}+2 d_{15}) \lambda _u$\\\hline
$  \overline B^0\to K^0\overline K^{*0}$  &
$- \lambda _t(A^{LA}+A_V^{LP})$ & $-(a_3+f_3) \lambda _t-(2 a_{15}+d_6+d_{15}-f_6+f_{15}) \lambda _u$ \\\hline
$  \overline B^0\to K^+K^{*-}$  &
$\lambda _uE_V-\lambda _tA^{LA} $ &$-a_3 \lambda _t+(a_6+3 a_{15}-d_6-d_{15}) \lambda _u$ \\\hline
$  \overline B^0\to \pi^0\omega_8$  &
$\frac{1}{2\sqrt{3}}\lambda_u(-C_P+C_V+E_P+E_V)
+\frac{1}{2\sqrt{3}}\lambda_t(A_P^{LP}+A_V^{LP})$ & $\frac{1}{2\sqrt{3}}(a_6+5 a_{15}-3 c_6-3 c_{15}+d_6+5 d_{15}$ \\ &&$+f_6+5 f_{15}) \lambda _u +\frac{1}{2\sqrt{3}}(d_3+f_3) \lambda _t$ \\\hline
$  \overline B^0\to \eta_8\omega_8$  &
$\frac{1}{6}\lambda_u(C_P+C_V+E_P+E_V)
$ &$\frac{1}{2} (-a_6-a_{15}+c_6+c_{15}-d_6-d_{15}+f_6+f_{15}) \lambda _u$ \\ &$-\frac{1}{6}\lambda_t(6 A^{LA}+A_P^{LP}+A_V^{LP})$&$-\frac{1}{6} (6 a_3+d_3+f_3) \lambda _t$ \\\hline
$  \overline B^0\to \eta_1\omega_8$  &
$\frac{1}{3\sqrt{2}}\lambda_u(C_P+C_V+E_P+E_V+3 T_V^{ES})
$ & $\frac{1}{\sqrt{2}}(a_6+a_{15}+c_6+c_{15}+d_6+d_{15}$\\ & $-\frac{1}{3\sqrt{2}}\lambda_t(A_P^{LP}+3 A_V^{LC}+A_V^{LP})$& $+3 (e_6+e_{15})) \lambda _u-\frac{1}{3\sqrt{2}}(d_3+3 e_3+f_3) \lambda _t$\\\hline
$  \overline B^0\to \pi^0\omega_1$  &
$\frac{1}{\sqrt{6}}\lambda_u(-C_P+C_V+E_P+E_V+3 T_P^{ES})$ & $\frac{1}{\sqrt{6}}(a_6+5 a_{15}+3 b_6+15 b_{15}+d_6+5 d_{15}$ \\
& $+\frac{1}{\sqrt{6}}\lambda_t(3 A_P^{LC}+A_P^{LP}+A_V^{LP})$ & $+f_6+5 f_{15}) \lambda _u+\frac{1}{\sqrt{6}}(3 c_3+d_3+f_3) \lambda _t$ \\\hline
$  \overline B^0\to \eta_8\omega_1$  &
$\frac{1}{3\sqrt{2}}\lambda_u(C_P+C_V+E_P+E_V+3 T_P^{ES})$ & $\frac{1}{\sqrt{2}}(a_6+a_{15}+3 b_6+3 b_{15}+d_6+d_{15}$\\ &
$-\frac{1}{3\sqrt{2}}\lambda_t(3 A_P^{LC}+A_P^{LP}+A_V^{LP})$ & $+f_6+f_{15}) \lambda _u-\frac{1}{3\sqrt{2}}(3 c_3+d_3+f_3) \lambda _t$ \\\hline
$  \overline B^0\to \eta_1\omega_1$  &
$\frac{1}{3}\lambda_u(C_P+C_V+E_P+E_V+3(T_P^{ES}+T_V^{ES}))$&$-\frac{1}{3} (3 a_3+9 b_3+3 c_3+d_3+3 e_3+f_3) \lambda _t$ \\
&$-\frac{1}{3}\lambda_t(3 A_P^{LC}+A_P^{LP}+A_V^{LP}+3(A^{LA}+3 A^{LS}+A_V^{LC}))$ & \\
\end{tabular}
\end{ruledtabular}
\end{table}

\begin{table}[t!]
\caption{Decay amplitudes for the $\overline B_{s}\to PV$ decays induced by the $b\to d$ transition.}\label{tab:bpvd2}
\begin{ruledtabular}
\scriptsize
\begin{tabular}{|c|c|c|}
{Channel} & {TDA}& {IRA}   \\\hline
$  \overline B^0_s\to K^0\rho^0$  &
$\frac{1}{\sqrt{2}}(\lambda_uC_P+\lambda_tA_P^{LP})$ & $\frac{1}{\sqrt{2}}(d_3 \lambda _t+(c_6+5 c_{15}-d_6+d_{15}) \lambda _u)$ \\\hline
$  \overline B^0_s\to K^+\rho^-$  &
$\lambda _uT_P -\lambda _tA_P^{LP} $ &$-d_3 \lambda _t-(c_6-3 c_{15}-d_6+d_{15}) \lambda _u$ \\\hline
$  \overline B^0_s\to \pi^-K^{*+}$  &
$\lambda _uT_V -\lambda _tA_V^{LP} $ &$-f_3 \lambda _t+(a_6-a_{15}-f_6+3 f_{15}) \lambda _u$ \\\hline
$  \overline B^0_s\to \pi^0K^{*0}$  &
$\frac{1}{\sqrt{2}}(\lambda_uC_V+\lambda_tA_V^{LP})$ &$\frac{1}{\sqrt{2}}(f_3 \lambda _t+(-a_6+a_{15}+f_6+5 f_{15}) \lambda _u)$ \\\hline
$  \overline B^0_s\to \eta_8K^{*0}$  &
$\frac{1}{\sqrt{6}}\lambda_uC_V+\frac{1}{\sqrt{6}}\lambda_t(2 A_P^{LP}-A_V^{LP})$ & $\frac{1}{\sqrt{6}}(a_6-a_{15}-2 c_6+2 c_{15}-2 d_6+2 d_{15}+3(f_6+f_{15}))\lambda _u~~~$\\&&$+\frac{1}{\sqrt{6}}(2 d_3-f_3) \lambda _t$\\\hline
$  \overline B^0_s\to \eta_1K^{*0}$  &
$\frac{1}{\sqrt{3}}\lambda_uC_V-\frac{1}{\sqrt{3}}\lambda_t(A_P^{LP}+3 A_V^{LC}+A_V^{LP})$~~~~ & $\frac{1}{\sqrt{3}}(a_6-a_{15}+c_6-c_{15}+d_6-d_{15}+3 e_6-3 e_{15}) \lambda _u$\\ &&$-\frac{1}{\sqrt{3}}(d_3+3 e_3+f_3) \lambda _t$\\\hline
$  \overline B^0_s\to K^0\omega_8$  &
$\frac{1}{\sqrt{6}}\lambda_uC_P-\frac{1}{\sqrt{6}}\lambda_t(A_P^{LP}-2 A_V^{LP})$ &$\frac{1}{\sqrt{6}}(-2 a_6+2 a_{15}+3 c_6+3 c_{15}+d_6-d_{15}-2 f_6+2 f_{15}) \lambda _u~~~$ \\ &&$-\frac{1}{\sqrt{6}}(d_3-2 f_3) \lambda _t$\\\hline
$  \overline B^0_s\to K^0\omega_1$  &
$\frac{1}{\sqrt{3}}\lambda_uC_P
-\frac{1}{\sqrt{3}}\lambda_t(3 A_P^{LC}+A_P^{LP}+A_V^{LP})$ & $\frac{1}{\sqrt{3}}(a_6-a_{15}+3 b_6-3 b_{15}+d_6-d_{15}+f_6-f_{15}) \lambda _u$\\ &&$-\frac{1}{\sqrt{3}}(3 c_3+d_3+f_3) \lambda _t$\\
\end{tabular}
\end{ruledtabular}
\end{table}
\begin{table}[t!]
\caption{Decay amplitudes of the $\overline B_{u,d}\to PV$ decays induced by the $b \to s$ transition.}\label{tab:bpvs1}
\begin{ruledtabular}
\scriptsize
\begin{tabular}{|c|c|c|}
{Channel} & {TDA}& {IRA}   \\\hline
$B^-\to K^- \rho^0  $  &  $\frac{1}{\sqrt{2}}\lambda^\prime_u(A_V+C_P+T_V)
-\frac{1}{\sqrt{2}}\lambda^\prime_tA_V^{LP}$ & $\frac{1}{\sqrt{2}}(-a_6+3 a_{15}+2 c_6+4 c_{15}-f_6+3 f_{15})\lambda _u'-\frac{1}{\sqrt{2}}f_3 \lambda _t'$\\\hline
$B^-\to \overline K^0 \rho^-  $  &  $\lambda^\prime_uA_V
-\lambda^\prime_tA_V^{LP}$ &$(-a_6+3 a_{15}+f_6-f_{15}) \lambda _u'-f_3 \lambda _t'$\\\hline
$B^-\to \pi^- K^{*0} $  &  $\lambda^\prime_uA_P
-\lambda^\prime_tA_P^{LP}$ &$(c_6-c_{15}-d_6+3 d_{15}) \lambda _u'-d_3 \lambda _t'$\\\hline
$B^-\to \pi^0 K^{*-} $  &  $\frac{1}{\sqrt{2}}\lambda^\prime_u(A_P+C_V+T_P)
-\frac{1}{\sqrt{2}}\lambda^\prime_tA_P^{LP}$ &$\frac{1}{\sqrt{2}}(-c_6+3 c_{15}-d_6+3 d_{15}+2 f_6+4 f_{15}) \lambda _u'-\frac{1}{\sqrt{2}}d_3 \lambda _t'$~~~~~~\\\hline
$B^-\to \eta_8 K^{*-} $  &  $\frac{1}{\sqrt{6}}\lambda^\prime_u(A_P-2 A_V+C_V+T_P)$ & $\frac{1}{\sqrt{6}}(2 a_6-6 a_{15}-c_6+3 c_{15}-d_6$\\ &$-\frac{1}{\sqrt{6}}\lambda^\prime_t(A_P^{LP}-2A_V^{LP})$& $+3 d_{15}+6 f_{15})\lambda _u'-\frac{1}{\sqrt{6}}(d_3-2 f_3) \lambda _t'$\\\hline
$B^-\to \eta_1 K^{*-} $  &  $\frac{1}{\sqrt{3}}\lambda^\prime_u(A_P+A_V+C_V+T_P+3 T_V^{AS})
$ &$\frac{1}{\sqrt{3}}(-a_6+3 a_{15}-c_6+3 c_{15}-d_6$\\ &$-\frac{1}{\sqrt{3}}\lambda^\prime_t(A_P^{LP}+A_V^{LP}+3A_V^{LC})$&$+3 (d_{15}-e_6+3 e_{15}))\lambda _u'-\frac{1}{\sqrt{3}}(d_3+3 e_3+f_3) \lambda _t'$\\\hline
$B^-\to K^{-}\omega_8 $  &  $\frac{1}{\sqrt{6}}\lambda^\prime_u(-2A_P+A_V+C_P+T_V)$ & $-\frac{1}{\sqrt{6}}(a_6-3 a_{15}-6 c_{15}-2d_6+6 d_{15}$\\ &$+\frac{1}{\sqrt{6}}\lambda^\prime_t(2 A_P^{LP}-A_V^{LP})$&$+f_6-3 f_{15})\lambda _u'+\frac{1}{\sqrt{6}}(2 d_3-f_3) \lambda _t'$\\\hline
$B^-\to K^{-}\omega_1 $  &  $\frac{1}{\sqrt{3}}\lambda^\prime_u(A_P+A_V+C_P+T_V+3T_P^{AS})
$ & $\frac{1}{\sqrt{3}}(-a_6+3 a_{15}-3 b_6+9 b_{15}-d_6+3 d_{15}$\\ & $-\frac{1}{\sqrt{3}}\lambda^\prime_t(A_P^{LP}+A_V^{LP}+3A_P^{LC})$&$-f_6+3 f_{15})\lambda _u'-\frac{1}{\sqrt{3}}(3 c_3+d_3+f_3) \lambda _t'$\\\hline
$\overline B^0\to K^{-}\rho^+ $  &  $\lambda^\prime_uT_V
-\lambda^\prime_tA_V^{LP}$ &$-f_3 \lambda _t'+(a_6-a_{15}-f_6+3 f_{15}) \lambda _u'$\\\hline
$\overline B^0\to \overline K^{0}\rho^0 $  &  $\frac{1}{\sqrt{2}}\lambda^\prime_uC_P
+\frac{1}{\sqrt{2}}\lambda^\prime_tA_V^{LP}$ &$\frac{1}{\sqrt{2}}(-a_6+a_{15}+2 c_6+4 c_{15}-f_6+f_{15}) \lambda _u'+\frac{1}{\sqrt{2}}f_3 \lambda _t'$\\\hline
$\overline B^0\to \pi^{0}\overline K^{*0} $  &  $\frac{1}{\sqrt{2}}\lambda^\prime_uC_V
+\frac{1}{\sqrt{2}}\lambda^\prime_tA_P^{LP}$ &$\frac{1}{\sqrt{2}}(-c_6+c_{15}-d_6+d_{15}+2 f_6+4 f_{15}) \lambda _u'+\frac{1}{\sqrt{2}}d_3 \lambda _t'$\\\hline
$\overline B^0\to \eta_8\overline K^{*0} $  &  $\frac{1}{\sqrt{6}}\lambda^\prime_uC_V
-\frac{1}{\sqrt{6}}\lambda^\prime_t(A_P^{LP}-2A_V^{LP})$ &$-\frac{1}{\sqrt{6}}(2 a_6-2 a_{15}-c_6+c_{15}-d_6$\\&&$+d_{15}-6 f_{15})\lambda _u'-\frac{1}{\sqrt{6}}(d_3-2 f_3) \lambda _t'$\\\hline
$\overline B^0\to \eta_1\overline K^{*0} $  &  $\frac{1}{\sqrt{3}}\lambda^\prime_uC_V
-\frac{1}{\sqrt{3}}\lambda^\prime_t(A_P^{LP}+A_V^{LP}+3A_V^{LC})$ &$-\frac{1}{\sqrt{3}}(-a_6+a_{15}-c_6+c_{15}-d_6+d_{15}$\\&&$+3 (-e_6+e_{15}))\lambda _u'-\frac{1}{\sqrt{3}}(d_3+3 e_3+f_3) \lambda _t'$\\\hline
$\overline B^0\to \pi^+K^{*-} $  &  $\lambda _u'T_P-\lambda _t'A_P^{LP}$ &$(-c_6+3 c_{15}+d_6-d_{15}) \lambda _u'-d_3 \lambda _t'$\\\hline
$\overline B^0\to \overline K^0\omega_8$  &  $\frac{1}{\sqrt{6}}\lambda^\prime_uC_P+\frac{1}{\sqrt{6}}\lambda _t'(2A_P^{LP}-A_V^{LP})$ &$-\frac{1}{\sqrt{6}}(-a_6+a_{15}-6 c_{15}+2 d_6-2 d_{15}$\\&&$-f_6+f_{15})\lambda _u'+\frac{1}{\sqrt{6}}(2 d_3-f_3) \lambda _t'$\\\hline
$\overline B^0\to \overline K^0\omega_1$  &  $\frac{1}{\sqrt{3}}\lambda^\prime_uC_P-\frac{1}{\sqrt{3}}\lambda _t'(3 A_P^{LC}+A_P^{LP}+A_V^{LP})$~~~~~~ &$-\frac{1}{\sqrt{3}}(-a_6+a_{15}-3 b_6+3 b_{15}-d_6+d_{15}$\\ &&$-f_6+f_{15})\lambda _u'-\frac{1}{\sqrt{3}}(3 c_3+d_3+f_3) \lambda _t'$\\
\end{tabular}
\end{ruledtabular}
\end{table}
\begin{table}[t!]
\caption{Decay amplitudes of the $\overline B_s\to PV$ decays induced by the $b\to s$ transition.}\label{tab:bpvs2}
\begin{ruledtabular}
\scriptsize
\begin{tabular}{|c|c|c|}
{Channel} & {TDA}& {IRA}   \\\hline
$\overline B^0_s\to  \pi^-\rho^+$  &  $\lambda _u'E_P-\lambda _t'A^{LA} $ &$(-a_6-a_{15}+d_6+3 d_{15}) \lambda _u'-a_3 \lambda _t'$\\\hline
$\overline B^0_s\to  \pi^0\rho^0$  &  $\frac{1}{2}\lambda _u'(E_P+E_V)-\lambda _t'A^{LA} $ &$(a_{15}+d_{15}) \lambda _u'-a_3 \lambda _t'$\\\hline
$\overline B^0_s\to  \eta_8\rho^0$  &  $\frac{1}{2\sqrt{3}}\lambda _u'(-2 C_P+E_P+E_V) $ &$-\frac{1}{\sqrt{3}}(-a_6-2 a_{15}+2 c_6+4 c_{15}-d_6-2 d_{15}) \lambda _u'$\\\hline
$\overline B^0_s\to  \eta_1\rho^0$  &  $\frac{1}{\sqrt{6}}\lambda _u'(C_P+E_P+E_V+3 T_V^{ES}) $ &$\sqrt{\frac{2}{3}} \lambda _u' (a_6+2 a_{15}+c_6+2 c_{15}+d_6$\\&&$+2 d_{15}+3 e_6+6 e_{15})$\\\hline
$\overline B^0_s\to  \pi^+\rho^-$  &  $\lambda _u'E_V-\lambda _t'A^{LA} $ &$(a_6+3 a_{15}-d_6-d_{15}) \lambda _u'-a_3 \lambda _t'$\\\hline
$\overline B^0_s\to  K^-K^{*+}$  &  $\lambda _u'(E_P+T_V)-\lambda _t'(A^{LA}+A_V^{LP}) $ &$(-2 a_{15}+d_6+3 d_{15}-f_6+3 f_{15}) \lambda _u'-(a_3+f_3) \lambda _t'$\\\hline
$\overline B^0_s\to  \overline K^0K^{*0}$  &  $-\lambda _t'(A^{LA}+A_V^{LP})  $ &$-(2 a_{15}+d_6+d_{15}-f_6+f_{15}) \lambda _u'-(a_3+f_3) \lambda _t'$\\\hline
$\overline B^0_s\to   K^0\overline K^{*0}$  &  $-\lambda _t'(A^{LA}+A_P^{LP})  $ &$(-a_6-a_{15}+c_6-c_{15}-2 d_{15}) \lambda _u'-(a_3+d_3) \lambda _t'$\\\hline
$\overline B^0_s\to  K^+K^{*-}$  &  $\lambda _u'(E_V+T_P)-\lambda _t'(A^{LA}+A_P^{LP}) $ &$(a_6+3 a_{15}-c_6+3 c_{15}-2 d_{15}) \lambda _u'-(a_3+d_3) \lambda _t'$~~\\\hline
$\overline B^0_s\to  \pi^0\omega_8$  &  $\frac{1}{2\sqrt{3}}\lambda _u'(-2 C_V+E_P+E_V)$ &$-\frac{1}{\sqrt{3}}(-a_6-2 a_{15}-2 (d_{15}-f_6-2 f_{15})-d_6) \lambda _u'$\\\hline
$\overline B^0_s\to  \eta_8\omega_8$  &  $\frac{1}{6}\lambda _u'(-2 C_P-2 C_V+E_P+E_V)$ & $- (a_{15}+2 c_{15}+d_{15}+2 f_{15})\lambda _u'$\\ &$-\frac{1}{3}\lambda _t' (3 A^{LA}+2 A_P^{LP}+2 A_V^{LP})$&$-\frac{1}{3} (3 a_3+2 (d_3+f_3)) \lambda _t'$\\\hline
$\overline B^0_s\to  \eta_1\omega_8$  &  $\frac{1}{3\sqrt{2}}\lambda _u'(C_P-2 C_V+E_P+E_V+3 T_V^{ES})$ &$\frac{\sqrt{2}}{3} (3 (a_{15}+c_{15}+d_{15}+3 e_{15})\lambda _u'$\\ &$+\frac{\sqrt{2}}{3}\lambda _t'(A_P^{LP}+3 A_V^{LC}+A_V^{LP})$&$+(d_3+3 e_3+f_3) \lambda _t')$\\\hline
$\overline B^0_s\to  \pi^0\omega_1$  &  $\frac{1}{\sqrt{6}}\lambda _u'(C_V+E_P+E_V+3 T_P^{ES})$ &$\sqrt{\frac{2}{3}} \lambda _u' (a_6+2 a_{15}+3 b_6+6 b_{15}+d_6+2 d_{15}$\\&&$+f_6+2 f_{15})$\\\hline
$\overline B^0_s\to  \eta_8\omega_1$  &  $\frac{1}{3\sqrt{2}}\lambda _u'(-2C_P+C_V+E_P+E_V+3 T_P^{ES})$ &$\frac{\sqrt{2}}{3}  (3  (a_{15}+3 b_{15}+d_{15}+f_{15})\lambda _u'$\\ &$+\frac{\sqrt{2}}{3}\lambda _t'(3 A_P^{LC}+A_P^{LP}+A_V^{LP})$&$+(3 c_3+d_3+f_3) \lambda _t')$ \\\hline
$\overline B^0_s\to  \eta_1\omega_1$  &  $\frac{1}{3}\lambda _u'(C_P+C_V+E_P+E_V+3 (T_P^{ES}+T_V^{ES}))$ &$-\frac{1}{3} (3 a_3+9 b_3+3 c_3+d_3+3 e_3+f_3)\lambda _t'$ \\ & $-\frac{1}{3}\lambda _t'(3 A_P^{LC}+A_P^{LP}+A_V^{LP}+3 (A^{LA}+3 A^{LS}+A_V^{LC}))$~~& \\
\end{tabular}
\end{ruledtabular}
\end{table}

The topological amplitude of $\overline B\to PV$ decay can be written as
\begin{align}\label{bhv}
{\cal A}^{\rm TDA}_{\overline B_\gamma \to P_\alpha V_\beta}& = T_P  (\overline B_\gamma)_i (H)^{jl}_k (P_\alpha)^{i}_j  (V_\beta)^k_l +T_V  (\overline B_\gamma)_i (H)^{jl}_k (V_\beta)^{i}_j  (P_\alpha)^k_l + C_P (\overline B_\gamma)_i(H)^{lj}_k  (P_\alpha)^{i}_j  (V_\beta)^k_l   \nonumber\\& + C_V (\overline B_\gamma)_i (H)^{lj}_k(V_\beta)^{i}_j   (P_\alpha)^k_l + E_P  (\overline B_\gamma)_i  (H)^{li}_j (P_\alpha)^j_k (V_\beta)^{k}_l + E_V  (\overline B_\gamma)_i (H)^{li}_j (V_\beta)^j_k (P_\alpha)^{k}_l   \nonumber\\& + A_P  (\overline B_\gamma)_i (H)^{il}_j   (P_\alpha)^j_k (V_\beta)^{k}_l + A_V  (\overline B_\gamma)_i (H)^{il}_j   (V_\beta)^j_k (P_\alpha)^{k}_l+T^{ES}_P (\overline B_\gamma)_i   (H)^{ji}_{l}   (P_\alpha)^{l}_j    (V_\beta)_k^k  \nonumber\\& +T^{ES}_V (\overline B_\gamma)_i   (H)^{ji}_{l} (V_\beta)^{l}_j (P_\alpha)_k^k  +T^{AS}_P (\overline B_\gamma)_i  (H)^{ij}_{l}  (P_\alpha)^{l}_j  (V_\beta)_k^k +T^{AS}_V (\overline B_\gamma)_i (H)^{ij}_{l} (V_\beta)^{l}_j  (P_\alpha)_k^k \nonumber\\&+T^{LP}_P (\overline B_\gamma)_i (H)^{lk}_{l} (P_\alpha)^{i}_j  (V_\beta)^j_k  +T^{LP}_V (\overline B_\gamma)_i (H)^{lk}_{l} (V_\beta)^{i}_j   (P_\alpha)^j_k  + T^{LC}_P (\overline B_\gamma)_i  (H)^{lj}_{l} (P_\alpha)^{i}_j  (V_\beta)^k_k\nonumber\\
&+ T^{LC}_V (\overline B_\gamma)_i (H)^{lj}_{l} (V_\beta)^{i}_j  (P_\alpha)^k_k+T^{QP}_P (\overline B_\gamma)_i  (H)^{kl}_{l}(P_\alpha)^{i}_j   (V_\beta)^j_k +T^{QP}_V (\overline B_\gamma)_i (H)^{kl}_{l} (V_\beta)^{i}_j   (P_\alpha)^j_k   \nonumber\\
&+ T^{QC}_P (\overline B_\gamma)_i (H)^{jl}_{l} (P_\alpha)^{i}_j  (V_\beta)^k_k + T^{QC}_V (\overline B_\gamma)_i (H)^{jl}_{l}  (V_\beta)^{i}_j   (P_\alpha)^k_k + T^{LA} (\overline B_\gamma)_i  (H)^{li}_{l}  (P_\alpha)^j_k (V_\beta)^{k}_j   \nonumber\\
& + T^{LS} (\overline B_\gamma)_i  (H)^{li}_{l}  (P_\alpha)_j^j (V_\beta)_{k}^{k} + T^{QA} (\overline B_\gamma)_i  (H)^{il}_{l}  (P_\alpha)^j_k (V_\beta)^{k}_j  + T^{QS} (\overline B_\gamma)_i (H)^{il}_{l}  (P_\alpha)_j^j (V_\beta)_{k}^{k}
\nonumber\\&+T^{LP}_{Pc} (\overline B_\gamma)_i (H)^{ck}_{c} (P_\alpha)^{i}_j  (V_\beta)^j_k  +T^{LP}_{Vc} (\overline B_\gamma)_i (H)^{ck}_{c} (V_\beta)^{i}_j   (P_\alpha)^j_k  + T^{LC}_{Pc} (\overline B_\gamma)_i  (H)^{cj}_{c} (P_\alpha)^{i}_j  (V_\beta)^k_k\nonumber\\
&+ T^{LC}_{Vc} (\overline B_\gamma)_i (H)^{cj}_{c} (V_\beta)^{i}_j  (P_\alpha)^k_k+T^{QP}_{Pc} (\overline B_\gamma)_i  (H)^{kc}_{c}(P_\alpha)^{i}_j   (V_\beta)^j_k +T^{QP}_{Vc} (\overline B_\gamma)_i (H)^{kc}_{c} (V_\beta)^{i}_j   (P_\alpha)^j_k   \nonumber\\
&+ T^{QC}_{Pc} (\overline B_\gamma)_i (H)^{jc}_{c} (P_\alpha)^{i}_j  (V_\beta)^k_k + T^{QC}_{Vc} (\overline B_\gamma)_i (H)^{jc}_{c}  (V_\beta)^{i}_j   (P_\alpha)^k_k + T^{LA}_c (\overline B_\gamma)_i  (H)^{ci}_{c}  (P_\alpha)^j_k (V_\beta)^{k}_j   \nonumber\\
& + T^{LS}_c (\overline B_\gamma)_i  (H)^{ci}_{c}  (P_\alpha)_j^j (V_\beta)_{k}^{k} + T^{QA}_c (\overline B_\gamma)_i  (H)^{ic}_{c}  (P_\alpha)^j_k (V_\beta)^{k}_j  \nonumber\\&+ T^{QS}_c (\overline B_\gamma)_i (H)^{ic}_{c}  (P_\alpha)_j^j (V_\beta)_{k}^{k}.
\end{align}
The $SU(3)$ irreducible amplitude of the $\overline B\to PV$ decay is
\begin{align}
&~~~~~~~~~~{\cal A}^{\rm IRA}_{\overline B_\gamma \to P_\alpha V_\beta} =\nonumber\\&~~~~
a_6(\overline B_\gamma)_i (H(\overline 6))^{ji}_k(P_\alpha)_j^l(V_\beta)_l^k +d_6(\overline B_\gamma)_i (H(\overline 6))^{ji}_k(V_\beta)_j^l(P_\alpha)_l^k+ b_6(\overline B_\gamma)_i (H(\overline 6))^{ji}_k(P_\alpha)_j^k(V_\beta)^l_l \nonumber\\ & + e_6(\overline B_\gamma)_i (H(\overline 6))^{ji}_k(V_\beta)_j^k(P_\alpha)^l_l +c_6(\overline B_\gamma)_i (H(\overline 6))^{lj}_k(P_\alpha)_j^i(V_\beta)_l^k +f_6(\overline B_\gamma)_i (H(\overline 6))^{lj}_k(V_\beta)_j^i(P_\alpha)_l^k \nonumber\\
  & + a_{15}(\overline B_\gamma)_i (H({15}))^{ij}_k(P_\alpha)_j^l(V_\beta)_l^k+ d_{15}(\overline B_\gamma)_i (H({15}))^{ij}_k(V_\beta)_j^l(P_\alpha)_l^k+ b_{15}(\overline B_\gamma)_i (H({15}))^{ij}_k(P_\alpha)_j^k(V_\beta)^l_l\nonumber\\
  & +e_{15}(\overline B_\gamma)_i (H({15}))^{ij}_k(V_\beta)_j^k(P_\alpha)^l_l  + c_{15}(\overline B_\gamma)_i (H({15}))^{jl}_k(P_\alpha)_j^i(V_\beta)_l^k + f_{15}(\overline B_\gamma)_i (H({15}))^{jl}_k(V_\beta)_j^i(P_\alpha)_l^k\nonumber\\ &
+a_3 (\overline B_\gamma)_i (H(3_p))^i (P_\alpha)^k_j(V_\beta)^j_k +b_3 (\overline B_\gamma)_i (H(3_p))^i (P_\alpha)_k^k(V_\beta)_j^j+c_{3} (\overline B_\gamma)_i (H(3_p))^k (P_\alpha)^i_k(V_\beta)_j^j\nonumber\\
  &+e_{3} (\overline B_\gamma)_i (H(3_p))^k (V_\beta)^i_k(P_\alpha)_j^j+d_3 (\overline B_\gamma)_i (H(3_p))^k (P_\alpha)^i_j(V_\beta)^j_k +f_3 (\overline B_\gamma)_i (H(3_p))^k (V_\beta)^i_j(P_\alpha)^j_k
\nonumber\\  &+a_3^\prime (\overline B_\gamma)_i (H(3_t))^i (P_\alpha)^k_j(V_\beta)^j_k +b_3^\prime (\overline B_\gamma)_i (H(3_t))^i (P_\alpha)_k^k(V_\beta)_j^j+c_{3}^{\prime} (\overline B_\gamma)_i (H(3_t))^k (P_\alpha)^i_k(V_\beta)_j^j
\nonumber\\  & +e_{3}^{\prime} (\overline B_\gamma)_i (H(3_t))^k (V_\beta)^i_k(P_\alpha)_j^j +d_3^\prime (\overline B_\gamma)_i (H(3_t))^k (P_\alpha)^i_j(V_\beta)^j_k +f_3^\prime (\overline B_\gamma)_i (H(3_t))^k (V_\beta)^i_j(P_\alpha)^j_k\nonumber\\ &
+a_3^{\prime\prime} (\overline B_\gamma)_i (H(3_p))^i (P_\alpha)^k_j(V_\beta)^j_k +b_3^{\prime\prime} (\overline B_\gamma)_i (H(3_p))^i (P_\alpha)_k^k(V_\beta)_j^j+c_{3}^{\prime\prime} (\overline B_\gamma)_i (H(3_p))^k (P_\alpha)^i_k(V_\beta)_j^j\nonumber\\
  &+e_{3}^{\prime\prime} (\overline B_\gamma)_i (H(3_p))^k (V_\beta)^i_k(P_\alpha)_j^j+d_3^{\prime\prime} (\overline B_\gamma)_i (H(3_p))^k (P_\alpha)^i_j(V_\beta)^j_k +f_3^{\prime\prime} (\overline B_\gamma)_i (H(3_p))^k (V_\beta)^i_j(P_\alpha)^j_k
\nonumber\\  &+a_3^{\prime\prime\prime} (\overline B_\gamma)_i (H(3_t))^i (P_\alpha)^k_j(V_\beta)^j_k +b_3^{\prime\prime\prime} (\overline B_\gamma)_i (H(3_t))^i (P_\alpha)_k^k(V_\beta)_j^j+c_{3}^{\prime\prime\prime} (\overline B_\gamma)_i (H(3_t))^k (P_\alpha)^i_k(V_\beta)_j^j
\nonumber\\  & +e_{3}^{\prime\prime\prime} (\overline B_\gamma)_i (H(3_t))^k (V_\beta)^i_k(P_\alpha)_j^j +d_3^{\prime\prime\prime} (\overline B_\gamma)_i (H(3_t))^k (P_\alpha)^i_j(V_\beta)^j_k \nonumber\\ & +f_3^{\prime\prime\prime} (\overline B_\gamma)_i (H(3_t))^k (V_\beta)^i_j(P_\alpha)^j_k.
\end{align}
The relations between topological amplitudes and the irreducible amplitudes are
\begin{align}
 a_6&  =\frac{E_V-A_V}{4},  \qquad b_6 = \frac{T^{ES}_P-T^{AS}_P}{4},  \qquad c_6 = \frac{-T_P+C_P}{4}, \qquad
 d_6  =\frac{E_P-A_P}{4}, \nonumber\\ \ e_6 & = \frac{T^{ES}_V-T^{AS}_V}{4},  \qquad f_6 = \frac{-T_V+C_V}{4}, \qquad
  a_{15}  =\frac{E_V+A_V}{8},  \qquad b_{15} = \frac{T^{ES}_P+T^{AS}_P}{8}, \nonumber\\ c_{15}& = \frac{T_P+C_P}{8},\qquad
  d_{15}  =\frac{E_P+A_P}{8},  \qquad e_{15} = \frac{T^{ES}_V+T^{AS}_V}{8},  \qquad f_{15} = \frac{T_V+C_V}{8},\nonumber\\
 a^\prime_3& = \frac{3}{8}E_P-\frac{1}{8}A_P+\frac{3}{8}E_V-\frac{1}{8}A_V+T^{LA},\qquad
 a_3 = -\frac{1}{8}E_P+\frac{3}{8}A_P-\frac{1}{8}E_V+\frac{3}{8}A_V+ T^{QA},\nonumber\\
  b^\prime_3 & = \frac{3}{8}T^{ES}_P-\frac{1}{8}T^{AS}_P+\frac{3}{8}T^{ES}_V-\frac{1}{8}T^{AS}_V+T^{LS},\qquad
 b_3 = -\frac{1}{8}T^{ES}_P+\frac{3}{8}T^{AS}_P-\frac{1}{8}T^{ES}_V+\frac{3}{8}T^{AS}_V + T^{QS},\nonumber\\
  c^\prime_3 & = -\frac{1}{8}T_P+\frac{3}{8}C_P-\frac{1}{8}T^{ES}_P + \frac{3}{8}T^{AS}_P+T^{LC}_P,\qquad
  c_3  = \frac{3}{8}T_P-\frac{1}{8}C_P+\frac{3}{8}T^{ES}_P -\frac{1}{8}T^{AS}_P+T^{QC}_P,\nonumber\\
   d^\prime_3 & = \frac{3}{8}T_P-\frac{1}{8}C_P-\frac{1}{8}E_P + \frac{3}{8}A_P+T^{LP}_P,\qquad
   d_3  = -\frac{1}{8}T_P+\frac{3}{8}C_P+\frac{3}{8}E_P - \frac{1}{8}A_P+T^{QP}_P,\nonumber\\
   e^\prime_3 & = -\frac{1}{8}T_V+\frac{3}{8}C_V-\frac{1}{8}T^{ES}_V + \frac{3}{8}T_V^{AS}+T^{LC}_V,\qquad
   e_3  = \frac{3}{8}T_V-\frac{1}{8}C_V+\frac{3}{8}T^{ES}_V - \frac{1}{8}T_V^{AS}+T^{QC}_V,\nonumber\\
   f^\prime_3 & = \frac{3}{8}T_V-\frac{1}{8}C_V-\frac{1}{8}E_V + \frac{3}{8}A_V+T^{LP}_V,\qquad
    f_3  = -\frac{1}{8}T_V+\frac{3}{8}C_V+\frac{3}{8}E_V - \frac{1}{8}A_V+T^{QP}_V,  \nonumber\\ a_3^{\prime\prime} & = T^{LA}_c,\qquad  b_3^{\prime\prime} = T^{LS}_c, \qquad  c_3^{\prime\prime} = T^{LC}_{Pc}, \qquad d_3^{\prime\prime} = T^{LP}_{Pc},\qquad  e_3^{\prime\prime} = T^{LC}_{Vc}, \qquad f_3^{\prime\prime} = T^{LP}_{Vc},\nonumber\\
   a_3^{\prime\prime\prime} & = T^{QA}_c,\qquad  b_3^{\prime\prime\prime} = T^{QS}_c, \qquad  c_3^{\prime\prime\prime} = T^{QC}_{Pc}, \qquad d_3^{\prime\prime\prime} = T^{QP}_{Pc}, \qquad  e_3^{\prime\prime\prime} = T^{QC}_{Vc}, \qquad f_3^{\prime\prime\prime} = T^{QP}_{Vc}.
\end{align}

Similarly to the $\overline B\to PP$ decays, all the penguin-operator-induced amplitudes of the $\overline B\to PV$ modes in the SM can be absorbed into six parameters in both IRA and TDA approaches with following redefinitions.\\
IRA ($b\to d$):
\begin{align}\label{cc1}
 & a_3  = -\frac{\lambda_u}{\lambda_t}a_3^\prime-\frac{\lambda_c}{\lambda_t}a_3^{\prime\prime}+Pa_3^\prime + 3Pa_3+Pa^{\prime\prime\prime}_3, \qquad  b_3  = -\frac{\lambda_u}{\lambda_t}b_3^\prime-\frac{\lambda_c}{\lambda_t}b_3^{\prime\prime}+Pb_3^\prime + 3Pb_3+Pb^{\prime\prime\prime}_3, \nonumber\\ &   c_3  = -\frac{\lambda_u}{\lambda_t}c_3^\prime-\frac{\lambda_c}{\lambda_t}c_3^{\prime\prime}+Pc_3^\prime + 3Pc_3+Pc^{\prime\prime\prime}_3, \qquad d_3  = -\frac{\lambda_u}{\lambda_t}d_3^\prime-\frac{\lambda_c}{\lambda_t}d_3^{\prime\prime}+Pd_3^\prime + 3Pd_3+Pd^{\prime\prime\prime}_3,\nonumber\\ &   e_3  = -\frac{\lambda_u}{\lambda_t}e_3^\prime-\frac{\lambda_c}{\lambda_t}e_3^{\prime\prime}+Pe_3^\prime + 3Pe_3+Pe^{\prime\prime\prime}_3, \qquad f_3  = -\frac{\lambda_u}{\lambda_t}f_3^\prime-\frac{\lambda_c}{\lambda_t}f_3^{\prime\prime}+Pf_3^\prime + 3Pf_3+Pf^{\prime\prime\prime}_3.
\end{align}
TDA ($b\to d$):
\begin{align}\label{cc2}
 & A^{LA}  = -\frac{\lambda_u}{\lambda_t}T^{LA}-\frac{\lambda_c}{\lambda_t}T^{LA}_c
 +PA_P+PA_V+P^{LA}+3P^{QA}+P^{QA}_c, \nonumber\\ &   A^{LS}  =  -\frac{\lambda_u}{\lambda_t}T^{LS}-\frac{\lambda_c}{\lambda_t}T^{LS}_c
 +P^{AS}_P+P^{AS}_V+P^{LS}+3P^{QS} + +P^{QS}_c, \nonumber\\ &  A^{LC}_P  = -\frac{\lambda_u}{\lambda_t}T^{LC}_P-\frac{\lambda_c}{\lambda_t}T^{LC}_{Pc}
 +PT_P+P^{ES}_P+P^{LC}_P+3P^{QC}_P+P^{QC}_{Pc}, \nonumber\\ &  A^{LP}_P  =  -\frac{\lambda_u}{\lambda_t}T^{LP}_P-\frac{\lambda_c}{\lambda_t}T^{LP}_{Pc}
 +PC_P+PE_P+P^{LP}_P+3P^{QP}_P+P^{QP}_{Pc},\nonumber\\ &  A^{LC}_V  = -\frac{\lambda_u}{\lambda_t}T^{LC}_V-\frac{\lambda_c}{\lambda_t}T^{LC}_{Vc}
 +PT_V+P^{ES}_V+P^{LC}_V+3P^{QC}_V+P^{QC}_{Vc}, \nonumber\\ &  A^{LP}_V  =  -\frac{\lambda_u}{\lambda_t}T^{LP}_V-\frac{\lambda_c}{\lambda_t}T^{LP}_{Vc}
 +PC_V+PE_V+P^{LP}_V+3P^{QP}_V+P^{QP}_{Vc}.
\end{align}
For $\Delta S =-1$ transition, $\lambda_{u,c,t}$ in Eqs.~\eqref{cc1} and \eqref{cc2} are replaced by  $\lambda^\prime_{u,c,t}$.
The tree- and penguin-operator-induced amplitudes of all the $\overline B\to PV$ modes are listed in Tables.~\ref{tab:bpvd1}$\sim$\ref{tab:bpvs2}.

\end{appendix}

\end{document}